\IfFileExists{venue_preamble.tex}{\def\VenueMode{1}
  \input{venue_preamble}
}{\documentclass[11pt]{article}
  \usepackage[margin=1in]{geometry}
}

\usepackage[T1]{fontenc}
\usepackage{microtype}
\usepackage{amsmath}
\usepackage{amssymb}
\usepackage{amsthm}
\usepackage{aliascnt}
\usepackage{mathtools}
\usepackage{graphicx}
\usepackage{booktabs}
\usepackage{array}
\usepackage{xcolor}
\usepackage{tikz}
\usetikzlibrary{arrows.meta,backgrounds,decorations.markings,decorations.pathreplacing}
\definecolor{DPBlue}{HTML}{0072B2}
\definecolor{DPOrange}{HTML}{D55E00}
\definecolor{DPPurple}{HTML}{7057A3}
\tikzset{
  dp one/.style={draw=DPBlue,line width=1.1pt},
  dp two/.style={draw=DPOrange,line width=1.1pt},
  dp edge/.style={draw=DPPurple,line width=1.6pt,-{Stealth[length=4pt]}},
  dp arrow/.style={postaction={decorate},decoration={markings,mark=at position 0.58 with {\arrow{Stealth[length=4pt]}}}},
  dp label/.style={inner sep=1.5pt,font=\normalsize}
}
\usepackage{algorithm}
\usepackage{algpseudocode}
\algrenewcommand\algorithmicindent{1.3em}
\algrenewcommand\algorithmiccomment[1]{\hfill$\triangleright$ #1}
\usepackage{hyperref}
\usepackage[nameinlink,capitalize,noabbrev]{cleveref}

\ifdefined\VenueMode
\else
  \hypersetup{
    colorlinks=true,
    citecolor=blue,
    linkcolor=blue,
    urlcolor=blue
  }
\fi

\graphicspath{{figures/}}
\allowdisplaybreaks

\newcommand{\OPT}{\mathrm{OPT}}
\newcommand{\UB}{\mathrm{UB}}

\newcommand{\Rtree}{\mathcal{T}}

\DeclarePairedDelimiter{\abs}{\lvert}{\rvert}
\DeclarePairedDelimiter{\ceil}{\lceil}{\rceil}
\DeclarePairedDelimiter{\floor}{\lfloor}{\rfloor}

\makeatletter
\newcommand{\manualcitation}[2]{\begingroup
  \def\@cite##1##2{[#1]}#2\endgroup}
\makeatother

\theoremstyle{plain}
\newtheorem{theorem}{Theorem}[section]
\newaliascnt{lemma}{theorem}
\newtheorem{lemma}[lemma]{Lemma}
\aliascntresetthe{lemma}
\newaliascnt{proposition}{theorem}
\newtheorem{proposition}[proposition]{Proposition}
\aliascntresetthe{proposition}
\newaliascnt{corollary}{theorem}
\newtheorem{corollary}[corollary]{Corollary}
\aliascntresetthe{corollary}

\theoremstyle{definition}
\newaliascnt{definition}{theorem}
\newtheorem{definition}[definition]{Definition}
\aliascntresetthe{definition}

\theoremstyle{remark}
\newaliascnt{remark}{theorem}
\newtheorem{remark}[remark]{Remark}
\aliascntresetthe{remark}
\newcommand{\lineref}[1]{line~\ref{#1}}
\newcommand{\Lineref}[1]{Line~\ref{#1}}
\newcommand{\linerange}[2]{lines~\ref{#1}--\ref{#2}}
\newcommand{\Linerange}[2]{Lines~\ref{#1}--\ref{#2}}

\algblockdefx[DATASTRUCTURE]{DataStructure}{EndDataStructure}[1]{\textbf{data structure} \textsc{#1}}{\textbf{end data structure}}
\algblockdefx[MEMBERS]{Members}{EndMembers}{\textbf{members}}{\textbf{end members}}
\newcommand{\PhaseComment}[1]{{\color{DPBlue}/* #1 */}}

\newlength{\figx}
\colorlet{figh1}{DPBlue!60}
\colorlet{figh2}{DPBlue!36}
\colorlet{figh4}{DPBlue!15}
\colorlet{figh8}{black!9}
\tikzset{
  figtxt/.style={font=\scriptsize,inner sep=1pt},
  figlab/.style={figtxt,anchor=east,text=black!75},
  figtitle/.style={font=\footnotesize,inner sep=0pt,anchor=south west,align=left},
  kblk/.style={draw=black!55,line width=0.5pt,rounded corners=1.5pt},
  kblk retire/.style={draw=black,line width=1.1pt,rounded corners=1.5pt,dash pattern=on 2.2pt off 1.3pt},
  kblk add/.style={draw=DPOrange,line width=1.3pt,rounded corners=1.5pt},
  sdot/.style={circle,fill=black!65,inner sep=0pt,minimum size=4.6pt},
  tdot/.style={circle,fill=DPOrange,inner sep=0pt,minimum size=5.4pt},
  tcol/.style={draw=DPOrange,line width=0.6pt},
  knode/.style={circle,inner sep=0pt,minimum size=10.2pt,font=\scriptsize,draw=black!50,line width=0.4pt},
  ksep/.style={rectangle,inner sep=0pt,minimum size=9.6pt,font=\scriptsize,draw=black,line width=1pt},
  kcur/.style={draw=DPOrange,line width=1.4pt},
  inedge/.style={draw=black!45,line width=0.5pt},
  sepedge/.style={draw=black,line width=1.2pt},
  spath/.style={draw=DPOrange,line width=1.6pt},
}
\newcommand{\figkeyaxis}[1]{\draw[black!45,line width=0.3pt] (0.5,#1) -- (32.5,#1);
  \foreach \k in {1,...,32}{\draw[black!45,line width=0.3pt] (\k,#1) -- (\k,#1-0.14);}}
\newcommand{\figkeynums}[1]{\foreach \k in {1,8,16,24,32}{\node[font=\tiny,text=black!60] at (\k,#1) {$\k$};}}
\newcommand{\figregion}[4]{\fill[figh#3] (#1-0.46,#4-0.3) rectangle (#2+0.46,#4+0.3);
  \node[figtxt,below] at ({(#1+#2)/2},#4-0.34) {$#1$--$#2$};}
\newcommand{\figcursor}[2]{\fill[DPOrange] (#1,#2) -- (#1-0.36,#2+0.5) -- (#1+0.36,#2+0.5) -- cycle;}

\newcommand{\PaperTitle}{A Binary Search Tree with the Unified Bound}
\newcommand{\AuthorCao}{Yang Cao}
\newcommand{\AuthorQi}{Haoran Qi}
\newcommand{\AuthorWang}{Hanzhi Wang}
\newcommand{\PaperAuthors}{\AuthorCao\thanks{\texttt{yangcao@psu.edu}. Pennsylvania State University.} \and \AuthorQi\thanks{Jilin University.} \and \AuthorWang\thanks{Tsinghua University.}}
\newcommand{\PaperPdfAuthors}{\AuthorCao, \AuthorQi, \AuthorWang}
\newcommand{\PaperDate}{}

\newif\ifPaperPrintReferences
\PaperPrintReferencestrue

\title{\PaperTitle}
\author{\PaperAuthors}
\date{\PaperDate}
\hypersetup{pdftitle={\PaperTitle},pdfauthor={\PaperPdfAuthors}}

\providecommand{\PaperMakeTitle}{\maketitle}
\providecommand{\PaperPrintBibliography}{\ifPaperPrintReferences
    \bibliographystyle{alpha}
    \bibliography{references}
  \fi}
\providecommand{\PaperEndMatter}{}

\begin{document}

\PaperMakeTitle

\begin{abstract}
We give a deterministic online binary search tree (BST) that satisfies the unified bound of Iacono [SODA 2001], which subsumes the working-set and dynamic-finger bounds: it serves every sequence $X$ of $m$ searches on $n$ keys at total cost $O(m+\mathrm{UB}(X))$, and every single search at cost $O(\log n)$. Previously, no BST was known to satisfy the unified bound, even offline. As a consequence, the dynamic optimality conjecture implies Iacono's conjecture that splay trees satisfy the unified bound. We also show that the unified bound is $O((m+n)(1+\log c_\sigma))$ on every sequence of searches that avoids a fixed permutation $\sigma$, where $c_\sigma$ is the Füredi--Hajnal limit of $\sigma$. Hence every BST that satisfies the unified bound, including ours, serves such sequences at this cost. For permutations this bound is tight, and it improves the previous upper bound $O(nc_\sigma^2)$ on the optimal cost, due to Berendsohn, Kozma and Opler [STOC 2024]. Finally, we extend the tree to support insertions and deletions: a sequence $X$ of $m$ operations, starting from $n_0$ keys, costs $O(n_0+m+\mathrm{UB}(X))$ in total, where $\mathrm{UB}$ is now the extension of the unified bound to insertions and deletions by Bădoiu, Cole, Demaine and Iacono [TCS 2007]. Hence, if splay trees are dynamically optimal for searches, insertions and deletions, then they serve every such sequence at this total cost as well.

\end{abstract}

\clearpage

\section{Introduction}
\label{sec:intro}

A binary search tree (BST) stores the keys $1,\dots,n$ and serves a sequence of searches $X=(x_1,\dots,x_m)$. In the standard model of dynamic BSTs \cite{st85,wil89}, which we state in the form of \cite{dhi+09} that is equivalent up to a constant factor, each search starts at the root, walks down to the searched key, and may rearrange the nodes it has visited by rotations, and its cost is the number of nodes it touches. The cost of the optimal BST, $\OPT(X)$, is the least total cost of serving $X$ when the whole sequence is known in advance. Sleator and Tarjan conjectured that splay trees, the BSTs that move each searched key to the root by rotations, serve every sequence at cost $O(\OPT(X))$ \cite{st85}. This dynamic optimality conjecture is still open, and much of the work around it asks which regularities of a search sequence an optimal tree can exploit. Each such regularity is expressed by a formula in $X$ that bounds $\OPT(X)$ from above.

Two regularities are classical. The \emph{working-set bound} charges a search about the logarithm of the number of distinct keys searched since the previous search for the same key (a first search for a key is charged about $\log_2n$). The \emph{dynamic-finger bound} charges a search $\log_2(\abs{x_t-x_{t-1}}+2)$. Splay trees satisfy both, the second by a long proof of Cole \cite{col00}, which builds on a special case proved by Cole, Mishra, Schmidt and Siegel \cite{cmss00}, so the optimum satisfies both. Iacono \cite{iac01} proposed a single bound that is at most both of them up to an additive $O(m)$.

To state it, write $\rho(u,t)$, for times $u<t$, for the number of distinct keys among $x_{u+1},\dots,x_{t-1}$. The \emph{unified bound} of $X$ is
\begin{align*}
\UB(X)\coloneqq\log_2(n+2)+\sum_{t=2}^{m}\min_{1\le u<t}\log_2\bigl(\abs{x_t-x_u}+\rho(u,t)+2\bigr).
\end{align*}
Choosing $u=t-1$ recovers the dynamic-finger term, and choosing $u$ to be the previous search for the key $x_t$ recovers the working-set term. Iacono's definition, in the form of \cite{bcdi07}, minimizes over keys instead of times, and the two definitions agree up to an additive $O(m)$ (\cref{rem:forms}). The minimum lets each search use whichever earlier search is closest. Consider two scans that move rightwards through two distant parts of the key space, one step at a time, with their searches interleaved. Every key is searched once and consecutive searches are far apart, so the working-set and dynamic-finger bounds are both about $m\log n$. Yet each search is adjacent to the search two steps earlier, so the unified bound is $O(m)$.

This paper answers the following question, which underlies the conjecture of Iacono that splay trees satisfy the unified bound \cite{iac01} and is stated as open in \cite{cgj+24}: \emph{is there a BST that satisfies the unified bound, that is, serves every sequence $X$ at cost $O(m+n+\UB(X))$?} Iacono stated his conjecture for searches in a static set of keys, as here. Bădoiu, Cole, Demaine and Iacono later stated it for searches, insertions and deletions \cite{bcdi07}; we return to this form below. A weaker, offline form of the question, whether the optimal BST satisfies the unified bound, was also open \cite{cgk+16,cgk+18}. Since every BST costs at least $\OPT(X)$, a negative answer to the offline question would rule out the unified bound for every BST.

Iacono \cite{iac01} gave a static comparison-based structure that serves $X$ in time $O(m+\UB(X))$, and Bădoiu and Demaine \cite{bd04} a simpler one that also supports insertions, and deletions at an additional cost of $O(\log\log n)$. The journal version \cite{bcdi07} supports searches, insertions and deletions within the unified bound. These structures store the keys in several trees and are not BSTs. In the BST model all known bounds pay an overhead. Skip-splay trees \cite{ds09}, the layered working-set trees of Bose, Douïeb, Dujmović and Howat combined with skip-splay \cite{bddh12}, the combination of trees of Demaine, Iacono, Langerman and Özkan \cite{dilo13}, and the group-access trees of Chalermsook, Gupta, Jiamjitrak, Pareek and Yingchareonthawornchai \cite{cgj+24} cost $O(\UB(X)+m\log\log n)$, with an additional $O(n\log n)$ in the case of \cite{dilo13}. Derryberry's thesis \cite{der09} proposed the cache-splay tree and claimed that it achieves the unified bound. Demaine, Iacono, Langerman and Özkan reported that this claim was withdrawn \cite[footnote~4 of the arXiv version]{dilo13}, see also \cite[Section~1]{him13}, and we know of no published account of the gap. Chalermsook, Goswami, Kozma, Mehlhorn and Saranurak \cite{cgk+18} proved that $\OPT(X)$ is at most a factor times $\UB_\ell(X)$ for a time-windowed bound $\UB_\ell$, in which only the last $\ell$ searches may serve as $u$, with a factor $O(\ell!\cdot\ell\log\ell)$, and \cite{cgj+24} obtained $O(\UB_\ell(X)+m\log\ell\log\log n)$ in expectation, with a randomized tree and for sufficiently long sequences.

The known BSTs lose in similar ways. Skip-splay trees cut the keys into about $\log\log n$ levels of nested blocks whose sizes grow doubly exponentially, and a search can pass through all of these levels, paying a constant at each, even when it is adjacent to the previous search. The multi-finger approach of \cite{cgk+18} lets a BST simulate $k$ pointers that move in the tree, called fingers, at a cost of $O(1+\log k)$ per finger move, and its bound for a window of the last $\ell$ searches uses up to $\ell!$ fingers, so covering every earlier search would require the number of fingers to grow with the sequence. In both cases the overhead comes from serving all scales, or all fingers, uniformly, although a given search needs only one of them.

\paragraph{Our results.} We answer the question positively, with a deterministic online BST. An \emph{online} BST decides how to serve each search knowing only the searches so far.

\begin{theorem}[Main theorem; see \cref{thm:scales,thm:lambda-ub}]
\label{thm:main}
There is a deterministic online BST algorithm that, starting from a fixed tree on the keys $1,\dots,n$, serves every sequence $X$ of $m$ searches at total cost
\begin{align*}
O\bigl(m+\UB(X)\bigr),
\end{align*}
and every single search at cost $O(\log n)$.
\end{theorem}

The bound has no additive term $n$: the initial tree is a fixed balanced tree, and the first search is paid for by the term $\log_2(n+2)$ of the unified bound. If the initial tree is prescribed, the first search can rebuild it at cost $n$, which gives $O(m+n+\UB(X))$, the form of the question above. The bound on a single search is optimal up to a constant factor, since every BST on $n$ keys has a key of depth at least $\log_2(n+1)-1$. Both bounds also hold for the running time on a word RAM, after an initialization of $O(n)$ time (\cref{thm:runtime}). Since $\OPT(X)$ is at most the cost of any execution on $X$, the offline question is answered as well.

\begin{corollary}
\label{cor:opt}
For every sequence $X$ of $m$ searches over the keys $1,\dots,n$, $\OPT(X)=O(m+\UB(X))$.
\end{corollary}

The difficulty is that a single tree must keep the neighbourhood of every recent search near its root, at every scale at once. Cut the keys into blocks of $2^s$ consecutive keys at every \emph{scale} $s$, and count a block as searched whenever one of its keys is searched. The \emph{locality scale} of a search is the least scale $s$ such that the block of scale $s$ that contains its key was searched before and fewer than about $2^s$ distinct keys were searched since, and in total the unified bound charges the searches about these scales (\cref{sec:overview-unified}). Say that a block of scale $s$ is in use while it is among the $2^s$ blocks of its scale that were searched most recently. At any single moment a tree in which the keys carry weights, and heavier keys are nearer the root, can place the keys of every block of scale $s$ that is in use within depth $O(s)$. But blocks keep coming into use and falling out of use, and moving a block of scale $s$ costs about $s$. A search can bring a block into use at every scale below its own, so a tree that tracks every scale pays about the square of what the bound allows (\cref{sec:overview-moment}).

The way out is that a tree need not track every scale. Our tree tracks blocks only at the scales $1,2,4,8,\dots$: at such a scale $s$, it \emph{keeps} the $2^s$ blocks of $2^s$ keys that were searched most recently, each with a weight that puts its keys near the root. The scales grow geometrically, so the blocks that a search brings in cost only about the largest of them, and a search with locality scale $s$ still finds a kept block of scale less than $2s$ around its key, namely at the least power of two that is at least $s$ (\cref{sec:overview-count-promoted}). The sizes $2,4,16,256,\dots$ of the blocks at these scales grow doubly exponentially, like those of the levels of skip-splay trees \cite{ds09}, and keeping at each level the blocks that were searched most recently is the rule of the cache-splay tree \cite{der09}. What is new is the tree that holds the blocks and the analysis. Every kept block has a weight that depends only on its size: a kept block of scale $s$ has the weight $4^{-s}$, whatever larger blocks contain it. At most $2^s$ blocks of scale $s$ are kept, so all weights together are $O(1)$, and a search tree for weighted items can place an item of weight $w$ at depth $O(\log(1/w))$. So a search reaches a kept block of scale $s$ at depth $O(s)$. It does not pass through the levels above the block, as a search in a skip-splay tree does. The tree that holds the weighted blocks is a biased search tree of Bent, Sleator and Tarjan \cite{bst85}, a balanced search tree for weighted items in which heavy items are near the root. In it a block of scale $s$ is inserted and removed in $O(s)$ time in the worst case, so no step is amortized over the tree. A search replaces at most one block at each scale, the least recently searched block by its own, and only at scales below its locality scale. So a search with locality scale $s$ costs $O(s)$ in the worst case, and the only amortized step of the proof is the comparison of the locality scales with the unified bound.

\paragraph{Pattern-avoiding sequences.} A search sequence \emph{avoids} a permutation $\sigma$ of $\{1,\dots,k\}$ if no $k$ of its searches have distinct keys in the relative order given by $\sigma$. It was conjectured that $\OPT(X)=O(n)$ for every permutation $X$ of $\{1,\dots,n\}$ that avoids a fixed pattern $\sigma$; Berendsohn, Kozma and Opler \cite{bko24} attribute the conjecture to \cite{cgk+15}. Berendsohn, Kozma and Opler proved it \cite[Section~3]{bko24}, with the bound $\OPT(X)=O(n c_\sigma^2)$. Here $c_\sigma$ is the Füredi--Hajnal limit of $\sigma$, the least constant such that every $0$-$1$ matrix with $M_1$ rows, $M_2$ columns and more than $c_\sigma\max(M_1,M_2)$ ones has $k$ ones, in $k$ different rows and $k$ different columns, that lie relative to one another as the ones of the permutation matrix of $\sigma$; it is finite by the theorem of Marcus and Tardos \cite{mt04}. They also noted a lower bound of $\Omega(n\log c_\sigma)$ for some $\sigma$-avoiding $X$, by counting the $\sigma$-avoiding permutations of $\{1,\dots,n\}$, whose number grows like $L^n$ for a constant $L$ that is polynomially related to $c_\sigma$ \cite{cib09,ck17}, and they called sharpening the dependence on $c_\sigma$ a challenging open question. We close the gap.

\begin{theorem}[see \cref{prop:ub-patterns}]
\label{thm:patterns}
There is a deterministic online BST algorithm on the keys $1,\dots,n$ that, for every permutation $\sigma$ of length at least $2$, serves every sequence $X$ of $m$ searches over $\{1,\dots,n\}$ that avoids $\sigma$ at cost
\begin{align*}
O\bigl((m+n)(1+\log c_\sigma)\bigr).
\end{align*}
The algorithm does not depend on $\sigma$, and the constant in the $O$-notation does not depend on $\sigma$.
\end{theorem}

For a permutation $X$, where $m=n$, this gives $\OPT(X)=O(n(1+\log c_\sigma))$, which matches the lower bound of \cite{bko24,cib09} for the worst $\sigma$-avoiding permutations. Since $c_\sigma=2^{O(k)}$ for patterns of length $k$ \cite{fox13}, the bound is also $O((m+n)k)$. Berendsohn, Kozma and Opler asked whether bounds of this kind hold for online algorithms such as splay trees or Greedy \cite{bko24}, the algorithm of Lucas and Munro \cite{luc88,mun00} in the online form of \cite{dhi+09}, which is the other leading candidate for dynamic optimality. Before this work, the best bound for any online BST on all $\sigma$-avoiding permutations was $n\cdot2^{(1+o(1))\alpha(n)}$ for fixed $\sigma$, where $\alpha$ is the inverse Ackermann function, for Greedy \cite{cpy24}. \Cref{thm:patterns} follows from the bound in terms of scales (\cref{thm:scales}) and the theorem of Marcus and Tardos, applied at every scale $s$ to the grid of blocks of $2^s$ keys and intervals of $2^s$ consecutive searches (\cref{sec:patterns}). The same count gives $\UB(X)=O((m+n)(1+\log c_\sigma))$ for every $\sigma$-avoiding sequence (\cref{prop:ub-patterns}). So the unified bound captures pattern avoidance, which answers, for permutations, the question of \cite{cgk+15} whether some parameter subsumes both pattern avoidance and the dynamic-finger bound.

\paragraph{Insertions and deletions.} The question above concerns searches in a fixed set of keys, and this is the setting of the earlier BSTs as well; skip-splay trees, for instance, do not support insertions and deletions \cite{bddh12}. The unified bound, however, also makes sense for a set that changes. Bădoiu, Cole, Demaine and Iacono \cite{bcdi07} extended it to insertions and deletions, with the distance between two keys counted among the keys that are present. We call this the \emph{unified bound with updates}. They gave a structure that satisfies it and is not a BST, and they conjectured that splay trees satisfy it for all three operations. We extend our tree to support insertions and deletions (\cref{thm:updates}): it serves every sequence $X$ of $m$ searches, insertions and deletions, starting from a set of $n_0$ keys, at total cost $O(n_0+m+\UB(X))$, and every single operation at a cost that is logarithmic in the current number of keys. Up to the additive term $n_0+m$, the unified bound with updates therefore bounds the cost of the best BST for every sequence of searches, insertions and deletions (\cref{cor:upd-opt}). The algorithm is the same, except that the blocks of keys are no longer those of a fixed grid. They are split and merged as the set changes, which happens rarely and changes the tree only near its root.

\paragraph{Consequences for dynamic optimality.} By \cref{cor:opt}, the unified bound is a sound bound in the sense of \cite{cgk+16}: it bounds the optimum. Consequently, the dynamic optimality conjecture implies the conjecture of Iacono that splay trees satisfy the unified bound, and in the same way, if Greedy is dynamically optimal \cite{dhi+09}, then it satisfies the unified bound. Iacono noted that his own structure, built from several search trees rather than one, gives no such implication \cite{iac01}, and the same is noted in \cite{bcdi07}. The implication also holds with insertions and deletions: if splay trees are dynamically optimal for searches, insertions and deletions, then by \cref{cor:upd-opt} they serve every such sequence at total cost $O(n_0+m+\UB(X))$. For sequences that start from the empty set this is the conjecture of \cite{bcdi07}, stated for the total cost and up to the convention for keys that were never accessed (\cref{sec:updates}). \Cref{cor:opt} also transfers to every competitive BST. A BST that is $c$-competitive, that is, serves every sequence at cost $O(c(\OPT(X)+n))$, therefore costs $O(c(m+n+\UB(X)))$; this applies to Tango trees \cite{dhip07} and multi-splay trees \cite{wds06} with $c=O(\log\log n)$, and, by recent preprints, to splay trees with $c=O(\log\log n\cdot(\log\log\log n)^2)$ \cite{chh+26} and to Greedy with $c=2^{O(\sqrt{\log\log n})}$ \cite{gpsw26}. Since $\UB(X)\le\UB_\ell(X)$ for every $\ell$, \cref{thm:main} also gives $O(m+\UB_\ell(X))$ for the time-windowed bound, with no dependence on $\ell$. For the cost of a single search, a general transformation \cite{bcfl12} turns every BST algorithm into one that performs $O(\log n)$ BST operations per search at the same total cost up to a constant factor.

\paragraph{Related work.} Our structure can be seen as a multi-finger tree \cite{cgk+18} with one finger for each block that it holds, kept in a weighted tree. It differs in which fingers it keeps and how they are weighted: a finger stands for a block of keys around a recent search at one of the scales $1,2,4,\dots$, and its weight depends only on the size of the block, which is what removes the logarithmic factor (\cref{sec:overview-count}). Storing a tree as a tree of search trees, one for each piece of a decomposition, goes back to the link-cut trees of Sleator and Tarjan \cite{st83} and is the basis of multi-splay trees \cite{wds06}, where the pieces are paths of a fixed balanced tree. Our tree has two levels only: a biased search tree \cite{bst85} whose leaves are the intervals into which the kept blocks cut the keys, and a biased search tree on the keys of each interval. The cache-splay tree of \cite{der09}, whose rule for keeping blocks we use, stores each of its $\log\log n$ levels in splay trees, and its analysis is amortized, with a potential function and a random shift of the block boundaries. Our analysis is worst-case for each search, and \cref{sec:unified} compares the locality scales with the unified bound by a deterministic charging argument for the fixed grid. We do not know where the analysis of \cite{der09} fails; our proof uses no step of it. The lazy-finger bound of Bose, Douïeb, Iacono and Langerman \cite{bdil14} and the weighted dynamic-finger bound of Iacono and Langerman \cite{il16} measure the distance between consecutive searches in a static reference tree or with key weights, rather than taking the best earlier witness with a charge for recency. Outside the BST model, Rutschmann \cite{rut26} recently gave a heap that satisfies the unified bound.

For pattern-avoiding inputs, \cite{cgk+15} proved linear bounds for special classes, such as $\OPT(X)=O(n\log k)$ for $k$-decomposable sequences, the permutations that can be reduced to a single point by repeatedly replacing a block of at most $k$ consecutive values at consecutive times by one point, and Goyal and Gupta \cite{gg19} obtained the same bound for Greedy with preprocessing. For Greedy, \cite{cgk+15} proved $n\cdot2^{\alpha(n)^{O(k)}}$ on every $\sigma$-avoiding permutation, which Chalermsook, Pettie and Yingchareonthawornchai \cite{cpy24} improved to $n\cdot2^{(1+o(1))\alpha(n)}$ for fixed $\sigma$, and \cite{cgj+23} treats further classes. The proof of \cite{bko24} builds an offline execution along a sequence of pairwise merges of rectangles that cover the permutation, provided by the twin-width decomposition of Guillemot and Marx \cite{gm14}, and pays the square of the number of rectangles that interact in a merge. Our proof uses no decomposition of the permutation: it counts, at every scale, the cells of the grid of keys and time that contain a search, and lets an online BST pay for each search at its own scale. Opler \cite{opl24} gave a sorting algorithm for pattern-avoiding sequences whose dependence on the pattern matches the information-theoretic lower bound, and \cref{thm:patterns} is the corresponding statement for online BSTs.

\paragraph{Open problems.} The conjecture of Iacono that splay trees satisfy the unified bound \cite{iac01} remains open, also in its form with insertions and deletions \cite{bcdi07}, and so does the corresponding question for Greedy. By \cref{prop:ub-patterns}, a proof for either of them would also show that it is linear on pattern-avoiding sequences, with the optimal dependence on $c_\sigma$, which would answer a question of \cite{cgk+15,bko24}. The unified bound itself is not a lower bound on the optimum. As Howat, Iacono and Morin observed \cite{him13}, it measures distances in the whole key space and not among the keys that are in use. They propose the fresh-finger property, which measures distances among the keys of a working set, and whether a BST satisfies a bound of this kind is open.

\paragraph{Roadmap.} \Cref{sec:overview} gives an overview of the proof. \Cref{sec:prelim} fixes the model and states the tools we use. \Cref{sec:algorithm} states the algorithm and the bound on its cost in terms of the locality scales of the searches (\cref{thm:scales}), \cref{sec:invariants} proves the facts about blocks and regions on which the algorithm relies, and \cref{sec:analysis} bounds its cost. \Cref{sec:unified} derives \cref{thm:main} by comparing the locality scales with the unified bound. \Cref{sec:patterns} proves \cref{thm:patterns}. \Cref{sec:runtime} states that the same bounds hold for the running time on a word RAM, with the implementation and the proof in \cref{app:runtime}, and \cref{sec:updates} extends the tree to insertions and deletions, with the algorithm and the proof in \cref{app:updates}. These two sections depend on \cref{sec:algorithm,sec:invariants,sec:analysis,sec:unified} but not on each other or on \cref{sec:patterns}.

\section{Technical overview}
\label{sec:overview}

The proof of \cref{thm:main} has two parts. The main part builds an online BST in which the cost of every search is proportional to a combinatorial quantity, the locality scale of the search, which is defined from a fixed grid of blocks of keys (\cref{thm:scales}). The second part shows that the sum of the locality scales is at most twice the unified bound plus $O(m)$. A reader who takes \cref{thm:scales} on trust can go from \cref{sec:algorithm} directly to \cref{sec:unified,sec:patterns}.

\subsection{Scales and the locality scale}

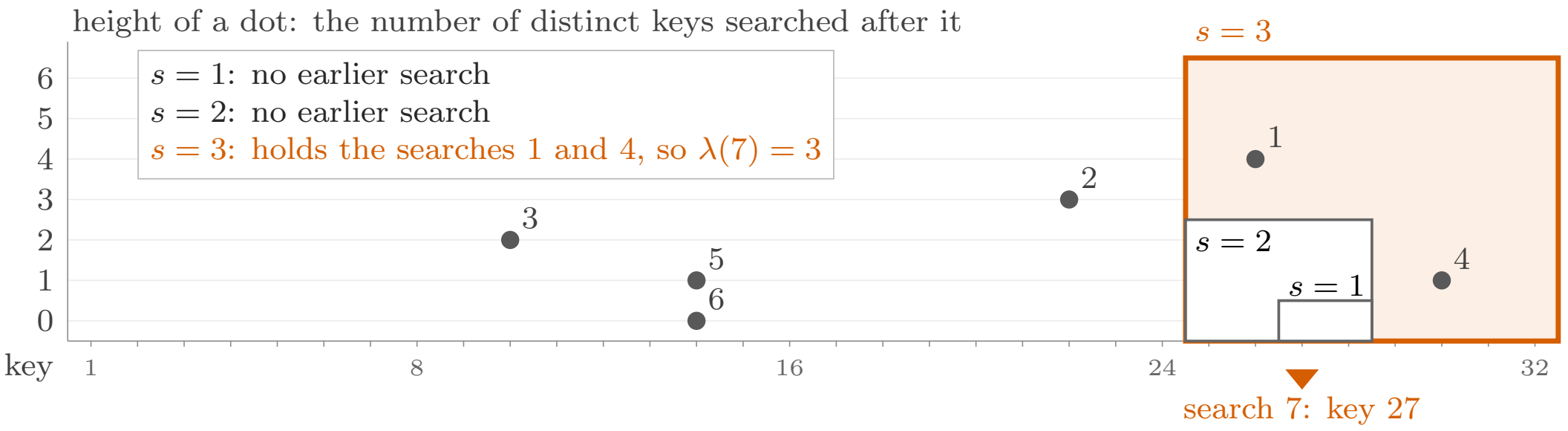
\begin{figure}[t]
\centering
\begin{tikzpicture}[x=\figx,y=0.36cm]
\path (-3.3,0) (36.3,0);
\foreach \r in {0,...,6}{\draw[black!8,line width=0.3pt] (0.5,\r) -- (32.5,\r); \node[figlab] at (0.3,\r) {$\r$};}
\draw[black!45,line width=0.3pt] (0.5,-0.5) -- (0.5,6.9);
\figkeyaxis{-0.5}\figkeynums{-1.15}
\node[figlab] at (0.3,-1.15) {key};
\node[figtxt,text=black!75,anchor=south west] at (0.5,6.9) {height of a dot: the number of distinct keys searched after it};
\draw[draw=DPOrange,line width=1.3pt,fill=DPOrange!10] (24.5,-0.5) rectangle (32.5,6.5);
\node[figtxt,anchor=south west,text=DPOrange] at (24.6,6.8) {$s=3$};
\draw[draw=black!60,line width=0.7pt,fill=white] (24.5,-0.5) rectangle (28.5,2.5);
\node[figtxt,anchor=south west] at (24.6,1.6) {$s=2$};
\draw[draw=black!60,line width=0.7pt,fill=white] (26.5,-0.5) rectangle (28.5,0.5);
\node[figtxt,anchor=south west] at (26.6,0.5) {$s=1$};
\figcursor{27}{-1.7}
\node[figtxt,text=DPOrange,anchor=north] at (27,-1.75) {search $7$: key $27$};
\node[figtxt,anchor=north west,align=left,text=black!85,fill=white,draw=black!30,line width=0.3pt,inner sep=3pt] at (2,6.7) {
  $s=1$: no earlier search\\
  $s=2$: no earlier search\\
  {\color{DPOrange}$s=3$: holds the searches $1$ and $4$, so $\lambda(7)=3$}}
  ;
\node[sdot] (p1) at (26,4) {}; \node[figtxt,text=black!75,anchor=south west] at (p1.north east) {$1$};
\node[sdot] (p2) at (22,3) {}; \node[figtxt,text=black!75,anchor=south west] at (p2.north east) {$2$};
\node[sdot] (p3) at (10,2) {}; \node[figtxt,text=black!75,anchor=south west] at (p3.north east) {$3$};
\node[sdot] (p4) at (30,1) {}; \node[figtxt,text=black!75,anchor=south west] at (p4.north east) {$4$};
\node[sdot] (p5) at (14,1) {}; \node[figtxt,text=black!75,anchor=south west] at (p5.north east) {$5$};
\node[sdot] (p6) at (14,0) {}; \node[figtxt,text=black!75,anchor=south west] at (p6.north east) {$6$};
\end{tikzpicture}
\caption{The locality scale of search $7$ in the running example of this section: seven searches on $32$ keys, for the keys $26,22,10,30,14,14,27$. Each earlier search, labelled by its time, is a dot above its key, at the number of distinct keys searched after it; search $1$, for instance, is followed by the four keys $22,10,30,14$. The keys are cut into blocks of $2^s$ consecutive keys for every $s$. The box of scale $s$ spans, horizontally, the block of $2^s$ keys that contains the searched key and, vertically, the numbers $0,\dots,2^s-2$. The locality scale is the least $s$ whose box holds a dot (\cref{def:locality}), here $\lambda(7)=3$.}
\label{fig:scales}
\end{figure}

We cut the keys into blocks. A \emph{block} of height $a$ is one of the intervals $\{j2^a+1,\dots,(j+1)2^a\}$ of $2^a$ consecutive keys. Blocks of different heights are nested. A block of scale $s$, in the words of the introduction, is a block of height $s$: we say height for a block, and scale for the number $s$ when it also counts distinct keys, as it does next. Let $K\coloneqq\ceil{\log_2n}$, and let $\hat K$ be the least power of two that is at least $K$. A \emph{scale} is an integer $s$ between $0$ and $\hat K$; the scales above $K$ are needed only in \cref{sec:overview-count-promoted}. We say that a search lies in a block, and that the block is searched, if the key of the search is in the block. At scale $s$ we look at the block of height $s$ that contains the key $x_t$ of a search $t$, and we ask when this block was searched last and how many distinct keys were searched since. The locality scale $\lambda(t)$ of the introduction is the least scale $s$ such that the block of height $s$ around $x_t$ contains the key of an earlier search $u$, and fewer than $2^s-1$ distinct keys were searched between $u$ and $t$; in the notation of \cref{sec:intro}, $\rho(u,t)+2\le2^s$. So one number $s$ measures two things, a block of $2^s$ keys and a set of at most $2^s$ distinct keys. The block of height $K$ has all keys, and no search lies between $t-1$ and $t$, so $\lambda(t)\le K$; for the first search we set $\lambda(1)\coloneqq K$. Draw every earlier search $u$ at its key $x_u$ and at the number $\rho(u,t)$ of distinct keys searched after it. The scale $s$ then stands for a box: the block of height $s$ around $x_t$, of $2^s$ keys, times the numbers $0,\dots,2^s-2$. The locality scale is the least $s$ whose box contains an earlier search (\cref{fig:scales}). We use the example of this figure, on $32$ keys, throughout this section.

The locality scale is a version of the unified bound on the grid: it is the term of search $t$, with the distance of the two keys replaced by the size of the smallest block that contains both. If the search $u$ shows that the scale $\lambda(t)$ has the property above, then $\abs{x_t-x_u}<2^{\lambda(t)}$ and $\rho(u,t)+2\le2^{\lambda(t)}$, so the term of $t$ in $\UB(X)$ is at most $\lambda(t)+1$. The converse holds in total over all searches, not for each search alone (\cref{sec:overview-unified}). The main part of the proof is an online BST in which every search $t$ costs $O(\lambda(t))$.

\subsection{A tree for one moment, and why moving it is expensive}
\label{sec:overview-moment}

Call a block of height $s$ \emph{recent} at time $t$, or in use, in the words of the introduction, if it is among the $2^s$ blocks of height $s$ that were searched most recently before $t$. Give one key of each recent block of height $s$ the weight $4^{-s}$, and every other key the weight $1/n$. Summed over all heights, the total weight is $O(1)$, and a BST in which every key of weight $w$ has depth $O(\log(1/w))$ exists for any weights \cite{meh75}. Such a tree brings one key of every recent block of height $s$ to depth $O(s)$, and the construction of \cref{sec:overview-fingers} arranges the other keys of the block within $O(s)$ further levels. A search $t$ is then served from its block of height $\lambda(t)$, at depth $O(\lambda(t))$. This block is recent: fewer than $2^{\lambda(t)}$ distinct keys were searched since the last search in it, so fewer than $2^{\lambda(t)}$ other blocks of its height were searched since. So at any single moment a tree with the right costs exists.

Such a tree can also be maintained. In the biased search trees of Bent, Sleator and Tarjan \cite{bst85}, an item of weight $w$ has depth $O(\log(W/w))$ at every moment, where $W$ is the total weight, and it can be inserted or removed in $O(\log(W/w))$ time in the worst case, by the split and join operations of these trees (\cref{sec:biased}). So when a search makes a block of height $s$ recent, the key of this block can be put into the tree with its weight, and the key of the block that is no longer recent can be removed, at a total cost of $O(s)$. The trouble is the number of such changes. At a height $s$ below $\lambda(t)$, the block of $x_t$ was not searched before, or $2^s-1$ or more distinct keys were searched since, and these keys may lie in as many different blocks. So search $t$ may make a new block recent at every height below $\lambda(t)$, and paying $O(s)$ for each gives
\begin{align*}
\sum_{s<\lambda(t)}O(s)=O\bigl(\lambda(t)^2\bigr)
\end{align*}
per search, against the $O(\lambda(t))$ we can afford. A random permutation shows that the gap is real. All its keys are distinct and they are spread out, so for small $s$ the block of height $s$ around $x_t$ is recent only if one of the last $2^s$ or so searches lies in it. A block of height $s$ holds $2^s$ of the $n$ keys, so this happens with probability about $4^s/n$, and the height must be about $\frac12\log_2n$ before the block of $x_t$ is likely to be recent. Hence most searches make a new block recent at about $\frac12\log_2n$ heights, and the tree above costs, heuristically, about $n\log^2n$, whereas $\OPT$ is $\Theta(n\log n)$ for most permutations \cite{cgk+15}. For the same reason $\lambda(t)$ is about $\frac12\log_2n$ for most searches.

\subsection{Keeping only geometrically spaced scales}
\label{sec:overview-count-promoted}

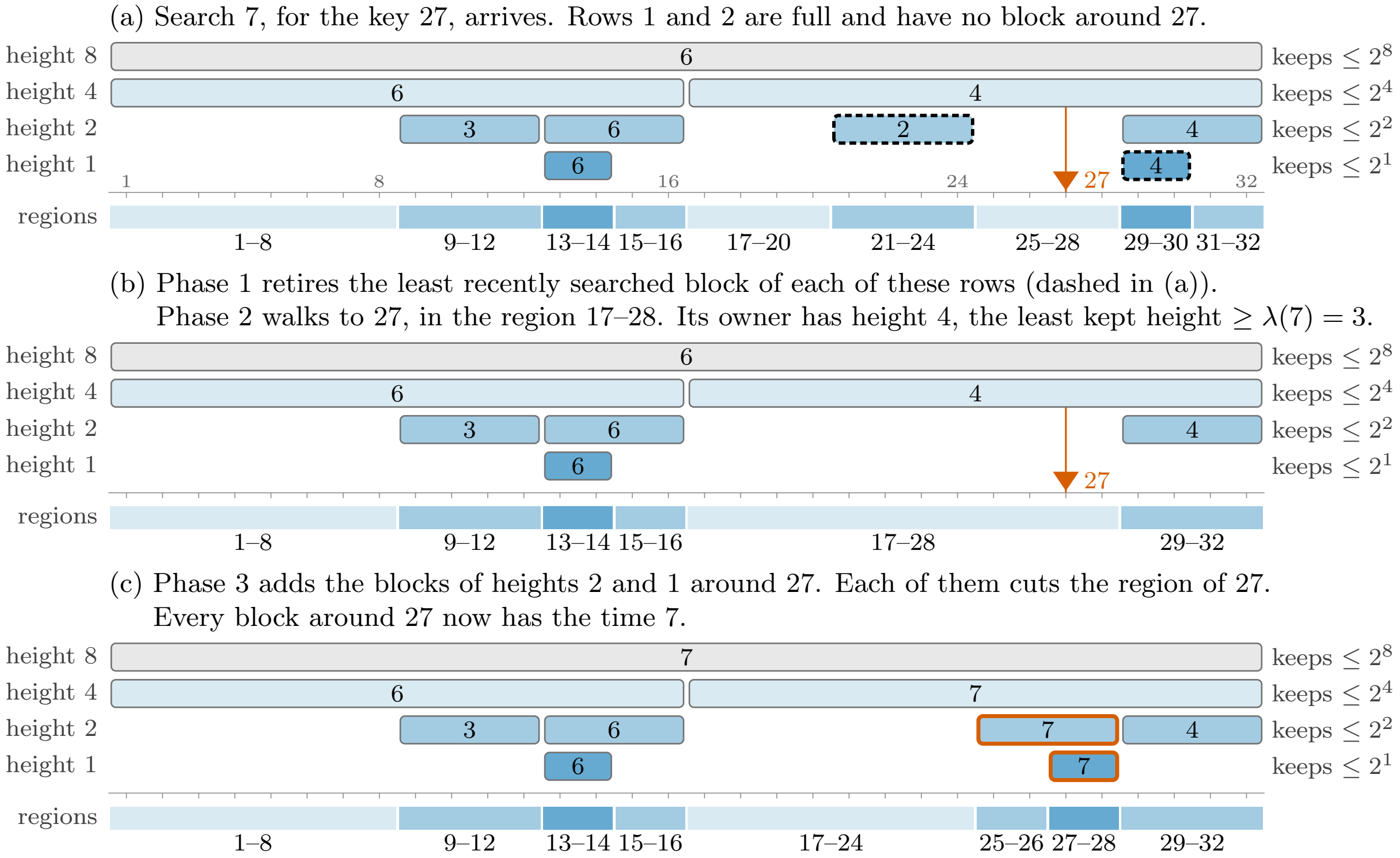
\begin{figure}[t]
\centering
\begin{tikzpicture}[x=\figx,y=0.44cm]
\path (-3.3,0) (36.3,0);
\node[figtitle] at (0.5,4.62) {(a) Search $7$, for the key $27$, arrives. Rows $1$ and $2$ are full and have no block around $27$.};
\node[figlab] at (0.3,4.00) {height $8$};
\node[figlab,anchor=west] at (32.6,4.00) {keeps $\le2^{8}$};
\node[figlab] at (0.3,3.05) {height $4$};
\node[figlab,anchor=west] at (32.6,3.05) {keeps $\le2^{4}$};
\node[figlab] at (0.3,2.10) {height $2$};
\node[figlab,anchor=west] at (32.6,2.10) {keeps $\le2^{2}$};
\node[figlab] at (0.3,1.15) {height $1$};
\node[figlab,anchor=west] at (32.6,1.15) {keeps $\le2^{1}$};
\draw[tcol] (27,0.45) -- (27,2.69);
\draw[kblk,fill=figh8] (0.58,3.64) rectangle (32.42,4.36); \node[figtxt] at (16.5,4.00) {$6$};
\draw[kblk,fill=figh4] (16.58,2.69) rectangle (32.42,3.41); \node[figtxt] at (24.5,3.05) {$4$};
\draw[kblk,fill=figh4] (0.58,2.69) rectangle (16.42,3.41); \node[figtxt] at (8.5,3.05) {$6$};
\draw[kblk,fill=figh2] (28.58,1.74) rectangle (32.42,2.46); \node[figtxt] at (30.5,2.10) {$4$};
\draw[kblk retire,fill=figh2] (20.58,1.74) rectangle (24.42,2.46); \node[figtxt] at (22.5,2.10) {$2$};
\draw[kblk,fill=figh2] (12.58,1.74) rectangle (16.42,2.46); \node[figtxt] at (14.5,2.10) {$6$};
\draw[kblk,fill=figh2] (8.58,1.74) rectangle (12.42,2.46); \node[figtxt] at (10.5,2.10) {$3$};
\draw[kblk retire,fill=figh1] (28.58,0.79) rectangle (30.42,1.51); \node[figtxt] at (29.5,1.15) {$4$};
\draw[kblk,fill=figh1] (12.58,0.79) rectangle (14.42,1.51); \node[figtxt] at (13.5,1.15) {$6$};
\figkeyaxis{0.45}
\figkeynums{0.75}
\figcursor{27}{0.5}
\node[figtxt,text=DPOrange,anchor=west] at (27.4,0.78) {$27$};
\node[figlab] at (0.3,-0.2) {regions};
\figregion{1}{8}{4}{-0.2}
\figregion{9}{12}{2}{-0.2}
\figregion{13}{14}{1}{-0.2}
\figregion{15}{16}{2}{-0.2}
\figregion{17}{20}{4}{-0.2}
\figregion{21}{24}{2}{-0.2}
\figregion{25}{28}{4}{-0.2}
\figregion{29}{30}{1}{-0.2}
\figregion{31}{32}{2}{-0.2}
\end{tikzpicture}
\par\vspace{5pt}
\begin{tikzpicture}[x=\figx,y=0.44cm]
\path (-3.3,0) (36.3,0);
\node[figtitle] at (0.5,4.62) {(b) Phase 1 retires the least recently searched block of each of these rows (dashed in (a)).\\\phantom{(b)} Phase 2 walks to $27$, in the region $17$--$28$. Its owner has height $4$, the least kept height $\ge\lambda(7)=3$.};
\node[figlab] at (0.3,4.00) {height $8$};
\node[figlab,anchor=west] at (32.6,4.00) {keeps $\le2^{8}$};
\node[figlab] at (0.3,3.05) {height $4$};
\node[figlab,anchor=west] at (32.6,3.05) {keeps $\le2^{4}$};
\node[figlab] at (0.3,2.10) {height $2$};
\node[figlab,anchor=west] at (32.6,2.10) {keeps $\le2^{2}$};
\node[figlab] at (0.3,1.15) {height $1$};
\node[figlab,anchor=west] at (32.6,1.15) {keeps $\le2^{1}$};
\draw[tcol] (27,0.45) -- (27,2.69);
\draw[kblk,fill=figh8] (0.58,3.64) rectangle (32.42,4.36); \node[figtxt] at (16.5,4.00) {$6$};
\draw[kblk,fill=figh4] (16.58,2.69) rectangle (32.42,3.41); \node[figtxt] at (24.5,3.05) {$4$};
\draw[kblk,fill=figh4] (0.58,2.69) rectangle (16.42,3.41); \node[figtxt] at (8.5,3.05) {$6$};
\draw[kblk,fill=figh2] (28.58,1.74) rectangle (32.42,2.46); \node[figtxt] at (30.5,2.10) {$4$};
\draw[kblk,fill=figh2] (12.58,1.74) rectangle (16.42,2.46); \node[figtxt] at (14.5,2.10) {$6$};
\draw[kblk,fill=figh2] (8.58,1.74) rectangle (12.42,2.46); \node[figtxt] at (10.5,2.10) {$3$};
\draw[kblk,fill=figh1] (12.58,0.79) rectangle (14.42,1.51); \node[figtxt] at (13.5,1.15) {$6$};
\figkeyaxis{0.45}
\figcursor{27}{0.5}
\node[figtxt,text=DPOrange,anchor=west] at (27.4,0.78) {$27$};
\node[figlab] at (0.3,-0.2) {regions};
\figregion{1}{8}{4}{-0.2}
\figregion{9}{12}{2}{-0.2}
\figregion{13}{14}{1}{-0.2}
\figregion{15}{16}{2}{-0.2}
\figregion{17}{28}{4}{-0.2}
\figregion{29}{32}{2}{-0.2}
\end{tikzpicture}
\par\vspace{5pt}
\begin{tikzpicture}[x=\figx,y=0.44cm]
\path (-3.3,0) (36.3,0);
\node[figtitle] at (0.5,4.62) {(c) Phase 3 adds the blocks of heights $2$ and $1$ around $27$. Each of them cuts the region of $27$.\\\phantom{(c)} Every block around $27$ now has the time $7$.};
\node[figlab] at (0.3,4.00) {height $8$};
\node[figlab,anchor=west] at (32.6,4.00) {keeps $\le2^{8}$};
\node[figlab] at (0.3,3.05) {height $4$};
\node[figlab,anchor=west] at (32.6,3.05) {keeps $\le2^{4}$};
\node[figlab] at (0.3,2.10) {height $2$};
\node[figlab,anchor=west] at (32.6,2.10) {keeps $\le2^{2}$};
\node[figlab] at (0.3,1.15) {height $1$};
\node[figlab,anchor=west] at (32.6,1.15) {keeps $\le2^{1}$};
\draw[kblk,fill=figh8] (0.58,3.64) rectangle (32.42,4.36); \node[figtxt] at (16.5,4.00) {$7$};
\draw[kblk,fill=figh4] (16.58,2.69) rectangle (32.42,3.41); \node[figtxt] at (24.5,3.05) {$7$};
\draw[kblk,fill=figh4] (0.58,2.69) rectangle (16.42,3.41); \node[figtxt] at (8.5,3.05) {$6$};
\draw[kblk,fill=figh2] (28.58,1.74) rectangle (32.42,2.46); \node[figtxt] at (30.5,2.10) {$4$};
\draw[kblk add,fill=figh2] (24.58,1.74) rectangle (28.42,2.46); \node[figtxt] at (26.5,2.10) {$7$};
\draw[kblk,fill=figh2] (12.58,1.74) rectangle (16.42,2.46); \node[figtxt] at (14.5,2.10) {$6$};
\draw[kblk,fill=figh2] (8.58,1.74) rectangle (12.42,2.46); \node[figtxt] at (10.5,2.10) {$3$};
\draw[kblk add,fill=figh1] (26.58,0.79) rectangle (28.42,1.51); \node[figtxt] at (27.5,1.15) {$7$};
\draw[kblk,fill=figh1] (12.58,0.79) rectangle (14.42,1.51); \node[figtxt] at (13.5,1.15) {$6$};
\figkeyaxis{0.45}
\node[figlab] at (0.3,-0.2) {regions};
\figregion{1}{8}{4}{-0.2}
\figregion{9}{12}{2}{-0.2}
\figregion{13}{14}{1}{-0.2}
\figregion{15}{16}{2}{-0.2}
\figregion{17}{24}{4}{-0.2}
\figregion{25}{26}{2}{-0.2}
\figregion{27}{28}{1}{-0.2}
\figregion{29}{32}{2}{-0.2}
\end{tikzpicture}
\caption{Search $7$ of \cref{fig:scales} in the algorithm. A box is a kept block, drawn over its keys in the row of its height; the number in it is the time of the last search in the block. Blocks are kept only at the heights $1$, $2$, $4$ and $8$. A block of height $a<8$ has $2^a$ keys, and at most $2^a$ blocks of this height are kept, those searched most recently; the block of height $8$ consists of all keys. The right margin gives this quota for each row: at height $4$ the quota of $16$ exceeds the $2$ blocks that exist, so a block of this height is never retired once it is searched, and only the rows of heights $1$ and $2$ ever retire a block. The bar below shows the regions, each shaded like the smallest kept block around it, which owns it. The darker a region, the smaller its owner and the nearer its keys are to the root (\cref{fig:realized}). The search retires the blocks $\{29,30\}$ and $\{21,\dots,24\}$, which were searched last at the times $4$ and $2$, is served from the block $\{17,\dots,32\}$, and adds the blocks $\{25,\dots,28\}$ and $\{27,28\}$.}
\label{fig:fingers}
\end{figure}

The remedy is to keep fewer blocks. Our tree tracks blocks only at the \emph{promoted scales} $1,2,4,8,\dots$, up to $\hat K$. They are dense enough: for every scale $s$ from $1$ to $K$, the least power of two $\hat s\ge s$ is a promoted scale, and $\hat s<2s$. At a promoted scale $s$, the tree \emph{keeps} the recent blocks of height $s$, that is, the $2^s$ blocks of height $s$ that were searched most recently (\cref{fig:fingers}(a)). At the scale $\hat K$ the only block is the block of all keys, so the tree always keeps it. In terms of single searches, every search \emph{promotes} its blocks at all promoted scales: it makes them kept if they are not kept already. We call a promoted scale \emph{missing} for a search if the block of the search at this scale is not kept when the search arrives. At a missing scale $s$ the search brings in its block, and if $2^s$ blocks of this height are kept already, the one of them that was searched least recently is no longer kept; we say that the search \emph{retires} it. At the other promoted scales nothing changes. The algorithm does this in three phases, which \cref{fig:fingers} shows for the running example: it first retires blocks, then walks to the searched key, and then adds its own blocks. The missing scales of search $t$ are below $\lambda(t)$, because the block of $x_t$ at every scale from $\lambda(t)$ on is recent: the count of \cref{sec:overview-moment} for the scale $\lambda(t)$ applies to every larger scale $s$ as well: the block of height $s$ contains the block of height $\lambda(t)$, and the number $2^s$ of blocks that may be kept is larger.

It matters here that the tree counts blocks and not time. Had it kept, at scale $s$, the blocks in which one of the last $2^s$ searches lies, a search could find its block gone although only a few distinct keys were searched since the last search in it, namely when these few keys were searched many times. Its cost would then not be bounded by its locality scale, and the analysis would have to amortize over the time axis as well.

A search $t$ thus replaces blocks at only $O(\log\lambda(t))$ scales, and the heights of these blocks grow geometrically, so adding and removing all of them costs $\sum_{s}O(s)=O(\lambda(t))$ over the promoted scales $s<\lambda(t)$. On a random permutation this gives $O(n\log n)$.

A search is still served at the right depth. Let $\hat s$ be the least power of two that is at least $\lambda(t)$, so that $\hat s<2\lambda(t)$; it is a promoted scale, because $\lambda(t)\le K$. This scale is not missing, so the block of height $\hat s$ around $x_t$ is kept when $t$ arrives, and $t$ is served from a block of height less than $2\lambda(t)$ (\cref{lem:serving}).

\subsection{From the idea to a binary search tree}
\label{sec:overview-fingers}
\label{sec:overview-count}

What remains is to build a BST that realizes this accounting. It has to make the keys of every kept block of height $a$ reachable at depth $O(a)$, to add the blocks of a search $t$ at cost $O(\lambda(t))$, and to remove a block at a cost proportional to its height. We describe the tree in four steps: how the kept blocks are nested, how they cut the keys into intervals, how a tree over these intervals stands for a BST on the keys, and what a change of the intervals costs. \Cref{fig:fingers} shows the running example.

The kept blocks form a nested family: two of them are disjoint, or one contains the other. Two facts describe this family (\cref{lem:kept}).
\begin{itemize}
\item If a block is kept, then so are the blocks around it at all larger promoted scales. The reason is a count. Let a kept block $g$ of height $s$ lie in a block $g'$ of a larger promoted height $s'$, and suppose that $g'$ is not kept. Then at least $2^{s'}$ blocks of height $s'$ were searched after the last search in $g'$, and the searches in them lie in as many different blocks of height $s$. So at least $2^{s'}$ blocks of height $s$, which is more than $2^s$, were searched after the last search in $g'$. The last search in $g$ is not later than the last search in $g'$, because $g$ lies in $g'$, so all these blocks were searched after the last search in $g$, and $g$ is not kept either. It follows that the next larger kept block around a kept block of height $a<\hat K$ is the kept block of height $2a$ around it.
\item The blocks that a search adds are nested, and no kept block lies inside the largest of them. Indeed, by the first fact a kept block inside it would force that block to be kept already.
\end{itemize}
The same count shows that a block that a search retires contains no kept block of a smaller promoted scale. Let the search retire the block $e$ of height $s$, and let $s'<s$ be a promoted scale. The other $2^s-1$ kept blocks of height $s$ were searched after the last search in $e$, and the searches in them lie in as many different blocks of height $s'$. Since $2^s-1\ge2^{s'}$, a block of height $s'$ inside $e$ is not among the $2^{s'}$ most recently searched ones.

So blocks appear together at the bottom of the family, nested inside one another, as the two blocks around the key $27$ do in \cref{fig:fingers}(c).

The kept blocks cut the keys into intervals, which we call \emph{regions}. The \emph{owner} of a key is the smallest kept block that contains it, and a region is a maximal interval of consecutive keys with the same owner. In \cref{fig:fingers}(a) there are nine regions. For instance, the block $\{17,\dots,32\}$ owns the keys that lie in it but in neither of the two kept blocks of height $2$ inside it, and these keys form the two regions $\{17,\dots,20\}$ and $\{25,\dots,28\}$. When a block is removed that contains no kept block, its keys form a single region. These keys pass to the next larger kept block around it, and the region is merged with the neighbouring regions of that block. Conversely, when a block is added, the region that contains it is split into at most three: the block, and what is left of the region on each side (\cref{lem:regions}).

Our data structure is a biased search tree whose leaves are the regions, in key order; we call it the \emph{region tree}. A region whose owner has height $a$ has the weight $4^{-a}$. Every leaf carries a balanced search tree whose leaves are the keys of its region, the \emph{inner tree} of the region; we use a biased search tree with equal weights. Together these trees stand for a BST on the keys (\cref{fig:realized}). Hang every inner tree at the leaf of its region. The result is one tree whose leaves are the keys, in order, and it has exactly one internal node between any two consecutive leaves. Let this node hold the smaller of the two keys, and remove the leaves, except the last one, which holds the largest key. The internal nodes and the last leaf then form a BST on all keys. So an internal node of the region tree holds the largest key of the region on its left, and the other keys of a region sit at the internal nodes of its inner tree.

\begin{figure}[t]
\centering
\begin{tikzpicture}[x=\figx,y=0.5cm]
\path (-3.3,0) (36.3,0);
\node[figtitle] at (0.5,0.75) {(a) When search $7$ arrives (\cref{fig:fingers}(a)): the key $27$ has depth $6$.};
\draw[black!8,line width=0.3pt] (0.5,0) -- (32.5,0); \node[figlab] at (0.3,0) {depth $0$};
\draw[black!8,line width=0.3pt] (0.5,-1) -- (32.5,-1); \node[figlab] at (0.3,-1) {$1$};
\draw[black!8,line width=0.3pt] (0.5,-2) -- (32.5,-2); \node[figlab] at (0.3,-2) {$2$};
\draw[black!8,line width=0.3pt] (0.5,-3) -- (32.5,-3); \node[figlab] at (0.3,-3) {$3$};
\draw[black!8,line width=0.3pt] (0.5,-4) -- (32.5,-4); \node[figlab] at (0.3,-4) {$4$};
\draw[black!8,line width=0.3pt] (0.5,-5) -- (32.5,-5); \node[figlab] at (0.3,-5) {$5$};
\draw[black!8,line width=0.3pt] (0.5,-6) -- (32.5,-6); \node[figlab] at (0.3,-6) {$6$};
\node[knode,fill=figh4] (k1) at (1,-5) {1};
\node[knode,fill=figh4] (k2) at (2,-4) {2};
\node[knode,fill=figh4] (k3) at (3,-5) {3};
\node[knode,fill=figh4] (k4) at (4,-3) {4};
\node[knode,fill=figh4] (k5) at (5,-5) {5};
\node[knode,fill=figh4] (k6) at (6,-4) {6};
\node[knode,fill=figh4] (k7) at (7,-5) {7};
\node[ksep,fill=figh4] (k8) at (8,-2) {8};
\node[knode,fill=figh2] (k9) at (9,-4) {9};
\node[knode,fill=figh2] (k10) at (10,-3) {10};
\node[knode,fill=figh2] (k11) at (11,-4) {11};
\node[ksep,fill=figh2] (k12) at (12,-1) {12};
\node[knode,fill=figh1] (k13) at (13,-2) {13};
\node[ksep,fill=figh1] (k14) at (14,0) {14};
\node[knode,fill=figh2] (k15) at (15,-5) {15};
\node[ksep,fill=figh2] (k16) at (16,-4) {16};
\node[knode,fill=figh4] (k17) at (17,-6) {17};
\node[knode,fill=figh4] (k18) at (18,-5) {18};
\node[knode,fill=figh4] (k19) at (19,-6) {19};
\node[ksep,fill=figh4] (k20) at (20,-3) {20};
\node[knode,fill=figh2] (k21) at (21,-6) {21};
\node[knode,fill=figh2] (k22) at (22,-5) {22};
\node[knode,fill=figh2] (k23) at (23,-6) {23};
\node[ksep,fill=figh2] (k24) at (24,-4) {24};
\node[knode,fill=figh4] (k25) at (25,-6) {25};
\node[knode,fill=figh4] (k26) at (26,-5) {26};
\node[knode,fill=figh4,kcur] (k27) at (27,-6) {27};
\node[ksep,fill=figh4] (k28) at (28,-2) {28};
\node[knode,fill=figh1] (k29) at (29,-3) {29};
\node[ksep,fill=figh1] (k30) at (30,-1) {30};
\node[knode,fill=figh2] (k31) at (31,-2) {31};
\node[knode,fill=figh2] (k32) at (32,-3) {32};
\begin{scope}[on background layer]
\draw[inedge] (k2.center) -- (k1.center);
\draw[inedge] (k2.center) -- (k3.center);
\draw[inedge] (k4.center) -- (k2.center);
\draw[inedge] (k4.center) -- (k6.center);
\draw[inedge] (k6.center) -- (k5.center);
\draw[inedge] (k6.center) -- (k7.center);
\draw[sepedge] (k8.center) -- (k4.center);
\draw[sepedge] (k8.center) -- (k10.center);
\draw[inedge] (k10.center) -- (k9.center);
\draw[inedge] (k10.center) -- (k11.center);
\draw[sepedge] (k12.center) -- (k8.center);
\draw[sepedge] (k12.center) -- (k13.center);
\draw[sepedge] (k14.center) -- (k12.center);
\draw[spath] (k14.center) -- (k30.center);
\draw[sepedge] (k16.center) -- (k15.center);
\draw[sepedge] (k16.center) -- (k18.center);
\draw[inedge] (k18.center) -- (k17.center);
\draw[inedge] (k18.center) -- (k19.center);
\draw[sepedge] (k20.center) -- (k16.center);
\draw[spath] (k20.center) -- (k24.center);
\draw[inedge] (k22.center) -- (k21.center);
\draw[inedge] (k22.center) -- (k23.center);
\draw[sepedge] (k24.center) -- (k22.center);
\draw[spath] (k24.center) -- (k26.center);
\draw[inedge] (k26.center) -- (k25.center);
\draw[spath] (k26.center) -- (k27.center);
\draw[spath] (k28.center) -- (k20.center);
\draw[sepedge] (k28.center) -- (k29.center);
\draw[spath] (k30.center) -- (k28.center);
\draw[sepedge] (k30.center) -- (k31.center);
\draw[inedge] (k31.center) -- (k32.center);
\end{scope}
\figkeyaxis{-6.55}
\node[figlab] at (0.3,-7.05) {regions};
\figregion{1}{8}{4}{-7.05}
\figregion{9}{12}{2}{-7.05}
\figregion{13}{14}{1}{-7.05}
\figregion{15}{16}{2}{-7.05}
\figregion{17}{20}{4}{-7.05}
\figregion{21}{24}{2}{-7.05}
\figregion{25}{28}{4}{-7.05}
\figregion{29}{30}{1}{-7.05}
\figregion{31}{32}{2}{-7.05}
\end{tikzpicture}
\par\vspace{6pt}
\begin{tikzpicture}[x=\figx,y=0.5cm]
\path (-3.3,0) (36.3,0);
\node[figtitle] at (0.5,0.75) {(b) After search $7$ (\cref{fig:fingers}(c)): the key $27$ has depth $3$.};
\draw[black!8,line width=0.3pt] (0.5,0) -- (32.5,0); \node[figlab] at (0.3,0) {depth $0$};
\draw[black!8,line width=0.3pt] (0.5,-1) -- (32.5,-1); \node[figlab] at (0.3,-1) {$1$};
\draw[black!8,line width=0.3pt] (0.5,-2) -- (32.5,-2); \node[figlab] at (0.3,-2) {$2$};
\draw[black!8,line width=0.3pt] (0.5,-3) -- (32.5,-3); \node[figlab] at (0.3,-3) {$3$};
\draw[black!8,line width=0.3pt] (0.5,-4) -- (32.5,-4); \node[figlab] at (0.3,-4) {$4$};
\draw[black!8,line width=0.3pt] (0.5,-5) -- (32.5,-5); \node[figlab] at (0.3,-5) {$5$};
\draw[black!8,line width=0.3pt] (0.5,-6) -- (32.5,-6); \node[figlab] at (0.3,-6) {$6$};
\node[knode,fill=figh4] (k1) at (1,-5) {1};
\node[knode,fill=figh4] (k2) at (2,-4) {2};
\node[knode,fill=figh4] (k3) at (3,-5) {3};
\node[knode,fill=figh4] (k4) at (4,-3) {4};
\node[knode,fill=figh4] (k5) at (5,-5) {5};
\node[knode,fill=figh4] (k6) at (6,-4) {6};
\node[knode,fill=figh4] (k7) at (7,-5) {7};
\node[ksep,fill=figh4] (k8) at (8,-2) {8};
\node[knode,fill=figh2] (k9) at (9,-4) {9};
\node[knode,fill=figh2] (k10) at (10,-3) {10};
\node[knode,fill=figh2] (k11) at (11,-4) {11};
\node[ksep,fill=figh2] (k12) at (12,-1) {12};
\node[knode,fill=figh1] (k13) at (13,-2) {13};
\node[ksep,fill=figh1] (k14) at (14,0) {14};
\node[knode,fill=figh2] (k15) at (15,-3) {15};
\node[ksep,fill=figh2] (k16) at (16,-2) {16};
\node[knode,fill=figh4] (k17) at (17,-6) {17};
\node[knode,fill=figh4] (k18) at (18,-5) {18};
\node[knode,fill=figh4] (k19) at (19,-6) {19};
\node[knode,fill=figh4] (k20) at (20,-4) {20};
\node[knode,fill=figh4] (k21) at (21,-6) {21};
\node[knode,fill=figh4] (k22) at (22,-5) {22};
\node[knode,fill=figh4] (k23) at (23,-6) {23};
\node[ksep,fill=figh4] (k24) at (24,-3) {24};
\node[knode,fill=figh2] (k25) at (25,-4) {25};
\node[ksep,fill=figh2] (k26) at (26,-1) {26};
\node[knode,fill=figh1,kcur] (k27) at (27,-3) {27};
\node[ksep,fill=figh1] (k28) at (28,-2) {28};
\node[knode,fill=figh2] (k29) at (29,-4) {29};
\node[knode,fill=figh2] (k30) at (30,-3) {30};
\node[knode,fill=figh2] (k31) at (31,-4) {31};
\node[knode,fill=figh2] (k32) at (32,-5) {32};
\begin{scope}[on background layer]
\draw[inedge] (k2.center) -- (k1.center);
\draw[inedge] (k2.center) -- (k3.center);
\draw[inedge] (k4.center) -- (k2.center);
\draw[inedge] (k4.center) -- (k6.center);
\draw[inedge] (k6.center) -- (k5.center);
\draw[inedge] (k6.center) -- (k7.center);
\draw[sepedge] (k8.center) -- (k4.center);
\draw[sepedge] (k8.center) -- (k10.center);
\draw[inedge] (k10.center) -- (k9.center);
\draw[inedge] (k10.center) -- (k11.center);
\draw[sepedge] (k12.center) -- (k8.center);
\draw[sepedge] (k12.center) -- (k13.center);
\draw[sepedge] (k14.center) -- (k12.center);
\draw[spath] (k14.center) -- (k26.center);
\draw[sepedge] (k16.center) -- (k15.center);
\draw[sepedge] (k16.center) -- (k24.center);
\draw[inedge] (k18.center) -- (k17.center);
\draw[inedge] (k18.center) -- (k19.center);
\draw[inedge] (k20.center) -- (k18.center);
\draw[inedge] (k20.center) -- (k22.center);
\draw[inedge] (k22.center) -- (k21.center);
\draw[inedge] (k22.center) -- (k23.center);
\draw[sepedge] (k24.center) -- (k20.center);
\draw[sepedge] (k24.center) -- (k25.center);
\draw[sepedge] (k26.center) -- (k16.center);
\draw[spath] (k26.center) -- (k28.center);
\draw[spath] (k28.center) -- (k27.center);
\draw[sepedge] (k28.center) -- (k30.center);
\draw[inedge] (k30.center) -- (k29.center);
\draw[inedge] (k30.center) -- (k31.center);
\draw[inedge] (k31.center) -- (k32.center);
\end{scope}
\figkeyaxis{-6.55}
\node[figlab] at (0.3,-7.05) {regions};
\figregion{1}{8}{4}{-7.05}
\figregion{9}{12}{2}{-7.05}
\figregion{13}{14}{1}{-7.05}
\figregion{15}{16}{2}{-7.05}
\figregion{17}{24}{4}{-7.05}
\figregion{25}{26}{2}{-7.05}
\figregion{27}{28}{1}{-7.05}
\figregion{29}{32}{2}{-7.05}
\end{tikzpicture}
\caption{The BST of the algorithm before and after search $7$, with every key drawn at its depth and shaded like its region. The square nodes and the thick edges below them are the region tree: its internal nodes, each of which separates two consecutive regions and holds the largest key of the region on its left, as the root $14$ does for the region $\{13,14\}$. The round nodes hold the other keys of each region, in a balanced tree of their own, the inner tree of the region. A region whose owner has height $a$ has the weight $4^{-a}$ in the region tree, which is a biased search tree, so its keys have depth $O(a+1)$. The orange path is the search for the key $27$. In (a) it passes the keys $14$, $30$, $28$, $20$ and $24$ of the region tree and then descends in the inner tree of the region $\{25,\dots,28\}$.}
\label{fig:realized}
\end{figure}

This tree has two properties, which reduce the cost of the BST to that of the region tree.
\begin{itemize}
\item {\bf Depth.} At most $2^a$ blocks of height $a$ are kept at any time. Every region begins at the first key of its owner or right after the last key of a lower kept block. Charge the region to its owner in the first case and to this lower block in the second. A kept block of height $a$ is then charged at most two regions, the one at its first key and the one after its last key, and the owner of each has a height of at least $a$, so each weighs at most $4^{-a}$. So the total weight of the regions is at most twice the sum of $4^{-a}$ over the kept blocks, which is $O(1)$ (\cref{lem:weights}). Hence the leaf of a region whose owner has height $a$ is at depth $O(a+1)$. The region has at most $2^a$ keys, so its inner tree adds $O(a+1)$ levels (\cref{lem:pathdepth}).
\item {\bf Changing regions.} When regions are merged or split, the leaves involved, at most three, are removed from the region tree and new leaves are inserted, and their inner trees are joined or split in the same way, all in time proportional to the largest height of their owners. These operations change the BST only at the nodes that they visit, and the visited nodes contain all their ancestors. In the BST model, a search may rearrange any set of nodes that contains the root and, with each of its nodes, the parent of this node, at a cost equal to the size of the set (\cref{sec:prelim}). The visited nodes are such a set, and their number is at most the running time (\cref{lem:wc-visited,lem:opcost}).
\end{itemize}

Serving a search costs $O(\lambda(t))$, because the region of $x_t$ is owned by a block of height less than $2\lambda(t)$. The blocks of a search are added one at a time, from the largest to the smallest. Each of them splits the region of $x_t$ once more, at a cost proportional to the height of the block that owns this region, which is the serving block for the first of them and, for each later one, the block that was added just before it. This is $O(\lambda(t))$ in all. Removing a block of height $a$ merges its region with regions of the next larger kept block around it, whose height is at most $2a$, so it costs $O(a+1)$. A search removes at most one block at each of its missing scales, which are below $\lambda(t)$, so this is $O(\lambda(t))$ in all as well. Hence search $t$ costs $O(\lambda(t))$ in the worst case, and no cost is passed to another search (\cref{sec:algorithm,sec:invariants,sec:analysis}).

\subsection{From scales to the unified bound}
\label{sec:overview-unified}

The grid is fixed in advance, so a search $t$ can lose against its term in the unified bound: the block that contains $x_t$ and the key of a nearby earlier search $u$ may be much higher than their distance suggests, when a boundary between two high blocks falls between the two keys. The loss can be of order $\log n$ for a single search. We show that it is small in total: a search that the grid serves badly returns to a block that no search has visited for a long time, and the searches that came in between have already paid for the loss in the unified bound.

Write $\tau_t$ for the term of search $t$ in the unified bound. We saw that $\tau_t\le\lambda(t)+1$ for every search. In the other direction we prove (\cref{lem:keyloss})
\begin{align*}
\sum_t\lambda(t)\le2\sum_t\tau_t+O(m).
\end{align*}
This gives $\sum_t\lambda(t)=O(m+\UB(X))$, and \cref{thm:main} follows from \cref{thm:scales}.

Call an earlier search that attains the term $\tau_t$ the \emph{partner} of $t$.

\begin{figure}[t]
\centering
\tikzset{
  kax/.style={draw=black!70,line width=0.6pt,-{Stealth[length=4pt]}},
  wline/.style={draw=black!70,line width=0.7pt,dash pattern=on 3pt off 2pt},
  cold/.style={draw=DPOrange,line width=0.9pt,fill=DPOrange!10},
  part/.style={draw=black!65,line width=0.7pt,-{Stealth[length=3.5pt]}},
  lab/.style={font=\footnotesize,inner sep=1.5pt},
  br/.style={decorate,decoration={brace,amplitude=3pt},black!70,line width=0.5pt},
}
\begin{tikzpicture}[x=0.55cm,y=0.5cm]
\draw[kax] (0,0) -- (12.4,0) node[lab,below left] {keys};
\draw[kax] (0,0) -- (0,-10.4) node[lab,above left,xshift=-2pt] {time};
\draw[cold] (6.5,-1.6) rectangle (11.5,-9);
\node[lab,text=DPOrange,align=center] at (9.3,-5.2) {no search\\in this block};
\draw[wline] (6.5,0) -- (6.5,-10.2) node[lab,below] {$\beta$};
\draw[br] (6.5,0.25) -- (11.5,0.25) node[lab,midway,above=3pt] {$B_J(x_t)$};
\node[sdot,label={[lab]above left:$v_J(t)$}] (r) at (9.6,-1.6) {};
\node[tdot,label={[lab,text=DPOrange]below right:$t$}] (t) at (7.3,-9) {};
\node[sdot,label={[lab]left:$u=s_0$}] (s0) at (5.6,-7.9) {};
\node[sdot,label={[lab]left:$s_1$}] (s1) at (4.4,-6.7) {};
\node[sdot,label={[lab]left:$s_2$}] (s2) at (5.3,-5.5) {};
\node[lab,text=black!65] (dots) at (4.7,-4.25) {$\vdots$};
\node[sdot,label={[lab]left:$s_{i^*}$}] (sj) at (4.2,-2.8) {};
\draw[part] (t) -- (s0); \draw[part] (s0) -- (s1); \draw[part] (s1) -- (s2);
\draw[part] (s2) -- (dots); \draw[part] (dots) -- (sj);
\draw[br] (-0.35,-9) -- (-0.35,-1.6) node[lab,midway,left=4pt,align=right] {at least\\$2^J-1$\\distinct\\keys};
\node[font=\small] at (6.2,-11.6) {(a) A crossing $t$ of level $J$ and its chain};
\end{tikzpicture}\hfill
\begin{tikzpicture}[x=0.55cm,y=0.5cm]
\draw[kax] (0,0) -- (12.4,0) node[lab,below left] {keys};
\draw[kax] (0,0) -- (0,-10.4) node[lab,above left,xshift=-2pt] {time};
\draw[cold] (6.5,-1.6) rectangle (11.5,-9);
\draw[wline] (6.5,0) -- (6.5,-10.2) node[lab,below] {$\beta$};
\draw[br] (6.5,0.25) -- (11.5,0.25) node[lab,midway,above=3pt] {$B_J(x_{t_2})$};
\node[sdot,label={[lab]above left:$v_J(t_2)$}] at (9.6,-1.6) {};
\node[tdot,label={[lab,text=DPOrange]below right:$t_2$}] (t2) at (7.3,-9) {};
\node[tdot,label={[lab,text=DPOrange]right:$t_1$}] (t1) at (8.1,-5.6) {};
\node[sdot,label={[lab]left:$u_{t_1}$}] (u1) at (5.5,-4.6) {};
\node[sdot,label={[lab]left:$s$}] (s) at (4.4,-3.2) {};
\draw[part] (t1) -- (u1); \draw[part] (u1) -- (s);
\node[font=\small] at (6.2,-11.6) {(b) Two crossings cannot share $s$};
\end{tikzpicture}
\caption{The charging in the proof of \cref{lem:keyloss}. Keys are horizontal and time runs downwards; a dot is a search, and the dashed line is the boundary $\beta$ between two blocks of height $J$. An arrow leads from a search to its partner, an earlier search that attains its term in the unified bound. A crossing of level $J$ is a search $t$ with $\lambda(t)>J$ whose partner $u$ is so near in key and in time that $x_u$ lies just across $\beta$, and $v_J(t)$ is the last search before $t$ in the block $B_J(x_t)$. In (a), no search lies in the block of the crossing $t$ between the times $v_J(t)$ and $t$, and the chain of partners of $t$ stays after $v_J(t)$ and near $\beta$. In (b), if a search $s$ were on the chains of two crossings $t_1<t_2$ of level $J$, then $t_1$ or its partner would be a search in the block of $t_2$ after $v_J(t_2)$.}
\label{fig:keyloss}
\end{figure}
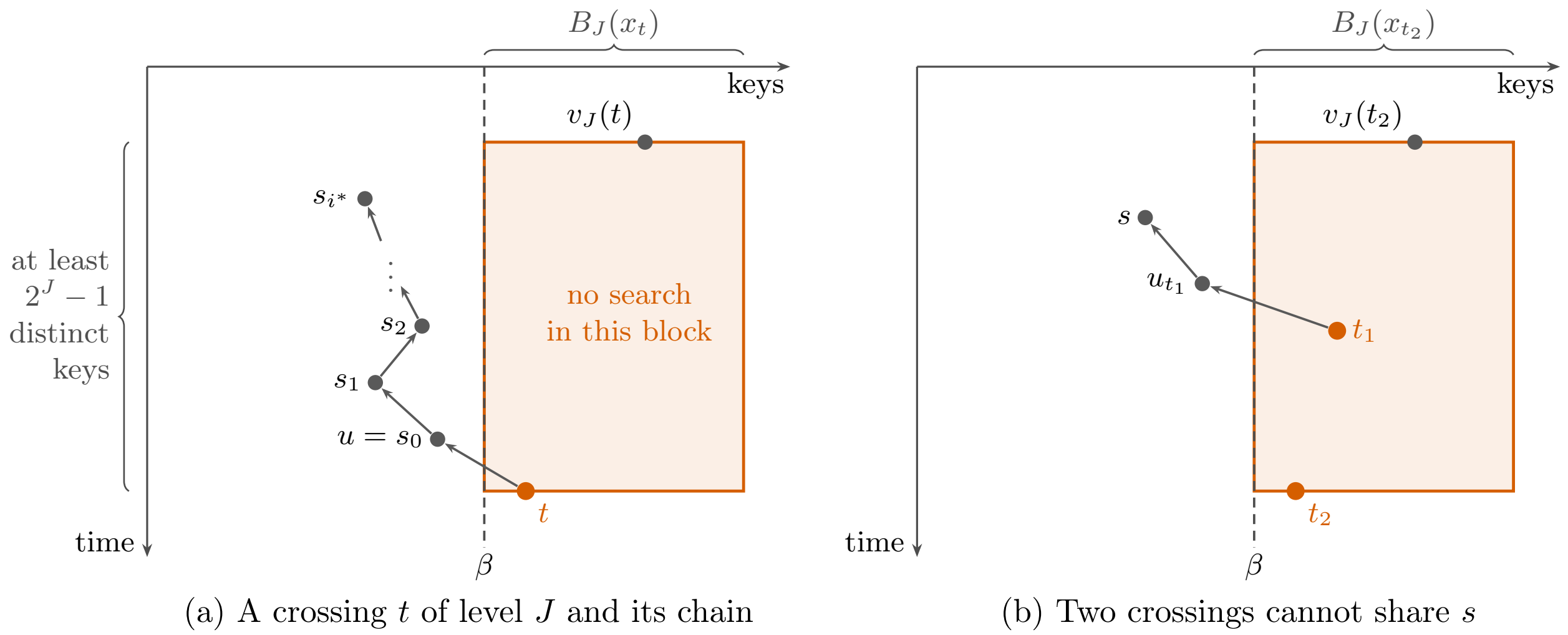

We count level by level, where a \emph{level} is a height $J$ of blocks. The sum $\sum_t\lambda(t)$ is the number of pairs $(t,J)$ with $\lambda(t)>J$, and we pay one unit for each such pair. If $\tau_t>J-2$, the search $t$ pays the unit itself, which happens at only $\tau_t+O(1)$ levels. Otherwise $\tau_t\le J-2$, which says that the partner $u$ of $t$ is close to $t$: their keys differ by less than $2^{J-2}$, and fewer than $2^{J-2}$ distinct keys were searched between them. Now $\lambda(t)>J$ means that the block of height $J$ around $x_t$ was never searched before, or that at least $2^J-1$ distinct keys were searched since the last search in it. So the search $u$, which is recent, does not lie in this block. Its key is less than a quarter of a block away from $x_t$, so it lies in the neighbouring block, just across the boundary $\beta$ between the two (\cref{fig:keyloss}(a)). We call such a search $t$ a \emph{crossing} of level $J$. Here others must pay for $t$, and they are the searches made near this boundary while the block of $t$ stayed unvisited.

To find them, follow partners backwards: from $u$ to the partner of $u$, from there to its partner, and so on. We call these searches the \emph{chain} of $t$; they are the dots joined by arrows in \cref{fig:keyloss}(a). A step from a search with term $\tau'$ to its partner moves by less than $2^{\tau'}$ in key and passes fewer than $2^{\tau'}$ distinct keys in time. We follow the chain until the numbers $2^{\tau'}$ of its searches add up to at least $2^{J-2}$, a quarter of the size of a block of height $J$. Each search of the chain then pays the share $2^{\tau'}/2^{J-2}$ of the unit, or the whole unit if this share is more than one. The shares add up to at least one, by the rule for ending the chain, so the unit is paid. The chain cannot run out before that. Only the first search of the sequence has no partner, and its term in the unified bound is $\log_2(n+2)$, which exceeds every level $J$, since a level that we count is less than $\lambda(t)\le\ceil{\log_2n}$. So if the chain reaches the first search, the number $2^{\tau'}$ of this search alone is more than $2^{J-2}$, and the chain ends there.

Two facts make this charging sound; both are proved in \cref{lem:keyloss}. The first is that the chain stays near the boundary and inside the period in which the block of $t$ was not visited. Indeed, before its last search the numbers $2^{\tau'}$ of the chain add up to less than $2^{J-2}$, and the first step, from $t$ to $u$, adds less than $2^{\tau_t}\le2^{J-2}$. So every search $s_i$ of the chain, in the notation of \cref{fig:keyloss}(a), satisfies
\begin{align*}
\abs{x_{s_i}-\beta}<2^{J-2}+2^{J-2}=2^{J-1},
\end{align*}
which is half a block. The same sum bounds the time: the distinct keys searched between $s_i$ and $t$ are those passed by the steps from $t$ back to $s_i$, together with the keys of the chain searches in between, and a step from a search with term $\tau'$ passes at most $2^{\tau'}-2$ distinct keys, so the same sum bounds their number by $2^{J-1}$ (\cref{lem:chain}). This is fewer than the $2^J-1$ distinct keys that were searched since the last search in the block of $t$, so $s_i$ is later than that search.

The second fact is that, at one level $J$, a search is on the chain of at most one such $t$ (\cref{fig:keyloss}(b)). Suppose that a search, called $s$ in the figure, is on the chains of $t_1<t_2$. Both chains stay within half a block of their boundaries, and the boundaries between blocks of height $J$ are $2^J$ keys apart, so the two boundaries are the same. The search $t_1$ and its partner have their keys in the two blocks next to this boundary, one on each side, and the block of $t_2$ is one of these two blocks. So $t_1$ or its partner is a search in the block of $t_2$, and it is before $t_2$. It is not earlier than $s$, because the chain of $t_1$ starts at the partner of $t_1$ and runs backwards. But by the first fact, $s$ is later than the last search in the block of $t_2$ before $t_2$. This is a contradiction.

Hence a search with term $\tau'$ pays a full unit only at the levels $J\le\tau'+2$, and at a higher level $J$ it pays at most the share $2^{\tau'}/2^{J-2}$; these shares sum to $O(1)$ over the levels. Each search thus pays about its term twice, once for itself and once for others, which gives the inequality.

\subsection{Pattern-avoiding sequences}

For \cref{thm:patterns} we run the same tree. Cut the time axis into \emph{epochs} of $2^s$ consecutive searches, and call a block of height $s$ together with such an epoch a \emph{cell} of scale $s$. A search with $\lambda(t)>s$ is the first search in its cell of scale $s$: an earlier search $u$ in the same cell would lie in the block of $x_t$, with at most $2^s-2$ searches between $u$ and $t$. View the grid of scale $s$ as a $0$-$1$ matrix with one row per block, one column per epoch, and a one wherever a search lies in the cell. If this matrix contained the pattern $\sigma$, choosing one search in each of the corresponding cells would give an occurrence of $\sigma$ in $X$. The matrix has $O(n2^{-s})$ rows and $O(1+m2^{-s})$ columns, so by the theorem of Marcus and Tardos it has $O(c_\sigma(m+n)2^{-s})$ ones. Hence $\#\{t:\lambda(t)>s\}\le\min(m,O(c_\sigma(m+n)2^{-s}))$, and summing over $s$ gives $\sum_t\lambda(t)=O((m+n)(1+\log c_\sigma))$.

\section{Preliminaries}
\label{sec:prelim}

This section fixes the model of computation and states two tools: a lemma that lets a search pay for a sequence of changes of the tree one by one, and the biased search trees of Bent, Sleator and Tarjan.

\subsection{The BST model}

We use the model of Demaine, Harmon, Iacono, Kane and Pătraşcu \cite{dhi+09}, which agrees up to a constant factor with the classical models of BSTs with rotations \cite{st85,wil89}. A BST stores the keys $1,\dots,n$. To serve a search for a key $x$, an algorithm chooses a set $U$ of nodes that contains $x$ and the root and is connected, and it replaces the subtree formed by $U$ by any BST on the same keys; the subtrees that hang off $U$ are reattached in the unique way that keeps a BST. The cost of the search is $\abs U$. Only these nodes are counted: the computation that decides which nodes to touch, including the records that our algorithm keeps about past searches, is free. A node counts as an ancestor of itself, and the depth of a node is the number of its other ancestors.

For a sequence $X=(x_1,\dots,x_m)$ of keys, $\OPT(X)$ is the least total cost of serving $X$, over all initial trees and all choices of the sets $U$. An algorithm is online if its choices while serving $x_t$ depend only on $x_1,\dots,x_t$. Our algorithms choose their initial tree, which depends only on $n$.

A set of nodes of a BST is a \emph{top set} if it contains every ancestor of each of its nodes. A search for $x$ may replace any top set that contains $x$, at a cost equal to its size. Our algorithm goes from the tree before a search to the tree after it through several intermediate trees, one for each of its steps, which we call \emph{operations}. The next lemma lets the search pay for them one by one.

\begin{lemma}[Composing operations]
\label{lem:transition}
Let $T^0,T^1,\dots,T^k$ be BSTs on the same keys. For $1\le j\le k$, let $U_j$ be a top set of $T^{j-1}$ such that every node outside $U_j$ has the same children in $T^{j-1}$ and in $T^j$. If some $U_j$ contains the key $x$, then a search for $x$ can transform $T^0$ into $T^k$ at cost at most $\sum_{j=1}^k\abs{U_j}$.
\end{lemma}

\begin{proof}
Let $V\coloneqq U_1\cup\dots\cup U_k$. A node $w$ outside $V$ has the same children in all the trees $T^0,\dots,T^k$, and these children are outside $V$ as well, because a child of $w$ in a top set $U_j$ of $T^{j-1}$ would put its parent $w$ into $U_j$. So the subtree of $w$ is the same in $T^0$ and in $T^k$, and it avoids $V$. Hence $V$ is a top set of both trees, with the same subtrees hanging off it, and $T^k$ arises from $T^0$ when one search for $x$ replaces the subtree formed by $V$, at cost $\abs V\le\sum_{j=1}^k\abs{U_j}$.
\end{proof}

We always use the lemma in the same way. The \emph{cost of an operation} that turns a tree $T$ into a tree $T'$ is the number of keys that change a child or are ancestors in $T$ of such a key; these keys form a top set of $T$, which serves as $U_j$. The walk from the root to the searched key is one more step, which changes nothing, with the searched key and its ancestors as top set. It may be placed before, between or after the operations. So a search costs at most the sum of the costs of its operations, plus one more than the depth of the searched key in whichever of the trees is convenient.

\subsection{Biased search trees}
\label{sec:biased}

We use the globally biased binary trees of Bent, Sleator and Tarjan \cite[Section~4]{bst85}. Such a tree is a full binary search tree whose leaves hold items with positive weights, in order; its internal nodes hold no items. Every node stores an integer, its \emph{rank}, which for the leaf of an item of weight $w$ is $\floor{\log_2w}$, and the rules of \cite{bst85} that relate the rank of a node to the ranks of its children keep heavy items near the root. We use these rules only through the three bounds below. We import three bounds, in which $W$ is the total weight of the trees involved; the last two hold in the worst case \cite[p.~563]{bst85}.
\begin{itemize}
\item {\bf Depth.} The leaf of an item of weight $w$ has depth $O(1+\log(W/w))$ \cite[Lemma~2 and p.~559]{bst85}.
\item {\bf Join.} Two trees, the items of the first preceding those of the second, can be joined in time $O(1+\log(W/(w_-+w_+)))$, where $w_-$ is the weight of the last item of the first tree and $w_+$ that of the first item of the second \cite[Theorem~8]{bst85}.
\item {\bf Split.} A tree can be split at the leaf of an item of weight $w$, into the tree of the items before it and the tree of the items after it, in time $O(1+\log(W/w))$ \cite[Theorem~10]{bst85}.
\end{itemize}
Weights below $1$ are allowed, because multiplying all weights by a power of two shifts all ranks equally, and where the algorithms leave a choice we fix one, so that they are deterministic. We use these trees in two ways: with weights, for a tree whose leaves are intervals of keys, and with all weights $1$, for a tree on the keys of one interval. With $N$ leaves of weight $1$, every leaf has depth $O(1+\log N)$, and a join or a split takes $O(1+\log N)$ time.

Our algorithm changes its trees by a few joins and splits at a time. The next lemma bounds the part of a tree that they can affect, and its proof uses how the two algorithms proceed \cite[pp.~560--563]{bst85}, not only their running times. Every node has a \emph{record}, which consists of its rank and of its pointers to its children and to its parent, and an operation \emph{visits} a node when it reads or writes its record. The two algorithms proceed as follows.
\begin{itemize}
\item {\bf Join.} The paper has two forms of join, the local join \cite[p.~560]{bst85} and the global join \cite[pp.~561--563]{bst85}; the joins that our algorithm calls are global joins. In either form, a join starts at the roots of its two trees and from then on moves only from a node to a child of this node, down the right spine of the first tree and the left spine of the second. The root of the tree that it returns is a new node or a node that the join has visited.
\item {\bf Split.} A split at a leaf \cite[pp.~560--561]{bst85} starts at this leaf and moves from each node to its parent until it reaches the root, and it removes the nodes of this path. On the way it collects the subtrees that hang off the path, whose roots are children of path nodes, into two trees, one for each side of the leaf. The first subtree on a side is taken as it is, and each later one is joined, by a local join, with the tree collected so far on its side.
\end{itemize}
We use two consequences of this description.
\begin{itemize}
\item {\bf Changes.} The children of a node are stored in its record, so an operation changes the children of a node, or removes the node, only by writing its record, which is a visit.
\item {\bf Moves.} An operation comes to a node that it has not visited before in one of three ways: from the parent of the node at that moment, in a join; from a child of the node, on the walk of a split to the root, which then goes on to the parent of the node; or as the root of a tree to which a join is applied. In the third case, if the join belongs to a split, the node is the root of a subtree that hung off the path of the split. Indeed, the join is applied to such a subtree and to the tree collected so far, which is again such a subtree, taken as it is, or was returned by an earlier join of the split, and then its root is new or was visited by that join.
\end{itemize}

\begin{lemma}[Visited nodes]
\label{lem:wc-visited}
Apply a sequence of joins and splits to some trees, the \emph{starting trees}, and let $V$ be the set of the internal nodes of the starting trees that the operations visit.
\begin{itemize}
\item {\bf Part 1.} In every starting tree, the nodes of $V$ form a top set.
\item {\bf Part 2.} Every internal node of a starting tree outside $V$ keeps its two children until the end.
\item {\bf Part 3.} $\abs V$ is at most a constant times the total running time of the operations.
\end{itemize}
\end{lemma}

\begin{proof}
Part~2 is the first consequence, and Part~3 holds because every visit takes constant time. For Part~1, let $z\in V$ have the parent $y$ in its starting tree. We show that $y\in V$; going up from $z$ then gives Part~1. Consider the first visit to $z$. If $y$ is no longer the parent of $z$ at this moment, then $y$ has been removed or has changed a child, so $y$ has been visited. Otherwise, by the second consequence, the operation comes to $z$ from $y$, or from a child of $z$ on the walk of a split, which goes on to $y$, or $z$ is the root of a tree to which a join is applied. In the last case $z$ still has the parent $y$. A join that is called on its own is applied to whole trees, whose roots have no parents, so this join belongs to a split, and the tree hangs off the path of this split. Then $y$ lies on this path, which the split visits.
\end{proof}

\section{The algorithm}
\label{sec:algorithm}

This section states our online BST completely and without proofs; the ideas behind it are those of \cref{sec:overview-count-promoted,sec:overview-fingers}. \Cref{sec:invariants,sec:analysis} prove the facts that its steps rely on and the bound on its cost.

\subsection{The grid and the locality scale}
\label{sec:scales}

We first fix the blocks of keys with which the algorithm works, and the quantity that bounds its cost. Throughout \cref{sec:algorithm,sec:invariants,sec:analysis}, $n\ge2$, the keys are $1,\dots,n$, and $X=(x_1,\dots,x_m)$ is a sequence of searches. Let $K\coloneqq\ceil{\log_2n}$. As in \cref{sec:intro}, $\rho(u,t)$ is the number of distinct keys among $x_{u+1},\dots,x_{t-1}$, for times $u<t$.

\paragraph{Blocks.} Let $\hat K$ be the least power of two with $\hat K\ge K$. For $0\le a\le\hat K$, the \emph{blocks of height} $a$ are the nonempty sets
\begin{align*}
\{j2^a+1,\dots,(j+1)2^a\}\cap\{1,\dots,n\},\qquad j\ge0.
\end{align*}
They partition the keys into intervals of $2^a$ consecutive keys, except that the last one may be shorter. We write $B_a(y)$ for the block of height $a$ that contains the key $y$ and $h(g)$ for the height of a block $g$. A block of height $0$ is a single key, and for $K\le a\le\hat K$ the only block of height $a$ consists of all keys. We call the block of height $\hat K$ the \emph{top block}. Since $2^a$ divides $2^{a'}$ for $a\le a'$, we have $B_a(y)\subseteq B_{a'}(y)$, so two blocks are either disjoint or nested. Two blocks of different heights can consist of the same keys, at the right end of the key range and at the heights from $K$ on; we still regard them as different blocks. The blocks of all heights form a \emph{grid} on the keys that is fixed in advance and does not depend on the searches. A search $u$ \emph{lies in} a block $g$ if $x_u\in g$, and $g$ is \emph{searched before} a time $t$ if some search $u<t$ lies in it.

\paragraph{Scales.} A \emph{scale} is an integer $s\in\{0,\dots,\hat K\}$. A scale stands for a height of blocks and at the same time for a number of keys: at scale $s$ we look at the block of $2^s$ keys around a key, and we count up to $2^s$ distinct keys in time.

\begin{definition}[Locality scale]
\label{def:locality}
The \emph{locality scale} $\lambda(t)$ of a search $t\ge2$ is the least scale $s$ such that some search $u<t$ satisfies
\begin{align*}
x_u\in B_s(x_t)\qquad\text{and}\qquad\rho(u,t)+2\le2^s.
\end{align*}
For the first search, $\lambda(1)\coloneqq K$.
\end{definition}

The scale $K$ has this property with $u=t-1$, because the block of height $K$ contains all keys, among them $x_{t-1}$, and $\rho(t-1,t)=0$ and $2^K\ge2$; the scale $0$ does not have it. So $1\le\lambda(t)\le K$ for every search $t$. If a scale $s$ has the property with a search $u$, then every larger scale $s'$ has it with the same $u$, because $B_s(x_t)\subseteq B_{s'}(x_t)$ and $2^s\le2^{s'}$. Hence, for every search $t\ge2$ and every scale $s$,
\begin{align}
\min\bigl\{\rho(u,t):u<t,\ x_u\in B_s(x_t)\bigr\}\le2^s-2\quad\Longleftrightarrow\quad s\ge\lambda(t),
\label{eq:lambda}
\end{align}
where the minimum of the empty set is $\infty$. Finally, $\rho(u,t)$ does not increase when $u$ increases, because the searches strictly between a later $u$ and $t$ are among those between an earlier $u$ and $t$. So the minimum in \eqref{eq:lambda} is attained by the last search before $t$ whose key lies in $B_s(x_t)$, and at a scale $s$ it suffices to test this one search.

\subsection{Kept blocks}
\label{sec:fingers}

The tree holds near its root the blocks that were searched most recently, at a sparse set of scales.

\paragraph{Promoted scales.} The tree keeps blocks only at the heights in
\begin{align*}
\mathcal S\coloneqq\{1,2,4,\dots,\hat K\}=\{2^j:j\ge0,\ 2^j\le\hat K\}.
\end{align*}
These \emph{promoted scales} are spaced geometrically, so that the promoted scales below any number $\lambda$ sum to less than $2\lambda$. They are also dense enough: for every scale $s$ with $1\le s\le K$, the least power of two $\hat s$ with $\hat s\ge s$ is a promoted scale, and $\hat s<2s$. The largest promoted scale is $\hat K$, and its only block is the top block.

\paragraph{The most recent blocks.} Fix a time $t\in\{1,\dots,m+1\}$, where the time $m+1$ stands for the moment after the last search. For a block $g$ that is searched before $t$, the latest search $u<t$ that lies in $g$ is the \emph{last search} in $g$ before $t$. Two blocks of the same height are disjoint, so they have different last searches. Of two blocks of the same height that are searched before $t$, the one with the later last search was searched \emph{more recently}.

\begin{definition}[Kept blocks]
\label{def:kept}
Let $s\in\mathcal S$ and $t\in\{1,\dots,m+1\}$. The \emph{kept blocks of height $s$ at time $t$} are the $2^s$ blocks of height $s$ that were searched most recently before $t$; if fewer than $2^s$ blocks of height $s$ are searched before $t$, all of these are kept. The top block is kept at all times. We write $\mathcal F_t$ for the set of the blocks that are kept at time $t$. A promoted scale $s$ is \emph{missing} for search $t$ if $B_s(x_t)\notin\mathcal F_t$.
\end{definition}

So $\mathcal F_1$ consists of the top block, and the rule is online: $\mathcal F_t$ is determined by $x_1,\dots,x_{t-1}$. After search $t$, the block $B_s(x_t)$ is the most recently searched block of its height, so it is kept at time $t+1$; in this sense search $t$ \emph{promotes} its blocks at all promoted scales. At a missing scale $s$, if $2^s$ blocks of height $s$ are kept, the one of them that was searched least recently is no longer kept after the search; we call it the \emph{victim} of the scale.

The algorithm maintains a set $\mathcal F$ of blocks, which is $\mathcal F_t$ when search $t$ begins and $\mathcal F_{t+1}$ when it ends (\cref{lem:kept}). It \emph{retires} a block when it removes the block from $\mathcal F$, and it \emph{adds} a block when it puts the block into $\mathcal F$. At a moment during a search, the kept blocks are the blocks of $\mathcal F$ at this moment.

\paragraph{Owners.} Let $\mathcal F$ be a set of blocks that contains the top block. The \emph{owner} $\operatorname{own}_{\mathcal F}(y)$ of a key $y$ is the block of least height in $\mathcal F$ that contains $y$. The \emph{parent} of a block $g\in\mathcal F$ other than the top block is the block of least height in $\mathcal F$ among those that contain $g$ and are higher than $g$. The \emph{serving block} $f_t$ of search $t$ is the owner of $x_t$ in $\mathcal F_t$, the smallest kept block around the searched key.

\subsection{Regions and the tree they stand for}
\label{sec:regions}
\label{sec:layout}

The kept blocks cut the keys into intervals.

\paragraph{Regions.} Let $\mathcal F$ be a set of blocks that contains the top block. A \emph{region} of $\mathcal F$ is a maximal interval of consecutive keys that have the same owner. We say that this block \emph{owns} the region, and the \emph{height} $h(Q)$ of a region $Q$ is the height of its owner. We write $\operatorname{Reg}(\mathcal F)$ for the set of the regions of $\mathcal F$. By definition the regions partition the keys into intervals. A block owns the keys that lie in it and in no block of $\mathcal F$ of smaller height, so it owns at most one region between any two consecutive such blocks inside it and at most one at each end. \Cref{fig:fingers}(a) shows the regions of the running example.

\paragraph{The region tree and the inner trees.} A \emph{region tree} is a full binary tree $\Rtree$ whose leaves are the regions, in key order. An \emph{inner tree} of a region $Q$ is a full binary tree whose leaves are the keys of $Q$, in order. Given a region tree and an inner tree for each of its regions, replace every leaf of $\Rtree$ by the inner tree of its region. The result is one full binary tree whose leaves are the keys $1,\dots,n$, in order. We call it the \emph{expanded tree}. Its internal nodes are those of $\Rtree$ and those of the inner trees. It stands for a BST on the keys in the following way.

\begin{definition}[Realization]
\label{def:layout}
The \emph{realization} $T(\Rtree)$ of a region tree $\Rtree$ with inner trees for its regions is obtained from the expanded tree in two steps:
\begin{itemize}
\item every internal node gets the key of the last leaf in its left subtree, and
\item the leaves are removed, except the last leaf, which keeps the key $n$.
\end{itemize}
\end{definition}

Every internal node of a full binary tree separates two consecutive leaves, the last leaf of its left subtree and the first leaf of its right subtree, and each pair of consecutive leaves is separated by exactly one internal node. So a key $y<n$ is the key of exactly one node, the node that separates the leaves of $y$ and $y+1$, and the key $n$ is at its own leaf. The node of $y$ is an ancestor of the leaf of $y$, so the depth of a key in $T(\Rtree)$ is at most the depth of its leaf in the expanded tree. The keys are also in search-tree order, so $T(\Rtree)$ is a BST on $\{1,\dots,n\}$ (\cref{lem:realize-bst}).
An internal node of $\Rtree$ gets the largest key of the last region in its left subtree, and the other keys of a region are at the internal nodes of its inner tree, except the key $n$. \Cref{fig:realized} in \cref{sec:overview-fingers} shows the realization in the situation of \cref{fig:fingers}(a).

The algorithm maintains the region tree and the inner trees as globally biased binary trees (\cref{sec:biased}). In the region tree, a region $Q$ has the weight
\begin{align*}
w(Q)\coloneqq4^{-h(Q)},
\end{align*}
and in an inner tree every key has the weight $1$. With these weights every key of a region $Q$ has depth $O(h(Q)+1)$ in $T(\Rtree)$ (\cref{lem:pathdepth}).

\subsection{Processing a search}
\label{sec:realized}
\label{sec:operations}

The procedure {\sc Search} of \cref{alg:search} processes search $t$ in three phases: it makes room at the missing scales by retiring their victims, walks down to $x_t$, and adds the blocks of $x_t$ at the missing scales. \Cref{alg:retire,alg:add,alg:change} give the procedures {\sc Retire}, {\sc Add} and {\sc ChangeRegions}, which change the region tree and the inner trees, and with them the BST $T(\Rtree)$, by the splits and joins of \cref{sec:biased}. Initially the top block is the only kept block, and all keys form one region, whose inner tree is a fixed globally biased tree on all keys. Which kept block of a height was searched least recently is part of the records about past searches, which are free in the model of \cref{sec:prelim}; \cref{app:runtime} maintains them.

\begin{algorithm}[!t]
\caption{The online BST: members and the procedures {\sc Initialize} and {\sc Search}.}
\label{alg:search}
\begin{algorithmic}[1]
\DataStructure{OnlineBST}
  \Members
    \State $\mathcal F$, a set of blocks \Comment{the kept blocks}
    \State $\Rtree$, a globally biased binary tree on the regions of $\mathcal F$ \Comment{the region tree; the BST is $T(\Rtree)$}
    \State for every region, a globally biased binary tree on its keys, with weights $1$ \Comment{the inner trees}
  \EndMembers
  \Procedure{Initialize}{$n$}
    \State $\mathcal F\gets\{\text{the top block}\}$
    \State $\Rtree\gets$ the tree with one leaf, the region $\{1,\dots,n\}$ \Comment{its inner tree holds all keys}
  \EndProcedure
  \Procedure{Search}{$t$} \Comment{search $t$, for the key $x_t$}
    \State $M\gets\{s\in\mathcal S:B_s(x_t)\notin\mathcal F\}$ \Comment{the missing scales of search $t$}\label{line:missing}
    \State \PhaseComment{Phase 1: make room at the missing scales}
    \For{each $s\in M$, by increasing $s$}\label{line:loop}
      \If{$\mathcal F$ has $2^s$ blocks of height $s$}\label{line:full}
        \State $e\gets$ the block of height $s$ in $\mathcal F$ that was searched least recently \Comment{the victim}\label{line:victim}
        \State \Call{Retire}{$e$} \Comment{cost $O(s+1)$ (\cref{lem:amortized}, Part~1)}\label{line:retire}
      \EndIf
    \EndFor
    \State \PhaseComment{Phase 2: serve the search}
    \State walk from the root of $T(\Rtree)$ down to $x_t$ \Comment{cost $O(\lambda(t))$ (\cref{lem:amortized}, Part~2)}\label{line:serve}
    \State \PhaseComment{Phase 3: promote the blocks of $x_t$ at the missing scales}
    \For{each $s\in M$, by decreasing $s$}\label{line:addloop}
      \State \Call{Add}{$B_s(x_t)$} \Comment{cost $O(\lambda(t))$ for all of them (\cref{lem:amortized}, Part~3)}\label{line:add}
    \EndFor
  \EndProcedure
\EndDataStructure
\end{algorithmic}
\end{algorithm}

\Cref{sec:inv-regions} verifies that the regions named in \cref{alg:retire,alg:add} exist and that after each step the leaves of $\Rtree$ are the regions of $\mathcal F$.

\emph{Make room.} When a victim retires, it contains no kept block of smaller height (\cref{lem:kept}), so its keys form a single region. When such a block $g$ is removed, its keys pass to its parent $p$, and they join the regions that $p$ owns on either side of $g$, if there are any; a neighbouring region that $p$ does not own stays as it is. This is the reverse of the addition of a block, described under \emph{Promote} below: at most three consecutive regions become one region again.

\begin{algorithm}[H]
\caption{The online BST, procedure {\sc Retire}: remove a block that contains no kept block of smaller height.}
\label{alg:retire}
\begin{algorithmic}[1]
\DataStructure{OnlineBST} \Comment{continues \cref{alg:search}}
  \Procedure{Retire}{$g$} \Comment{the keys of $g$ form one region; the parent of $g$ is $p$}
    \State $\mathcal Q\gets$ the list of the region $g$ and of those of its two neighbouring regions that $p$ owns\label{line:list}
    \State $\mathcal F\gets\mathcal F\setminus\{g\}$
    \State \Call{ChangeRegions}{$\mathcal Q$, the union of the regions of $\mathcal Q$}\label{line:merge}
  \EndProcedure
\EndDataStructure
\end{algorithmic}
\end{algorithm}

\emph{Serve.} The search only walks down; the region tree does not change. The serving block $f_t$ owns the region of $x_t$, so $x_t$ has depth $O(h(f_t)+1)$, and the serving block is low: $h(f_t)<2\lambda(t)$ (\cref{lem:serving}).

\emph{Promote.} The blocks of $x_t$ at the missing scales are added one at a time, from the highest to the lowest. A block $g$ that is added contains no kept block of smaller height, so all its keys have the same owner and lie in one region $Q$, the region of $x_t$ at this moment (\cref{lem:regions}). Afterwards $g$ owns its keys, and the old owner still owns what is left of $Q$ on the two sides of $g$. So $Q$ is cut into at most three regions: the keys $Q_L$ of $Q$ that are smaller than those of $g$, the block $g$, and the keys $Q_R$ of $Q$ that are larger, where $Q_L$ and $Q_R$ may be empty. The next block is added inside the region $g$.

\begin{algorithm}[H]
\caption{The online BST, procedure {\sc Add}: add a block that contains no kept block of smaller height.}
\label{alg:add}
\begin{algorithmic}[1]
\DataStructure{OnlineBST} \Comment{continues \cref{alg:search}}
  \Procedure{Add}{$g$} \Comment{the keys of $g$ lie in one region}
    \State $Q\gets$ the region that contains $g$\label{line:region}
    \State $\mathcal F\gets\mathcal F\cup\{g\}$
    \State \Call{ChangeRegions}{$Q$, the nonempty ones of $Q_L,g,Q_R$} \Comment{the regions of $\mathcal F$ inside $Q$}\label{line:replace}
  \EndProcedure
\EndDataStructure
\end{algorithmic}
\end{algorithm}

\emph{Change of regions.} {\sc Retire} and {\sc Add} only decide which regions are replaced by which, and {\sc ChangeRegions} changes the trees. For the inner trees, it joins those of the old regions into one tree $I$ and then \emph{cuts} $I$ into the inner trees of the new regions, where to cut a tree after a key $y$ is to split it at the leaf of $y$ and to join this leaf to the tree of the smaller keys. All joins of several trees go from left to right. A call costs $O(a+1)$, where $a$ is the largest height of a region that it removes or creates (\cref{lem:opcost}).

\begin{algorithm}[H]
\caption{The online BST, procedure {\sc ChangeRegions}: replace the consecutive leaves $\mathcal Q$ of $\Rtree$ by the regions $\mathcal Q'$, which have the same union. Each of the two lists has at most three regions. The comments give the time of each line, where $a$ is the largest height of a region of $\mathcal Q$ or $\mathcal Q'$ (\cref{lem:opcost}).}
\label{alg:change}
\begin{algorithmic}[1]
\DataStructure{OnlineBST} \Comment{continues \cref{alg:search}}
  \Procedure{ChangeRegions}{$\mathcal Q,\mathcal Q'$}
    \State split $\Rtree$ at the leaves of $\mathcal Q$, into $\Rtree_1$, $\mathcal Q$, $\Rtree_2$ \Comment{$O(a+1)$}\label{line:split}
    \State $I\gets$ the join of the inner trees of $\mathcal Q$ \Comment{$O(a+1)$}\label{line:joininner}
    \State cut $I$ into the inner trees of $\mathcal Q'$ \Comment{$O(a+1)$}\label{line:cut}
    \State $\Rtree\gets$ the join of $\Rtree_1$, the regions of $\mathcal Q'$ as single leaves, and $\Rtree_2$ \Comment{$O(a+1)$}\label{line:joinall}
  \EndProcedure
\EndDataStructure
\end{algorithmic}
\end{algorithm}

The main technical result of the paper is that this algorithm is an online BST in which every search is paid for by its own locality scale.

\begin{theorem}[Online bound in terms of scales]
\label{thm:scales}
There is an absolute constant $\gamma$ with the following property. For every $n\ge2$, \cref{alg:search} is a deterministic online BST algorithm on the keys $1,\dots,n$ that serves every search $t$ of every sequence $X$ of searches at cost at most
\begin{align*}
\gamma\cdot\lambda(t).
\end{align*}
In particular, every search costs at most $\gamma\ceil{\log_2n}$.
\end{theorem}
\section{Kept blocks and regions}
\label{sec:invariants}

A BST always finds the searched key, and by \cref{lem:transition} any sequence of operations on a BST is a legal execution, so \cref{alg:search} needs no proof of correctness in the usual sense. Its steps, however, rely on facts about how the kept blocks and the regions change. The set $\mathcal F$ must follow the kept blocks of \cref{def:kept}, and there must be few of them. A retiring block must be a single region. A new block must lie inside one region. And the realization of a region tree must be a BST. This section proves these facts, and \cref{sec:analysis} uses the same facts to bound the cost.

\subsection{The kept blocks}
\label{sec:inv-fingers}

The next lemma is what the rest of the paper uses about the kept blocks.

\begin{lemma}[Kept blocks]
\label{lem:kept}
Parts 1 and 2 hold for every time $t\in\{1,\dots,m+1\}$, and Parts 3 to 5 for every search $t$.
\begin{itemize}
\item {\bf Part 1 (count).} For every $s\in\mathcal S$, at most $2^s$ blocks of $\mathcal F_t$ have height $s$.
\item {\bf Part 2 (nesting).} If a block $g\in\mathcal F_t$ has height $s\in\mathcal S$, then for every promoted scale $s'>s$ the block of height $s'$ that contains $g$ is in $\mathcal F_t$.
\item {\bf Part 3 (the blocks of $x_t$).} The missing scales of search $t$ are the promoted scales below some bound, and each of them is less than $\lambda(t)$.
\item {\bf Part 4 (one search).} The family $\mathcal F_{t+1}$ arises from $\mathcal F_t$ as follows. For every missing scale $s$ of search $t$, the block $B_s(x_t)$ is added, and if $\mathcal F_t$ has $2^s$ blocks of height $s$, then the one among them that was searched least recently, the victim of the scale, is removed. Nothing else changes.
\item {\bf Part 5 (the steps of the algorithm).} Suppose that $\mathcal F=\mathcal F_t$ when {\sc Search}$(t)$ begins. Then $\mathcal F=\mathcal F_{t+1}$ when it ends. Every block that it retires or adds contains, at that moment, no block of $\mathcal F$ of smaller height. And at every moment of the procedure, $\mathcal F$ has the nesting property of Part~2 and at most $2^s$ blocks of height $s$ for every $s\in\mathcal S$.
\end{itemize}
\end{lemma}

The proof works with one number per block, its rank. Let $g$ be a block of height $s$, and let $\ell_t(g)$ be the last search in $g$ before $t$. Then
\begin{align*}
R_t(g)\coloneqq\bigl\{B_s(x_u):\ell_t(g)<u<t\bigr\},\qquad r_t(g)\coloneqq\abs{R_t(g)},
\end{align*}
and $r_t(g)\coloneqq\infty$ if $g$ is not searched before $t$. So the \emph{rank} $r_t(g)$ is the number of blocks of height $s$ that were searched since the last search in $g$. The most recently searched block of a height has the rank $0$. This rank of a block has nothing to do with the ranks of the nodes of a biased search tree (\cref{sec:biased}), which do not occur in this section.

\begin{proof}
Each part is a short computation with ranks. At the height $\hat K$ every claim holds trivially, because the top block is the only block there and is always kept. So let $s\in\mathcal S$ with $s<\hat K$ throughout, and write $\mathcal F_t^s\coloneqq\{g\in\mathcal F_t:h(g)=s\}$.

We first restate \cref{def:kept} in terms of ranks. Let $g\ne g'$ be blocks of height $s$. Then $\ell_t(g')>\ell_t(g)$ if and only if $g'$ contains some $x_u$ with $\ell_t(g)<u<t$, that is, if and only if $g'\in R_t(g)$. So $r_t(g)$ is the number of blocks of height $s$ that were searched more recently than $g$, and the blocks of height $s$ that are searched before $t$ have the ranks $0,1,2,\dots$, each rank once. Since the kept blocks are the $2^s$ most recently searched ones,
\begin{align}
\mathcal F_t^s=\bigl\{g:h(g)=s,\ r_t(g)<2^s\bigr\}.
\label{eq:recent}
\end{align}

\emph{Part 1.} By \eqref{eq:recent}, the blocks of $\mathcal F_t^s$ have distinct ranks in $\{0,\dots,2^s-1\}$.

\emph{Part 2.} We show that ranks do not increase upwards: if $g\subseteq g'$ and $h(g)=s<h(g')=s'$, then
\begin{align}
r_t(g')\le r_t(g).
\label{eq:rank-up}
\end{align}
This is trivial if $r_t(g)=\infty$, so let $g$ be searched before $t$. Let $\pi$ map a block of height $s$ to the block of height $s'$ that contains it. Every search in $g$ is a search in $g'$, so $\ell_t(g)\le\ell_t(g')$, and
\begin{align*}
R_t(g')=\pi\bigl(\{B_s(x_u):\ell_t(g')<u<t\}\bigr)\subseteq\pi\bigl(R_t(g)\bigr),\qquad\text{so}\qquad r_t(g')\le\abs{\pi(R_t(g))}\le r_t(g).
\end{align*}
Now let $g\in\mathcal F_t$. If $s'=\hat K$, then $g'$ is the top block, which is kept. Otherwise $r_t(g')\le r_t(g)<2^s<2^{s'}$ by \eqref{eq:rank-up} and \eqref{eq:recent}, so $g'\in\mathcal F_t$ by \eqref{eq:recent}.

\emph{Part 3.} Write $g_s\coloneqq B_s(x_t)$, so that $s$ is missing if and only if $r_t(g_s)\ge2^s$. If $s$ is not missing and $s'>s$ is promoted, then $s'$ is not missing by Part~2. So the missing scales are the promoted scales below some bound.

Let $t\ge2$ and $s\ge\lambda(t)$. By \eqref{eq:lambda}, some search $w<t$ lies in $g_s$ and has $\rho(w,t)\le2^s-2$, and $\ell_t(g_s)\ge w$. Hence
\begin{align*}
r_t(g_s)\overset{(a)}{\le}\rho\bigl(\ell_t(g_s),t\bigr)\overset{(b)}{\le}\rho(w,t)\le2^s-2<2^s,
\end{align*}
so $s$ is not missing. Here (a) holds because the blocks of $R_t(g_s)$ are disjoint and each contains a key searched between $\ell_t(g_s)$ and $t$, and (b) because $\rho(\cdot,t)$ does not increase. For $t=1$, only the top block is kept, so the missing scales are the promoted scales below $\hat K$, and these are less than $K=\lambda(1)$: the largest of them is $\hat K/2$, and $\hat K/2<K$ because $\hat K$ is the least power of two with $\hat K\ge K$.

\emph{Part 4.} Let $g\coloneqq g_s$. Search $t$ becomes the last search in $g$. For every other block $b$ of height $s$, it adds the time $t$, whose block is $g$, to the range of $u$ in $R_t(b)$. And $g$ is already in $R_t(b)$ exactly when $g$ was searched more recently than $b$. So
\begin{align*}
R_{t+1}(g)=\emptyset,\qquad R_{t+1}(b)=R_t(b)\cup\{g\},\qquad g\in R_t(b)\iff r_t(g)<r_t(b).
\end{align*}
Taking sizes, with $[P]\in\{0,1\}$ the truth value of $P$ and $\infty+1=\infty$,
\begin{align}
r_{t+1}(g)=0,\qquad r_{t+1}(b)=r_t(b)+\bigl[r_t(b)<r_t(g)\bigr]\quad\text{for }b\ne g.
\label{eq:rank-update}
\end{align}
So $g\in\mathcal F_{t+1}$, and by \eqref{eq:recent}, for $b\ne g$ of height $s$,
\begin{align*}
b\in\mathcal F_{t+1}\quad\Longleftrightarrow\quad r_t(b)+\bigl[r_t(b)<r_t(g)\bigr]<2^s.
\end{align*}
If $s$ is not missing, that is, $r_t(g)<2^s$, the right side is equivalent to $r_t(b)<2^s$, because $r_t(b)<r_t(g)$ implies $r_t(b)+1\le r_t(g)<2^s$. So $\mathcal F_{t+1}^s=\mathcal F_t^s$. If $s$ is missing, that is, $r_t(g)\ge2^s$, then every $b$ with $r_t(b)<2^s$ has $r_t(b)<r_t(g)$, and the right side is equivalent to $r_t(b)\le2^s-2$. So
\begin{align*}
\mathcal F_{t+1}^s=\bigl(\mathcal F_t^s\setminus\{e\}\bigr)\cup\{g\},\qquad e\coloneqq\text{the block with }r_t(e)=2^s-1,\text{ if there is one}.
\end{align*}
By Part~1, $e$ exists if and only if $\abs{\mathcal F_t^s}=2^s$, and then it has the largest rank in $\mathcal F_t^s$: it is the kept block searched least recently, the victim.

For example, let $s=1$, so that two blocks of height $1$ are kept, and let the blocks $a,b,c$ have the ranks $0,1,2$, so that $\mathcal F_t^1=\{a,b\}$. A search in $c$ is missing at the scale $1$, and by \eqref{eq:rank-update} the ranks become $c\colon0$, $a\colon1$, $b\colon2$. So $c$ is added and the victim $b$, of rank $2^1-1$, is removed.

\emph{Part 5.} Search $t$ makes the changes of Part~4 one block at a time: the victims leave from the lowest to the highest, and then the new blocks arrive from the highest to the lowest. A victim of height $1$ may lie inside the victim of height $2$, so it must leave first, and a new block of height $1$ lies inside the new block of height $2$, so it must arrive second.

For a block $b$ and a set $\mathcal G$ of blocks, let
\begin{align*}
{\uparrow}b\coloneqq\{B_{s'}(y):y\in b,\ s'\in\mathcal S,\ s'>h(b)\},\qquad{\downarrow}_{\mathcal G}b\coloneqq\{c\in\mathcal G:c\subseteq b,\ h(c)<h(b)\}
\end{align*}
be the blocks above $b$ at the promoted scales and the blocks of $\mathcal G$ below $b$. If $h(b)\in\mathcal S$, then $c\in{\downarrow}_{\mathcal G}b$ if and only if $c\in\mathcal G$ and $b\in{\uparrow}c$. Call $\mathcal G$ \emph{closed} if ${\uparrow}b\subseteq\mathcal G$ for every $b\in\mathcal G$; this is the nesting property, so $\mathcal F_t$ and $\mathcal F_{t+1}$ are closed by Part~2. Part~5 claims that every intermediate family $\mathcal F$ is closed with at most $2^s$ blocks of each height $s$, and that every block $b$ that is retired or added has ${\downarrow}_{\mathcal F}b=\emptyset$.

The set $M$ of \lineref{line:missing} consists of the missing scales, because $\mathcal F=\mathcal F_t$ there. Let
\begin{align*}
V\coloneqq\text{the victims of the scales in }M,\qquad A\coloneqq\{B_s(x_t):s\in M\},\qquad U\coloneqq\mathcal F_t\setminus V
\end{align*}
be the blocks that leave, the blocks that arrive, and the blocks that stay. Write $V_{\ge a}$, $V_{<a}$, $A_{\ge a}$ and $A_{>a}$ for the blocks of $V$ or $A$ whose height satisfies the condition. By Part~4, and since $A\cap\mathcal F_t=\emptyset$,
\begin{align}
\mathcal F_{t+1}=U\cup A,\qquad U=\mathcal F_t\cap\mathcal F_{t+1}\text{ is closed}.
\label{eq:kept-step}
\end{align}
When the first phase reaches a scale $s\in M$, only lower blocks have been retired, so the blocks of height $s$ in $\mathcal F$ are those of $\mathcal F_t$, and \linerange{line:full}{line:retire} retire the victim of height $s$. The third phase adds $A$ by decreasing height. So, with
\begin{align*}
\mathcal F^{(1)}_a\coloneqq U\cup V_{\ge a}=\mathcal F_t\setminus V_{<a},\qquad\mathcal F^{(3)}_a\coloneqq U\cup A_{\ge a},
\end{align*}
the family $\mathcal F$ runs through
\begin{align}
\begin{split}
&\text{first phase:}\qquad\mathcal F_t=\mathcal F^{(1)}_0\to\dots\to\mathcal F^{(1)}_s\to\mathcal F^{(1)}_{s+1}\to\dots\to U,\\
&\text{third phase:}\qquad U\to\dots\to\mathcal F^{(3)}_{s+1}\to\mathcal F^{(3)}_s\to\dots\to U\cup A=\mathcal F_{t+1},
\end{split}
\label{eq:kept-path}
\end{align}
where the index $a$ runs through all integers although the heights in $M$ are powers of two: the blocks of $V$ and $A$ have their heights in $M$, so the family changes only when $a$ passes a height $s\in M$. There the step $\mathcal F^{(1)}_s\to\mathcal F^{(1)}_{s+1}$ retires the victim of height $s$, if the scale has one, and the step $\mathcal F^{(3)}_{s+1}\to\mathcal F^{(3)}_s$ adds $B_s(x_t)$. Every $\mathcal F^{(1)}_a$ lies in $\mathcal F_t$ and every $\mathcal F^{(3)}_a$ in $\mathcal F_{t+1}$, which gives the count by Part~1. The rest follows from two facts:
\begin{align*}
&\text{(V)}\quad e\in V\ \Longrightarrow\ {\downarrow}_Ue=\emptyset,\qquad&&\text{since }c\in{\downarrow}_Ue\text{ gives }e\in{\uparrow}c\subseteq\mathcal F_{t+1},\text{ but }e\notin\mathcal F_{t+1};\\
&\text{(A)}\quad b\in A\ \Longrightarrow\ {\uparrow}b\subseteq A\cup U,\qquad&&\text{since }B_{s'}(x_t)\in A\text{ if }s'\in M,\text{ and }B_{s'}(x_t)\in\mathcal F_t\setminus V\text{ if }s'\notin M.
\end{align*}
In (V), ${\uparrow}c\subseteq\mathcal F_{t+1}$ because $c\in U\subseteq\mathcal F_{t+1}$ and $\mathcal F_{t+1}$ is closed. In (A), $B_{s'}(x_t)$ is not a victim because victims have heights in $M$.

The families in \eqref{eq:kept-path} are closed. Let $b\in\mathcal F^{(1)}_a$. Then ${\uparrow}b\subseteq\mathcal F_t$. If some $e\in{\uparrow}b$ were in $V_{<a}$, then $h(b)<h(e)<a$, so $b\in U$ and $b\in{\downarrow}_Ue$, contrary to (V). So ${\uparrow}b\subseteq\mathcal F_t\setminus V_{<a}=\mathcal F^{(1)}_a$. Let $b\in\mathcal F^{(3)}_a$. If $b\in U$, then ${\uparrow}b\subseteq U$. If $b\in A_{\ge a}$, then ${\uparrow}b\subseteq U\cup A_{>a}\subseteq\mathcal F^{(3)}_a$ by (A).

Finally, the victim $e$ of height $s$ is retired from $\mathcal F^{(1)}_s$, whose blocks of height less than $s$ lie in $U$, so ${\downarrow}_{\mathcal F^{(1)}_s}e={\downarrow}_Ue=\emptyset$ by (V). The block $g=B_s(x_t)$ is added to $\mathcal F^{(3)}_{s+1}$, whose blocks of height less than $s$ lie in $U\subseteq\mathcal F_t$. If $c\in{\downarrow}_{\mathcal F_t}g$, then $g\in{\uparrow}c\subseteq\mathcal F_t$, contrary to $s\in M$. So ${\downarrow}_{\mathcal F^{(3)}_{s+1}}g=\emptyset$.
\end{proof}

Initially $\mathcal F$ consists of the top block, which is $\mathcal F_1$. So by Part~5 and induction over the searches, $\mathcal F=\mathcal F_t$ when search $t$ begins, for every $t$, and the conclusions of Part~5 hold throughout every execution of \cref{alg:search}. Three consequences are used below. First, the blocks that search $t$ adds are the blocks of $x_t$ at its missing scales, which are promoted scales below $\lambda(t)$ by Part~3. Second, a victim of search $t$ does not contain $x_t$: it is a block of height $s$ in $\mathcal F_t$ for a missing scale $s$, hence a block other than $B_s(x_t)$. So the owner of $x_t$ is the serving block $f_t$ from the beginning of the search to its third phase. Third, by the nesting property, the parent of a kept block $g$ of height $s<\hat K$ is the kept block of height $2s$ that contains $g$, at the next promoted scale. So, for every kept block $g$ other than the top block, with parent $p$,
\begin{align}
h(p)=2h(g).
\label{eq:parent}
\end{align}

\subsection{How the regions change}
\label{sec:inv-regions}

The regions change only when one block is added or removed, and by Part~5 of \cref{lem:kept} the block then contains no kept block of smaller height. The next lemma describes this change: read forwards it is {\sc Add} (\cref{alg:add}), and read backwards it is {\sc Retire} (\cref{alg:retire}).

\begin{lemma}[Adding or removing a block]
\label{lem:regions}
Let $\mathcal F$ be a set of blocks that contains the top block, and let $g$ be a block that is not in $\mathcal F$ and contains no block of $\mathcal F$ of smaller height. Let $p$ be the block of least height in $\mathcal F$ among those that contain $g$ and are higher than $g$; this is the parent of $g$ in $\mathcal F\cup\{g\}$.
\begin{itemize}
\item {\bf Part 1 (owners).} In $\mathcal F$, every key of $g$ has the owner $p$. In $\mathcal F\cup\{g\}$, every key of $g$ has the owner $g$, and every other key has the same owner as in $\mathcal F$.
\item {\bf Part 2 (adding $g$).} The block $g$ lies in one region $Q$ of $\mathcal F$, which $p$ owns. The regions of $\mathcal F\cup\{g\}$ are those of $\mathcal F$, except that $Q$ is replaced by $g$ and by the at most two intervals of $Q\setminus g$, which $p$ still owns.
\item {\bf Part 3 (removing $g$).} The keys of $g$ form one region of $\mathcal F\cup\{g\}$. The regions of $\mathcal F$ are those of $\mathcal F\cup\{g\}$, except that $g$ and those of its two neighbouring regions that $p$ owns are merged into one region, owned by $p$.
\item {\bf Part 4 (the two procedures).} In every execution of \cref{alg:search}, the region $Q$ in \lineref{line:region} of \cref{alg:add} and the list $\mathcal Q$ in \lineref{line:list} of \cref{alg:retire} exist as stated there, every call of {\sc ChangeRegions} replaces at most three consecutive regions by at most three regions with the same union, and after each call the leaves of the region tree are the regions of $\mathcal F$.
\end{itemize}
\end{lemma}

\begin{proof}
A region is a maximal interval on which the owner is constant, so all four parts follow from one formula for the owners, \eqref{eq:own} below. For a set $\mathcal G$ of blocks that contains the top block and a key $z$, the blocks that contain $z$ form the chain $B_0(z)\subseteq B_1(z)\subseteq\dots\subseteq B_{\hat K}(z)$, and the owner of $z$ is its lowest block in $\mathcal G$:
\begin{align}
\operatorname{own}_{\mathcal G}(z)=B_{a}(z),\qquad a=\min J_{\mathcal G}(z),\qquad J_{\mathcal G}(z)\coloneqq\{b:B_b(z)\in\mathcal G\}.
\label{eq:block-chain}
\end{align}
Throughout, $\mathcal F'\coloneqq\mathcal F\cup\{g\}$, $a\coloneqq h(g)$, and $g=\{g^-,\dots,g^+\}$.

\emph{Part 1.} Let $z\in g$. For $b\le a$ we have $B_b(z)\subseteq B_a(z)=g$. So $B_b(z)\notin\mathcal F$ for $b<a$, by the assumption that $g$ contains no block of $\mathcal F$ of smaller height, and $B_a(z)=g\notin\mathcal F$. For $b>a$, the block $B_b(z)$ is the block $B_b(g)$ of height $b$ that contains $g$, the same for all $z\in g$. Hence, for every $z\in g$,
\begin{align*}
J_{\mathcal F}(z)=J\coloneqq\{b>a:B_b(g)\in\mathcal F\},\qquad J_{\mathcal F'}(z)=J\cup\{a\}.
\end{align*}
The set $J$ contains $\hat K$, because the top block is in $\mathcal F$ and $g\notin\mathcal F$ is not the top block. So $B_{\min J}(g)$ is the block of least height in $\mathcal F$ that contains $g$ and is higher than $g$, which is $p$. For $z\notin g$, the block $g$ is not in the chain of $z$, so $J_{\mathcal F'}(z)=J_{\mathcal F}(z)$. With \eqref{eq:block-chain},
\begin{align}
\operatorname{own}_{\mathcal F}(z)=p\quad\text{for }z\in g,\qquad\operatorname{own}_{\mathcal F'}(z)=\begin{cases}g,&z\in g,\\\operatorname{own}_{\mathcal F}(z),&z\notin g.\end{cases}
\label{eq:own}
\end{align}

\emph{Part 2.} By \eqref{eq:own}, $\operatorname{own}_{\mathcal F}$ is constant on the interval $g$, so $g$ lies in one region $Q=\{q^-,\dots,q^+\}$ of $\mathcal F$, and $p$ owns it. Cut $Q$ at $g$:
\begin{align*}
Q=Q_L\cup g\cup Q_R,\qquad Q_L\coloneqq\{q^-,\dots,g^--1\},\qquad Q_R\coloneqq\{g^++1,\dots,q^+\}.
\end{align*}
By \eqref{eq:own}, $\operatorname{own}_{\mathcal F'}$ is $p$ on $Q_L$, $g$ on $g$, $p$ on $Q_R$, and equal to $\operatorname{own}_{\mathcal F}$ outside $Q$. The owner changes at each cut, because $p\ne g$. It also changes at both ends of $Q$: the key $q^--1$, if it exists, has
\begin{align*}
\operatorname{own}_{\mathcal F'}(q^--1)=\operatorname{own}_{\mathcal F}(q^--1)\notin\{p,g\},
\end{align*}
where $\ne p$ holds because $Q$ is a maximal interval with owner $p$, and $\ne g$ because $g\notin\mathcal F$; and the same holds for $q^++1$. Hence the nonempty ones among $Q_L,g,Q_R$ are regions of $\mathcal F'$. Every region of $\mathcal F$ other than $Q$ is a region of $\mathcal F'$ as well: its keys keep their owner, and so does a key next to it, unless this key is $q^-$ or $q^+$, whose owner in $\mathcal F'$ is $p$ or $g$ and so is still different. Hence, with $\operatorname{Reg}(\mathcal G)$ the set of the regions of $\mathcal G$,
\begin{align}
\operatorname{Reg}(\mathcal F')=\bigl(\operatorname{Reg}(\mathcal F)\setminus\{Q\}\bigr)\cup\bigl(\{Q_L,g,Q_R\}\setminus\{\emptyset\}\bigr).
\label{eq:reg-change}
\end{align}

\emph{Part 3.} This is \eqref{eq:reg-change} read from right to left. By \eqref{eq:reg-change}, $g\in\operatorname{Reg}(\mathcal F')$. Its left neighbour in $\operatorname{Reg}(\mathcal F')$ is $Q_L$ if $Q_L\ne\emptyset$, and otherwise the region of $\mathcal F$ that ends at $q^--1$, whose owner is not $p$ by Part~2; similarly on the right. So
\begin{align*}
\{\text{neighbours of $g$ in $\operatorname{Reg}(\mathcal F')$ owned by }p\}=\{Q_L,Q_R\}\setminus\{\emptyset\},\qquad Q_L\cup g\cup Q_R=Q\in\operatorname{Reg}(\mathcal F),
\end{align*}
and $\operatorname{Reg}(\mathcal F)$ arises from $\operatorname{Reg}(\mathcal F')$ by replacing $g$ and these neighbours by their union $Q$, which $p$ owns.

\emph{Part 4.} We show by induction over the calls of {\sc ChangeRegions} that the leaves of the region tree are $\operatorname{Reg}(\mathcal F)$. Initially both are $\{\{1,\dots,n\}\}$. Consider a call of {\sc Add}$(g)$ on the current set $\mathcal F$. By Part~5 of \cref{lem:kept}, $g$ contains no block of $\mathcal F$ of smaller height, so Part~2 applies: the region $Q\supseteq g$ of \lineref{line:region} exists, and by \eqref{eq:reg-change} the regions of $\mathcal F\cup\{g\}$ inside $Q$ are the at most three sets $\{Q_L,g,Q_R\}\setminus\{\emptyset\}$, whose union is $Q$. This is the replacement of \lineref{line:replace}. Now consider a call of {\sc Retire}$(g)$, and let $\mathcal F^-\coloneqq\mathcal F\setminus\{g\}$. By Part~5 of \cref{lem:kept}, $g$ contains no block of $\mathcal F$, hence of $\mathcal F^-$, of smaller height, so the lemma applies to $\mathcal F^-$ and $g$, with $p$ the parent of $g$ in $\mathcal F$. By Part~3, the keys of $g$ form one region of $\mathcal F$, the list $\mathcal Q$ of \lineref{line:list} is $g$ together with $\{Q_L,Q_R\}\setminus\{\emptyset\}$, and these at most three consecutive regions are replaced by their union, which is the only region of $\mathcal F^-$ that is not a region of $\mathcal F$. In both cases a region that the call does not replace keeps its owner (Part~1), hence its height and its weight in the region tree.
\end{proof}

\subsection{The realized tree}
\label{sec:inv-layout}

It remains to check that the tree of \cref{def:layout} is a search tree.

\begin{lemma}[The realized tree is a BST]
\label{lem:realize-bst}
For every region tree $\Rtree$ with its inner trees, the realization $T(\Rtree)$ is a BST on $\{1,\dots,n\}$.
\end{lemma}

\begin{proof}
Every key occurs exactly once, as noted after \cref{def:layout}. For the order of the keys, let $v$ be an internal node of the expanded tree, with key $y$, so that the leaf of $y$ is the last leaf in the left subtree of $v$. A node of $T(\Rtree)$ in the left subtree of $v$ is an internal node, because the last leaf lies in no left subtree, and its key is the key of a leaf of this subtree. This key is not $y$, which is the key of $v$, so it is smaller than $y$. A node in the right subtree of $v$ is an internal node or the last leaf, and its key is the key of a leaf of the right subtree, so it is larger than $y$.
\end{proof}

\section{The cost of the algorithm}
\label{sec:analysis}

This section proves \cref{thm:scales}. The budget is the one noted in the comments of \cref{alg:search}: retiring a block of height $s$ costs $O(s+1)$, and the walk and the promotions of search $t$ cost $O(\lambda(t))$ each. A search retires at most one block at each of its missing scales, which are promoted scales below $\lambda(t)$ and sum to less than $2\lambda(t)$, so search $t$ costs $O(\lambda(t))$. \Cref{sec:cost-serving} shows that a search is served by a block of height less than $2\lambda(t)$, \cref{sec:cost-depth,sec:cost-operations} that a change of regions costs a constant times the largest height of its regions, and \cref{sec:cost-amortized} adds up.

\subsection{Searches are served low}
\label{sec:cost-serving}

The density of the promoted scales enters here: the promoted scale just above $\lambda(t)$ is not missing, so the block of $x_t$ at this scale is kept.

\begin{lemma}[The serving block]
\label{lem:serving}
For every search $t$, the serving block satisfies $h(f_t)\le2\lambda(t)-1$.
\end{lemma}

\begin{proof}
Let $\hat s$ be the least power of two with $\hat s\ge\lambda(t)$. We show that
\begin{align*}
h(f_t)\overset{(a)}{\le}\hat s\overset{(b)}{\le}2\lambda(t)-1.
\end{align*}
For (b), if $\hat s\ge2$, then the power of two $\hat s/2$ is less than $\lambda(t)$, so $\hat s\le2(\lambda(t)-1)$, and if $\hat s=1$, then $\hat s\le2\lambda(t)-1$ because $\lambda(t)\ge1$. For (a), $\hat s\le\hat K$ because $\lambda(t)\le K\le\hat K$ and $\hat K$ is a power of two, so $\hat s\in\mathcal S$, and $\hat s$ is not missing for search $t$, because the missing scales are less than $\lambda(t)$ (Part~3 of \cref{lem:kept}). So $B_{\hat s}(x_t)\in\mathcal F_t$. This block contains $x_t$, and $f_t=\operatorname{own}_{\mathcal F_t}(x_t)$ is the lowest block of $\mathcal F_t$ that does.
\end{proof}

The missing scales of one search have a small sum, and with them the heights of the blocks that the search retires and adds. Let $M_t$ be the set of the missing scales of search $t$, which is the set $M$ of \lineref{line:missing} of \cref{alg:search}. They are promoted scales $s<\lambda(t)$, by Part~3 of \cref{lem:kept}. If there are promoted scales below $\lambda(t)$, let $2^i$ be the largest of them. Then there are $i+1\le2^i$ of them and they sum to $2^{i+1}-1$, so with $2^i\le\lambda(t)-1$,
\begin{align}
\sum_{s\in M_t}(s+1)\le\sum_{s\in\mathcal S,\ s<\lambda(t)}(s+1)\le3\cdot2^i\le3\lambda(t).
\label{eq:heights}
\end{align}
If no promoted scale is below $\lambda(t)$, then $M_t$ is empty and \eqref{eq:heights} holds as well. Search $t$ retires at most one block of each height $s\in M_t$ and adds exactly one, and it retires and adds no other blocks (Parts 4 and 5 of \cref{lem:kept}).

\subsection{Weights and depths}
\label{sec:cost-depth}

The bounds of \cref{sec:biased} are in terms of the total weight of a tree. This subsection shows that the total weight of the regions is bounded by a constant, and concludes that the keys of a region of height $a$ have depth $O(a+1)$.

\begin{lemma}[Total weight]
\label{lem:weights}
At every moment of the execution, the total weight $W$ of all regions of the current set $\mathcal F$ is less than $2$.
\end{lemma}

\begin{proof}
We charge every region to a kept block that is not higher than its owner, in such a way that no block is charged more than twice. Let $\mathcal F$ be the set of the kept blocks at this moment, and write $\operatorname{own}\coloneqq\operatorname{own}_{\mathcal F}$. A region begins where its owner begins, or just after the last key of a lower kept block. So for a region $Q$ with first key $z$ and owner $f=\operatorname{own}(z)$, let
\begin{align*}
\varphi(Q)\coloneqq\begin{cases}f,&\text{if }z=\min f,\\\operatorname{own}(z-1),&\text{if }z>\min f.\end{cases}
\end{align*}
We show that
\begin{align}
h\bigl(\varphi(Q)\bigr)\le h(Q)\quad\text{for every region }Q,\qquad\abs{\varphi^{-1}(c)}\le2\quad\text{for every }c\in\mathcal F.
\label{eq:charge}
\end{align}
Let $z>\min f$, and let $c\coloneqq\operatorname{own}(z-1)$. The interval $f$ contains $z-1$, and $c\ne f$ because $Q$ is a maximal interval with the owner $f$. Both $c$ and $f$ are blocks of $\mathcal F$ that contain $z-1$, and $c$ is the lowest of these, so $h(c)<h(f)=h(Q)$. Moreover $z\notin c$, because $f$ is the lowest block of $\mathcal F$ that contains $z$. So
\begin{align*}
z-1=\max c.
\end{align*}
Hence a block $c\in\mathcal F$ is $\varphi(Q)$ only for the region $Q$ that contains $\min c$ and for the region $Q$ that begins at $\max c+1$, which proves \eqref{eq:charge}. Now
\begin{align*}
W=\sum_{Q\in\operatorname{Reg}(\mathcal F)}4^{-h(Q)}&\overset{(a)}{\le}\sum_{Q\in\operatorname{Reg}(\mathcal F)}4^{-h(\varphi(Q))}\overset{(b)}{\le}2\sum_{c\in\mathcal F}4^{-h(c)}\\
&\overset{(c)}{\le}2\sum_{s\in\mathcal S}2^s\cdot4^{-s}\le2\sum_{j\ge0}2^{-2^j}\overset{(d)}{<}2\cdot\frac78<2.
\end{align*}
Here (a) and (b) are the two parts of \eqref{eq:charge}, and (c) holds because at most $2^s$ blocks of $\mathcal F$ have height $s$, for every $s\in\mathcal S$ (Part~5 of \cref{lem:kept}). For (d), the terms with $j\ge3$ sum to less than $2\cdot2^{-8}$, so the sum is less than $\frac12+\frac14+\frac1{16}+\frac1{128}<\frac78$.
\end{proof}

\begin{lemma}[Size and depth of a region]
\label{lem:pathdepth}
Let $Q$ be a region of height $a$ at some moment between two steps of \cref{alg:search}. Then $Q$ has at most $2^a$ keys, and the leaf of every key of $Q$ has depth $O(a+1)$ in the expanded tree. In particular, the leaf of $Q$ has depth $O(a+1)$ in $\Rtree$, and every key of $Q$ has depth $O(a+1)$ in $T(\Rtree)$.
\end{lemma}

\begin{proof}
The region lies in its owner, a block of height $a$, so $\abs Q\le2^a$. In the expanded tree, the path to the leaf of a key $y\in Q$ goes through $\Rtree$ to the leaf of $Q$ and then through the inner tree of $Q$. So by the depth bound of \cref{sec:biased}, applied to both trees, the leaf of $y$ has depth
\begin{align*}
\underbrace{O\Bigl(1+\log\frac{W}{w(Q)}\Bigr)}_{\text{in }\Rtree}+\underbrace{O\bigl(1+\log\abs Q\bigr)}_{\text{in the inner tree}}\le O\bigl(1+\log(2\cdot4^a)\bigr)+O\bigl(1+\log2^a\bigr)=O(a+1),
\end{align*}
where $W<2$ by \cref{lem:weights} and $w(Q)=4^{-a}$, and the inner tree has $\abs Q$ leaves of weight $1$. The first term alone bounds the depth of the leaf of $Q$ in $\Rtree$. The statement about $T(\Rtree)$ follows, because the depth of a key in $T(\Rtree)$ is at most the depth of its leaf in the expanded tree (\cref{sec:regions}).
\end{proof}

\subsection{The cost of changing regions}
\label{sec:cost-operations}

Throughout, the tree is $T(\Rtree)$ for the current region tree, and each call of {\sc ChangeRegions} is one operation in the sense of \cref{lem:transition}: its cost is the number of keys that change a child or are ancestors, in the tree before the call, of such a key.

\begin{lemma}[Cost of changing regions]
\label{lem:opcost}
In every execution of \cref{alg:search}, consider a call of {\sc ChangeRegions} (\cref{alg:change}) that removes the regions of the list $\mathcal Q$ and creates the regions of the list $\mathcal Q'$. Let $a$ be the largest height of these regions, where the height of a region of $\mathcal Q$ is taken when the procedure that makes the call begins, before it changes $\mathcal F$, and that of a region of $\mathcal Q'$ after the call. Then the call has cost $O(a+1)$.
\end{lemma}

\begin{proof}
The cost of the call counts keys of the BST $T(\Rtree)$, while the bounds of \cref{sec:biased} count the nodes that the splits and joins visit in the region tree and in the inner trees. The two are related as follows. By \cref{def:layout}, the key of an internal node of the expanded tree, and the children of this key in $T(\Rtree)$, are determined by the subtree below the node: the key is that of the last leaf in the left subtree, and a child of the node in the expanded tree gives a child of this key in $T(\Rtree)$ exactly if it is an internal node or the leaf of the key $n$, which is then the last leaf of the subtree. So a node whose subtree the call leaves alone keeps its key, and this key keeps its children. For example, when {\sc Add} cuts a region $Q$ into three, the splits and joins work near the path from the root to the leaf of $Q$. A node of $\Rtree$ away from this path has the same regions below it before and after the call, with the same inner trees, so nothing changes at its key.

Let $E$ be the expanded tree when the call begins, and let $V$ consist of the internal nodes of $E$ that a split or a join of the call visits, together with the nodes of $\Rtree$ that lie above a leaf of $\mathcal Q$.
We show that
\begin{align}
\begin{split}
\text{cost of the call}\ \overset{(a)}{\le}\ \abs V\ &\overset{(b)}{\le}\ O\bigl(\text{time of the splits and joins}\bigr)+\sum_{Q\in\mathcal Q}\bigl(\text{depth of the leaf of $Q$ in }\Rtree\bigr)\\
&\overset{(c)}{\le}\ O(a+1).
\end{split}
\label{eq:opcost}
\end{align}

We prove the three steps in the order (b), (c), (a), because (a) uses the set $V$ that (b) and (c) bound.

\emph{Step (b).} We apply \cref{lem:wc-visited} with the starting trees $\Rtree$ and the inner trees of the regions of $\mathcal Q$; the single leaves of the new regions have no internal nodes. An operation on the region tree touches no record of an inner tree, and conversely. By Part~3 of the lemma, the first set in the definition of $V$ has $O(\text{time})$ nodes, and the second set has at most as many nodes as the sum in \eqref{eq:opcost}.

\emph{Step (c).} The lists $\mathcal Q$ and $\mathcal Q'$ have at most three regions each. So the four lines of \cref{alg:change} make at most $3$ splits, $2$ joins, $2$ cuts and $4$ joins, which are thirteen splits and joins in all, since a cut is a split and a join. Each takes $O(a+1)$ time:
\begin{align*}
\text{in the region tree (\lineref{line:split} and \lineref{line:joinall}):}&\quad O\Bigl(1+\log\frac Ww\Bigr)\le O\bigl(1+\log(2\cdot4^a)\bigr),\\
\text{in the inner trees (\lineref{line:joininner} and \lineref{line:cut}):}&\quad O(1+\log N)\le O\bigl(1+\log(3\cdot2^a)\bigr).
\end{align*}
In the first line, $W<2$, because every tree in \lineref{line:split} holds some of the leaves that $\Rtree$ has when the call begins, which are the regions at the beginning of the procedure that makes the call, and every tree in \lineref{line:joinall} holds some of the regions after the call (\cref{lem:weights}). And $w\ge4^{-a}$. A split is at a leaf of $\mathcal Q$. The joins of \lineref{line:joinall} go from left to right: $\Rtree_1$ is joined with the first leaf of $\mathcal Q'$, the result with the next leaf of $\mathcal Q'$, and at the end with $\Rtree_2$. So in every join, the last leaf of the first tree or the first leaf of the second tree is a region of $\mathcal Q'$, and the bound for a join in \cref{sec:biased} has the sum of the weights of these two leaves in its denominator. If $\Rtree_1$ or $\Rtree_2$ is empty, its join is omitted. In the second line, the inner trees of the call have $N\le3\cdot2^a$ leaves of weight $1$ in all (\cref{lem:pathdepth}). Finally, each of the at most three leaves of $\mathcal Q$ has depth $O(a+1)$ in $\Rtree$ (\cref{lem:pathdepth}).

\emph{Step (a).} This step rests on two properties of $V$. First, $V$ is a top set of $E$. Second, every internal node $v\notin V$ of $E$ has the same subtree in the expanded tree after the call.
For the first property, the nodes of $V$ in $\Rtree$ form a top set of $\Rtree$, because the visited ones do (Part~1 of \cref{lem:wc-visited}) and the nodes above a leaf do. The nodes of $V$ in the inner tree of a region $Q\in\mathcal Q$ form a top set of this tree, by the same Part~1, and all nodes of $\Rtree$ above the leaf of $Q$ are in $V$. For the second property, there are three cases.
\begin{itemize}
\item {\bf $v$ lies in the inner tree of a region outside $\mathcal Q$.} The call does not touch this tree.
\item {\bf $v$ lies in the inner tree of a region of $\mathcal Q$.} No node below $v$ is visited, because $V$ is a top set, so $v$ and the nodes below it keep their children (Part~2 of \cref{lem:wc-visited}).
\item {\bf $v$ is a node of $\Rtree$.} Again $v$ and the nodes of $\Rtree$ below it keep their children. The leaves of $\Rtree$ below $v$ are regions outside $\mathcal Q$, because $v$ does not lie above a leaf of $\mathcal Q$, and these regions keep their inner trees.
\end{itemize}
Now let $y$ be a key that changes a child in $T(\Rtree)$. It is not the key $n$, which is at a leaf before and after the call. So $y$ is the key of an internal node $v$ of $E$, and $v\in V$: otherwise $v$ would keep its subtree by the second property, and with it its key $y$ and the children of $y$. The ancestors of $y$ in $T(\Rtree)$ are the keys of the nodes above $v$ in $E$, which are in $V$ by the first property. So every key that the cost counts is the key of a node of $V$, and different keys have different nodes.
\end{proof}

\subsection{The cost of a search}
\label{sec:cost-amortized}

It remains to add up. The next lemma bounds the three kinds of steps of a search, and \cref{thm:scales} follows from it.

\begin{lemma}[Cost of the steps]
\label{lem:amortized}
In every execution of \cref{alg:search},
\begin{itemize}
\item {\bf Part 1.} retiring a block of height $s$ (\lineref{line:retire}) costs $O(s+1)$;
\item {\bf Part 2.} the walk of search $t$ (\lineref{line:serve}) costs $O(\lambda(t))$;
\item {\bf Part 3.} the promotions of search $t$ (\linerange{line:addloop}{line:add}) cost $O(\lambda(t))$ in all.
\end{itemize}
\end{lemma}

\begin{proof}
Each part combines \cref{lem:opcost} or the depth bound with the regions that the step changes.

In a retirement or an addition of a block $g$ with parent $p$, all regions that the call of {\sc ChangeRegions} removes or creates are owned by $g$ or by $p$ (Parts~2 and~3 of \cref{lem:regions}). Their heights are at most $h(p)$, so by \cref{lem:opcost}, retiring or adding $g$ costs $O(h(p)+1)$.

\emph{Retiring.} Let $g$ retire, with height $s$ and parent $p$. Then $h(p)=2s$ by \eqref{eq:parent}, so the cost is $O(2s+1)=O(s+1)$.

\emph{Serving.} The first phase does not change the owner of $x_t$, as noted after \cref{lem:kept}. So when the search is served, the region of $x_t$ has the height $h(f_t)$, and the walk costs one more than the depth of $x_t$, which is
\begin{align*}
O\bigl(h(f_t)+1\bigr)\le O\bigl(2\lambda(t)\bigr)
\end{align*}
by \cref{lem:pathdepth,lem:serving}.

\emph{Promoting.} Let $g_1,\dots,g_k$ be the blocks that search $t$ adds, in this order, with heights $s_1>\dots>s_k$; if the search adds no block, there is nothing to show. Let $p_i$ be the parent of $g_i$ when it is added. By Part~1 of \cref{lem:regions}, $p_i$ is the owner that the keys of $g_i$, among them $x_t$, have just before $g_i$ is added. Before the first addition this owner is the serving block. After the addition of $g_{i-1}$ it is $g_{i-1}$, the lowest kept block that contains $x_t$ at that moment, because the missing scales are consecutive promoted scales (Part~3 of \cref{lem:kept}) and the blocks of $x_t$ at the lower ones are added later. So
\begin{align*}
p_1=f_t,\qquad p_i=g_{i-1}\quad\text{for }2\le i\le k.
\end{align*}
The heights $s_1,\dots,s_k$ are the missing scales of the search. So the promotions cost
\begin{align*}
\sum_{i=1}^kO\bigl(h(p_i)+1\bigr)=O\bigl(h(f_t)+1\bigr)+\sum_{i=1}^{k-1}O(s_i+1)\overset{(a)}{\le}O\bigl(2\lambda(t)\bigr)+\sum_{s\in M_t}O(s+1)\overset{(b)}{=}O(\lambda(t)),
\end{align*}
where (a) is \cref{lem:serving} and (b) is \eqref{eq:heights}.
\end{proof}

\begin{proof}[Proof of \cref{thm:scales}]
By \cref{lem:transition}, the cost of a search is at most the sum of the costs of its steps, where the walk is one step. The steps of search $t$ are its retirements, its walk and its promotions. The search retires at most one block of each height $s\in M_t$ and no other block, so by \cref{lem:amortized} and \eqref{eq:heights} its cost is at most
\begin{align*}
\underbrace{\sum_{s\in M_t}O(s+1)}_{\text{retirements}}+\underbrace{O(\lambda(t))}_{\text{walk}}+\underbrace{O(\lambda(t))}_{\text{promotions}}=O(\lambda(t)).
\end{align*}
The constants do not depend on $n$, on $X$ or on $t$, which gives the constant $\gamma$, and $\lambda(t)\le K=\ceil{\log_2n}$. The algorithm is deterministic, and every step of search $t$ uses only $x_1,\dots,x_t$, so it is online.
\end{proof}

\section{From scales to the unified bound}
\label{sec:unified}

This section proves \cref{thm:main} from \cref{thm:scales}, by showing that $\sum_t\lambda(t)\le2\UB(X)+O(m)$. The locality scale measures the distance of two keys by the blocks of a grid that is fixed in advance, so it can exceed the term of a search in the unified bound by a factor of order $\log n$: two adjacent keys can lie on the two sides of a boundary between two high blocks. We show that in total the fixed grid loses only a factor $2$ and $O(m)$ (\cref{lem:keyloss}).

Throughout, $X=(x_1,\dots,x_m)$ is a sequence over $\{1,\dots,n\}$ with $n\ge2$, and $K$, the blocks $B_a(\cdot)$ and the locality scales $\lambda(t)$ are as in \cref{sec:scales}. For $t\ge2$ and $0\le J\le K$, let $v_J(t)$ be the last search before $t$ in the block of height $J$ around $x_t$, that is, the latest time $v<t$ with $x_v\in B_J(x_t)$, if there is one. By the remark after \eqref{eq:lambda}, some search satisfies the condition of \cref{def:locality} at the scale $J$ exactly if $v_J(t)$ exists and satisfies it, and by \eqref{eq:lambda} this is the case exactly if $J\ge\lambda(t)$. So for every search $t\ge2$ and every $J\in\{0,\dots,K\}$,
\begin{align}
\lambda(t)>J\quad\Longleftrightarrow\quad v_J(t)\text{ does not exist, or }\rho\bigl(v_J(t),t\bigr)\ge2^J-1.
\label{eq:cold}
\end{align}
In words, $\lambda(t)>J$ means that the block of height $J$ around $x_t$ was never searched before, or that at least $2^J-1$ distinct keys were searched since the last search in it.

Write $d(u,t)\coloneqq\abs{x_t-x_u}$ for $u<t$ and let
\begin{align*}
\tau(u,t)\coloneqq\log_2\bigl(d(u,t)+\rho(u,t)+2\bigr),\qquad \tau_t\coloneqq\min_{u<t}\tau(u,t)\quad(t\ge2),\qquad\tau_1\coloneqq\log_2(n+2),
\end{align*}
so that $\UB(X)=\sum_{t\ge1}\tau_t$ and $\tau_t\ge1$. For every $t\ge2$ we fix a time $u_t<t$ that attains $\tau_t$ and call it the \emph{partner} of $t$; any choice will do, and the same choice is used throughout.

The proof of the next lemma, and that of \cref{lem:upd-keyloss} for insertions and deletions, charge a time $t$ whose block was not visited for a long while to a \emph{chain} of earlier times, obtained by following partners backwards from $t$. The following lemma contains what the two proofs have in common: the chain is long enough to pay for $t$, and it lies within the period in which the block of $t$ was not visited. It uses only the numbers $\tau_r$, the partners, and the inequality $\rho(u,r)\le2^{\tau_r}-2$ between a time $r$ and its partner $u$, which holds because $\tau_r=\tau(u,r)$.

\begin{lemma}[Chains of partners]
\label{lem:chain}
Let every time $r$ have a number $\tau_r\ge1$, and let some times $r$ have a \emph{partner} $p(r)<r$ with $\rho(p(r),r)\le2^{\tau_r}-2$. Let $J$ and $c\ge2$ be integers, let $t$ be a time with a partner and $\tau_t\le J-c$, and let $v<t$ be a time with $v=0$ or $\rho(v,t)\ge2^J-1$. Suppose that every time $r$ with $v<r<t$ and $\tau_r<J-c$ has a partner. Let $s_0\coloneqq p(t)$ and $s_{i+1}\coloneqq p(s_i)$, up to the least index $i^*$ with
\begin{align*}
\sum_{i\le i^*}2^{\tau_{s_i}}\ge2^{J-c},
\end{align*}
and let $P(s_i)\coloneqq2^{\tau_t}+\sum_{i'<i}2^{\tau_{s_{i'}}}$. Then the times $s_0>s_1>\dots>s_{i^*}$ are defined and later than $v$, every $s$ among them satisfies
\begin{align}
\rho(s,t)<P(s)<2^{J-c+1},
\label{eq:chain}
\end{align}
and $\sum_{i\le i^*}\min\bigl(1,2^{\tau_{s_i}+c-J}\bigr)\ge1$.
\end{lemma}

\begin{proof}
The last claim is the choice of $i^*$. For the others we show by induction on $i\le i^*$ that $s_i$ is defined, later than $v$, and satisfies \eqref{eq:chain}. For $i=0$, the partner $s_0$ of $t$ is defined, and $\rho(s_0,t)\le2^{\tau_t}-2<2^{\tau_t}=P(s_0)\le2^{J-c}$. It is later than $v$: this is clear if $v=0$, and if $v\ge1$ and $s_0\le v$, then $\rho(v,t)\le\rho(s_0,t)<2^{J-c}<2^J-1$, because $\rho(\cdot,t)$ does not increase when its first argument increases, contrary to the choice of $v$. Let $i<i^*$. Then $\sum_{i'\le i}2^{\tau_{s_{i'}}}<2^{J-c}$, so $\tau_{s_i}<J-c$, and $v<s_i<t$, so $s_i$ has a partner $s_{i+1}$. Every time strictly between $s_{i+1}$ and $t$ lies in one of the pieces $(s_{i+1},s_i)$, $\{s_i\}$, $(s_i,s_{i-1})$, $\dots$, $\{s_0\}$, $(s_0,t)$. The piece $(s_{i'+1},s_{i'})$ holds at most $2^{\tau_{s_{i'}}}-2$ distinct keys, the piece $(s_0,t)$ at most $2^{\tau_t}-2$, and the $i+1$ singletons hold $i+1$ keys, which the terms $-2$ absorb. Hence
\begin{align*}
\rho(s_{i+1},t)<2^{\tau_t}+\sum_{i'\le i}2^{\tau_{s_{i'}}}=P(s_{i+1})<2^{J-c}+2^{J-c}=2^{J-c+1}.
\end{align*}
If $v\ge1$ and $s_{i+1}\le v$, then $\rho(v,t)\le\rho(s_{i+1},t)<2^{J-1}\le2^J-1$ as for $i=0$, which contradicts the choice of $v$. So $s_{i+1}>v$. Finally, the times $s_i$ decrease and stay later than $v\ge0$, so the sum reaches $2^{J-c}$ after finitely many steps, and $i^*$ exists.
\end{proof}

\begin{lemma}[Block boundaries cost little in total]
\label{lem:keyloss}
For every sequence $X$,
\begin{align*}
\sum_{t=2}^m\lambda(t)\le2\UB(X)+4m.
\end{align*}
\end{lemma}

\begin{proof}
We count level by level; in this proof a level is a height $J$ of blocks. Since $0\le\lambda(t)\le K$,
\begin{align}
\sum_{t=2}^m\lambda(t)=\sum_{J=0}^{K-1}\#\{t\ge2:\lambda(t)>J\}.
\label{eq:layers}
\end{align}
At each level $J$, a search with $\lambda(t)>J$ either has $\tau_t>J-2$ and pays for itself, or it is charged to earlier searches in such a way that every search receives at most one charge per level. The sum over the levels of what a search $s$ receives is then $\tau_s+O(1)$. \Cref{fig:keyloss} in \cref{sec:overview-unified} shows the second case.

Fix a level $J$ and a search $t\ge2$ with $\lambda(t)>J$. By \eqref{eq:cold}, either $v_J(t)$ does not exist or $\rho(v_J(t),t)\ge2^J-1$. A search $t$ has $\tau_t>J-2$ at the levels $J$ with $0\le J<\tau_t+2$, which are at most $\tau_t+3$ levels, so these searches contribute at most $\sum_{t\ge2}(\tau_t+3)$ to \eqref{eq:layers}. From now on let $\tau_t\le J-2$; we call $t$ a \emph{crossing} of level $J$. Its partner $u\coloneqq u_t$ satisfies
\begin{align}
d(u,t)+\rho(u,t)+2=2^{\tau_t}\le2^{J-2}.
\label{eq:crossing}
\end{align}

\emph{The chain of a crossing.} We apply \cref{lem:chain} with $c=2$, the partners $u_r$, and $v\coloneqq v_J(t)$, or $v\coloneqq0$ if $v_J(t)$ does not exist. Its hypotheses hold: $\rho(v,t)\ge2^J-1$ by \eqref{eq:cold} if $v\ge1$, and a time $r$ with $\tau_r<J-2$ is not the first search, whose term $\log_2(n+2)$ exceeds $K-1\ge J$, so it has a partner. We call the times $s_0=u,s_1,\dots,s_{i^*}$ of the lemma the \emph{chain} of the crossing; in the rest of the proof the letters $s$ and $s_i$ denote searches, not scales. The chain lies after $v_J(t)$, if that exists, and no search after $v_J(t)$ and before $t$ has its key in $B_J(x_t)$, no search before $t$ at all if $v_J(t)$ does not exist. So $x_u\notin B_J(x_t)$, and since $d(u,t)<2^{J-2}$, the keys $x_u$ and $x_t$ lie in two consecutive blocks of height $J$. Let $\beta$ be the largest key of the lower one, a multiple of $2^J$; we call $\beta$ the \emph{boundary} of the crossing. Moreover $d(u,t)\ge1$, so \eqref{eq:crossing} forces $J\ge4$. The chain stays near $\beta$, because it reaches a search $s_i$ from $\beta$ by short hops: from $\beta$ to $x_u=x_{s_0}$, and from each search of the chain to its partner. For every $s=s_i$ in the chain,
\begin{align}
\abs{x_s-\beta}\le\abs{x_{s_0}-\beta}+\sum_{i'<i}d(s_{i'+1},s_{i'})\overset{(a)}{<}2^{\tau_t}+\sum_{i'<i}2^{\tau_{s_{i'}}}=P(s)\overset{(b)}{<}2^{J-1}.
\label{eq:near}
\end{align}
Here (a) holds because $\beta$ lies between $x_u$ and $x_t$, so that $\abs{x_u-\beta}\le d(u,t)<2^{\tau_t}$ by \eqref{eq:crossing}, and because $d(s_{i'+1},s_{i'})<2^{\tau_{s_{i'}}}$ for the partner $s_{i'+1}$ of $s_{i'}$; and (b) is \eqref{eq:chain} with $c=2$.

\emph{Charging.} We charge the crossing to the searches of its chain: the search $s_i$ is charged $\min(1,2^{\tau_{s_i}-J+2})$. By \cref{lem:chain}, these charges sum to at least $1$.

\emph{A search is on one chain per level.} Let $t_1<t_2$ be crossings of the same level $J$, with boundaries $\beta_1$ and $\beta_2$, and suppose that a search $s$ is on both chains. By \eqref{eq:near},
\begin{align*}
\abs{\beta_1-\beta_2}\le\abs{\beta_1-x_s}+\abs{x_s-\beta_2}<2^{J-1}+2^{J-1}=2^J,
\end{align*}
and both are multiples of $2^J$, so $\beta_1=\beta_2$. The keys of $t_1$ and of its partner $u_{t_1}$ lie in the two blocks of height $J$ next to this boundary, one on each side, and $B_J(x_{t_2})$ is one of these two blocks. So one of the two searches $t_1$ and $u_{t_1}$ has its key in $B_J(x_{t_2})$; let $t'$ be its time. Then $t'\le t_1<t_2$, so $v_J(t_2)$ exists and is at least $t'$. The chain of $t_1$ starts at $u_{t_1}$ and goes backwards, so $s\le u_{t_1}\le t'$, and the chain of $t_2$ stays after $v_J(t_2)$. Hence
\begin{align*}
v_J(t_2)<s\le u_{t_1}\le t'\le v_J(t_2),
\end{align*}
a contradiction.

\emph{Charge received.} Fix a search $s$. At each level it is on the chain of at most one crossing, and crossings have level at least $4$. At a level $J$ it receives at most $\min(1,2^{\tau_s+2-J})$. So in total it receives at most
\begin{align*}
\sum_{J\ge4}\min\bigl(1,2^{\tau_s+2-J}\bigr)\le\underbrace{\#\{J:4\le J\le\tau_s+2\}}_{\le\tau_s-1}+\underbrace{\sum_{J>\tau_s+2}2^{\tau_s+2-J}}_{<2}<\tau_s+1.
\end{align*}
The second sum is less than $2$ because its first term is less than $1$ and each term is half the previous one. So the crossings contribute at most $\sum_{s\ge1}(\tau_s+1)$ to \eqref{eq:layers}.

\emph{Summing up.} By \eqref{eq:layers}, $\sum_{t\ge2}\lambda(t)\le\sum_{t\ge2}(\tau_t+3)+\sum_{s\ge1}(\tau_s+1)\le2\UB(X)+4m$.
\end{proof}

\begin{theorem}[Locality scales and the unified bound]
\label{thm:lambda-ub}
For every sequence $X$ of $m$ searches over $\{1,\dots,n\}$, $n\ge2$,
\begin{align*}
\UB(X)-m\ \le\ \sum_{t=1}^m\lambda(t)\ \le\ 2\UB(X)+4m+K.
\end{align*}
Consequently, \cref{alg:search} serves $X$ at total cost $O(m+\UB(X))$.
\end{theorem}

\begin{proof}
The upper bound is \cref{lem:keyloss} together with $\lambda(1)=K$.

For the lower bound we show that $\tau_t\le\lambda(t)+1$ for every $t$. Let $t\ge2$, and let $u<t$ be a search that satisfies the condition of \cref{def:locality} at the scale $\lambda(t)$. The keys $x_u$ and $x_t$ lie in one block of at most $2^{\lambda(t)}$ keys, so $d(u,t)\le2^{\lambda(t)}-1$, and $\rho(u,t)+2\le2^{\lambda(t)}$. So
\begin{align*}
\tau_t\le\tau(u,t)\le\log_2\bigl(2^{\lambda(t)}-1+2^{\lambda(t)}\bigr)<\lambda(t)+1.
\end{align*}
For the first search, $\tau_1=\log_2(n+2)\le\log_2(2^K+2)\le K+1=\lambda(1)+1$, because $K\ge1$.

For the total cost, \cref{alg:search} serves search $t$ at cost at most $\gamma\lambda(t)$ by \cref{thm:scales}, so by the upper bound it serves $X$ at cost at most $\gamma(2\UB(X)+4m+K)$. This is $O(m+\UB(X))$, because $K<\log_2n+1<\UB(X)+1$.
\end{proof}

Together with the bound of $\gamma\ceil{\log_2n}$ on the cost of a single search in \cref{thm:scales}, this is \cref{thm:main}.

The factor $2$ of \cref{lem:keyloss} cannot be improved for the fixed grid. Take $2^j$ pairs of adjacent keys, each pair on the two sides of a different boundary between two blocks of height $j+2$, and search the $2^{j+1}$ keys round robin, pair by pair. After the first round, the first search of a pair has the term $\log_2(2^{j+1}+1)$ and the second the term $\log_23$, while both have the locality scale $j+2$. So as $j$ grows, the locality scales sum to about twice the unified bound.

\begin{remark}[Two forms of the unified bound]
\label{rem:forms}
Iacono's definition of the unified bound, in the form of \cite{bcdi07}, minimizes over keys: its term for search $t$ is $\min_y\log_2(\abs{x_t-y}+w_t(y)+2)$, where $w_t(y)$ is the number of distinct keys searched since the last search for $y$; for a key $y$ that was never searched, a case that \cite{bcdi07} leaves open, we take $w_t(y)=n$. In \cite{iac01} the question does not arise: the keys are inserted first, insertions count as accesses, and the number of distinct keys accessed since the last access to a key is at most $n$. It agrees with the definition of \cref{sec:intro} up to an additive $O(m)$. For a key $y$ that was searched, taking $u$ to be its last search gives a term of ours that is at most Iacono's, and a key that was never searched gives a term of at least $\log_2(n+2)$, which no term of ours exceeds by more than one, since $\abs{x_t-x_u}+\rho(u,t)+2\le2(n+2)$. Conversely, for a time $u$, taking $y=x_u$ gives a term of Iacono's that is at most ours plus one.
\end{remark}

\section{Pattern-avoiding sequences}
\label{sec:patterns}

This section proves \cref{thm:patterns} and shows that the unified bound of a pattern-avoiding sequence is small. Both follow from one count: a search whose locality scale exceeds $s$ is the first search in its cell of scale $s$, a block of $2^s$ keys together with an interval of $2^s$ searches, and in a sequence that avoids $\sigma$, few cells of each scale contain a search.

We first fix the definitions. Let $\sigma$ be a permutation of $\{1,\dots,k\}$. A sequence $X$ \emph{contains} $\sigma$ if there are times $t_1<\dots<t_k$ such that $x_{t_i}<x_{t_j}$ exactly when $\sigma(i)<\sigma(j)$, and it \emph{avoids} $\sigma$ otherwise. The $k$ keys of an occurrence are distinct, so repeated searches for a key create no occurrence by themselves, and for a permutation $X$ this is the usual notion.

All $0$-$1$ matrices used as patterns are nonzero. A $0$-$1$ matrix \emph{contains} a pattern $P$ if some submatrix of it, with rows and columns in order, can be turned into $P$ by changing ones into zeros; the zeros of $P$ impose no requirement. We identify $\sigma$ with its permutation matrix, which has a one in row $\sigma(l)$ and column $l$.

The \emph{Füredi--Hajnal limit} $c_\sigma$ is the supremum over $M$ of $1/M$ times the largest number of ones in an $M\times M$ matrix that avoids $\sigma$. It is finite by the theorem of Marcus and Tardos \cite{mt04}, and $c_\sigma=2^{O(k)}$ by a result of Fox \cite{fox13}. Padding a rectangular matrix with zero rows or columns gives the form we use \cite[Lemma~2.1]{bko24}: every $0$-$1$ matrix with $M_1$ rows, $M_2$ columns and more than $c_\sigma\max(M_1,M_2)$ ones contains $\sigma$.

Throughout, $X$ is a sequence of $m$ searches over $\{1,\dots,n\}$, $n\ge2$, and the blocks and the locality scales are those of \cref{sec:scales}. Time is cut in the same way as the keys. For $0\le s\le K$, the \emph{epochs of level} $s$ are the intervals $\{j2^s+1,\dots,(j+1)2^s\}$, $j\ge0$, intersected with $\{1,\dots,m\}$; an epoch of level $s$ consists of at most $2^s$ consecutive times. A \emph{cell} of scale $s$ is a pair $(g,E)$ of a block $g$ of height $s$ and an epoch $E$ of level $s$. A search $u$ \emph{lies in} the cell if $x_u\in g$ and $u\in E$, and the cell is \emph{occupied} if a search lies in it. Every search lies in exactly one cell of each scale. At scale $s$ there are $r_s\coloneqq\ceil{n/2^s}$ blocks and $e_s\coloneqq\ceil{m/2^s}$ epochs. Since $2^s<2n$, we have $r_s\le2n/2^s$; and $e_s<2m/2^s$ when $2^s<m$, while $e_s=1$ when $2^s\ge m$.

\begin{lemma}[Scales and occupied cells]
\label{lem:count}
For every scale $s<K$, the number of searches $t$ with $\lambda(t)>s$ is at most the number $o_s$ of occupied cells of scale $s$, and $o_s\le\min(m,r_se_s)$.
\end{lemma}

\begin{proof}
We show that a search $t$ with $\lambda(t)>s$ is the first search in its cell of scale $s$. Suppose that an earlier search $u$ lies in the same cell. Then $x_u\in B_s(x_t)$, and $u$ and $t$ lie in one epoch of at most $2^s$ consecutive times, so
\begin{align*}
\rho(u,t)\le t-u-1\le2^s-2.
\end{align*}
So the scale $s$ satisfies the condition of \cref{def:locality}, and $\lambda(t)\le s$ by \eqref{eq:lambda}, contrary to the assumption. For $t=1$ there is no earlier search. Every occupied cell has one first search, so the searches with $\lambda(t)>s$ lie in different occupied cells. Finally, every occupied cell contains a search and there are $r_se_s$ cells of scale $s$, so $o_s\le\min(m,r_se_s)$.
\end{proof}

Since $1\le\lambda(t)\le K$, summing over the scales and using the lower bound of \cref{thm:lambda-ub} gives
\begin{align}
\UB(X)-m\le\sum_{t=1}^m\lambda(t)=\sum_{s=0}^{K-1}\#\{t:\lambda(t)>s\}\le\sum_{s=0}^{K-1}o_s.
\label{eq:occupied}
\end{align}

\begin{proposition}[Unified bound of pattern-avoiding sequences]
\label{prop:ub-patterns}
If $X$ avoids a permutation $\sigma$ of length $k\ge2$, then $\sum_t\lambda(t)\le m\log_2c_\sigma+4m+\frac{11}2n$ and $\UB(X)\le m\log_2c_\sigma+5m+\frac{11}2n=O((m+n)(1+\log c_\sigma))$.
\end{proposition}

\begin{proof}
Fix a scale $s$ and consider the $0$-$1$ matrix with one row for each of the $r_s$ blocks of height $s$, in key order, one column for each of the $e_s$ epochs of level $s$, in time order, and a one wherever the cell is occupied. Suppose that this matrix contains $\sigma$, with ones in rows $i_1,\dots,i_k$ and columns $j_1<\dots<j_k$, where the rows are ordered as $\sigma$. Choose a search $t_l$ in the cell of row $i_l$ and column $j_l$, for every $l$. The times increase with $l$, because the epochs are disjoint intervals, and the keys are distinct and ordered as $\sigma$, because the blocks are disjoint intervals. So $X$ contains $\sigma$.

Hence if $X$ avoids $\sigma$, the matrix avoids $\sigma$, and by the padded form of the Marcus--Tardos bound it has at most $c_\sigma\max(r_s,e_s)\le c_\sigma(r_s+e_s)$ ones, that is, $o_s\le c_\sigma(r_s+e_s)$. Since $\sigma$ has length at least $2$, an $M\times M$ matrix whose ones fill one row avoids it, so $c_\sigma\ge1$.

It remains to add up over the scales. By \eqref{eq:occupied}, $\sum_t\lambda(t)\le\sum_{s=0}^{K-1}o_s$, where $o_s\le m$. For the scales with $2^s\ge m$ there is one epoch, so $o_s\le r_s\le2n/2^s$, and these scales contribute at most $4n/m\le4n$ in total. For the scales with $2^s<m$, the bounds on $r_s$ and $e_s$ give
\begin{align*}
o_s\le\min\Bigl(m,\ \frac{2c_\sigma(n+m)}{2^s}\Bigr).
\end{align*}
Let $Z\coloneqq2c_\sigma(n+m)/m\ge2$. The scales with $2^s<Z$ contribute at most $m$ each, and there are at most $\log_2Z+1$ of them; the remaining scales form a geometric series with sum at most $2m$. Since $\log_2Z=1+\log_2c_\sigma+\log_2(1+n/m)$ and $m\log_2(1+n/m)\le n/\ln2<\frac32n$, altogether $\sum_t\lambda(t)\le m\log_2c_\sigma+4m+\frac{11}2n$. The lower bound of \cref{thm:lambda-ub} gives the bound on $\UB(X)$.

This proves \cref{thm:patterns} as well. By \cref{thm:scales}, \cref{alg:search}, which does not depend on $\sigma$, serves $X$ at cost $O(\sum_t\lambda(t))=O((m+n)(1+\log c_\sigma))$. With $c_\sigma=2^{O(k)}$ the cost is $O((m+n)k)$, and for a permutation, where $m=n$, it is $O(n(1+\log c_\sigma))$.
\end{proof}
\section{Running time}
\label{sec:runtime}

The model of \cref{sec:prelim} charges only the nodes that a search touches. The computation that decides which nodes to touch, including the records that the algorithm keeps about past searches, is free in the model. This section states that it fits in the same bounds on a word RAM and explains the one point that needs an argument. \Cref{app:runtime} gives the implementation and the proof.

\begin{proposition}[Running time]
\label{thm:runtime}
On a word RAM with words of $\Omega(\log(n+m))$ bits, \cref{alg:search} can be implemented deterministically with $O(n)$ words of memory so that, after an initialization of $O(n)$ time, search $t$ takes $O(\lambda(t))$ time. Hence every single search takes $O(\log n)$ time, and all searches together take $O(m+\UB(X))$ time.
\end{proposition}

The region tree and the inner trees are stored as they are, so a walk, a split or a join takes the time that bounds its cost in the model. Apart from these, {\sc Search} in \cref{alg:search} asks three questions about the kept blocks. Is a given block kept? Are $2^s$ blocks of height $s$ kept? And which kept block of height $s$ was searched least recently? The first two are answered by an array and a counter for every promoted scale. The third is the point that needs an argument.

\paragraph{Why the third question is not as easy.} The obvious answer is a list of the kept blocks of each height in the order of their last searches: a search moves its blocks to the front of their lists, and the victim of a scale is the last block of its list. This is too slow. A search has a kept block at every promoted scale, so it would touch a list at each of the about $\log_2\log_2n$ promoted scales, and a search with a small locality scale cannot pay for this.

\paragraph{The idea.} A victim never contains a kept block of smaller height (Part~5 of \cref{lem:kept}). So the list of a scale needs only the kept blocks that contain no kept block of smaller height; we call such a block \emph{childless}. After a search for $x_t$, all blocks of $x_t$ at the promoted scales are kept, so the only childless one among them is the lowest, $B_1(x_t)$. Hence, apart from a constant number of list operations for every block that it retires or adds, which it pays for anyway, a search has to touch one list only, that of the scale $1$. A higher block enters its list later, at the moment when the last kept block inside it retires, and it then needs to know when it was last searched. For this, a retiring block hands the time of its last search to its parent. The lists are of the same kind as those that the cache-splay tree keeps for its levels \cite[Section~6.3]{der09}. What has to be proved is that a block which enters a list in this way belongs at its front, so that the last block of a list is always the victim of its scale (\cref{lem:lists}).

\section{Insertions and deletions}
\label{sec:updates}

We now let the set of keys change. Bădoiu, Cole, Demaine and Iacono \cite{bcdi07} extended the unified bound to insertions and deletions, with the distance between two keys counted among the keys that are present; we call it the \emph{unified bound with updates}. Our tree extends to it (\cref{thm:updates}).

\subsection{The model and the result}
\label{sec:upd-model}

We are given a set $S_1$ of $n_0$ keys from an ordered set and a sequence $X$ of $m$ \emph{operations}. Operation $t$ has a key $x_t$ and is a search, an insertion or a deletion. Let $S_t$ be the set of keys before it, and $n_t\coloneqq\abs{S_t}$. A search or a deletion has $x_t\in S_t$, and an insertion has $x_t\notin S_t$. The \emph{successor} of a key in a set is the next larger key of the set, and its \emph{predecessor} the next smaller one.

Before operation $t$ the tree is a BST on $S_t$. An operation is carried out in \emph{steps}. A step replaces the subtree formed by a top set $U$ by another BST and reattaches the subtrees that hang off $U$, at cost $\abs U$.
\begin{itemize}
\item {\bf Search.} Every step keeps the keys of $U$, and some step has $x_t\in U$. By \cref{lem:transition} this is the model of \cref{sec:prelim}.
\item {\bf Insertion.} Exactly one step changes the keys. Its set $U$ contains the \emph{search path} of $x_t$, the nodes with which a search for $x_t$ compares it, and its new tree has the keys of $U$ and $x_t$. The predecessor and the successor of $x_t$ are on the search path, so no subtree that hangs off $U$ has keys on both sides of $x_t$.
\item {\bf Deletion.} Exactly one step changes the keys. Its set $U$ contains $x_t$, and its predecessor or its successor if $x_t$ has two children, and its new tree has the keys of $U$ other than $x_t$. Then at most one subtree hangs off $U$ between the two keys of $U$ next to $x_t$, so it can be attached between them.
\end{itemize}
So the subtrees that hang off $U$ can be reattached in every step. A step is carried out by $O(\abs U+1)$ pointer moves and rotations that start at the root, together with the insertion of a leaf or the deletion of a node with at most one child, and conversely the nodes that such an operation touches form a top set. So up to a constant factor and one unit per operation, these are the costs of a BST with rotations \cite{st85,dhi+09}.

\paragraph{The unified bound with updates.} A time $u<t$ is a \emph{witness} for operation $t$ if operation $u$ is a search or an insertion and $x_u$ is not deleted before $t$, that is, $x_u\in S_r$ for $u<r\le t$. For $y\in S_t$ and a key $x$, let $d_t(y,x)$ be the number of keys $z\in S_t$ with $y<z\le x$ or $x\le z<y$. As before, $\rho(u,t)$ is the number of distinct keys of the operations strictly between $u$ and $t$. Let
\begin{align*}
\tau(u,t)&\coloneqq\log_2\bigl(d_t(x_u,x_t)+\rho(u,t)+2\bigr),\\
\tau_t&\coloneqq\min\Bigl(\log_2(n_t+2),\ \min\bigl\{\tau(u,t):u\text{ is a witness for }t\bigr\}\Bigr),
\end{align*}
and $\UB(X)\coloneqq\sum_{t=1}^m\tau_t$. For searches in a fixed set this is the definition of \cref{sec:intro}, up to less than $1$ per term. The bound of \cite{bcdi07} for operation $t$ is the minimum over $y\in S_t$ of $\log_2(w_t(y)+d_t(y,x_t)+2)$, where $w_t(y)$ is the number of distinct keys of the operations since the last operation with the key $y$. That operation is a witness $u$ with $\rho(u,t)\le w_t(y)$, so $\tau_t$ is at most this bound, if a key that had no operation is given $w_t(y)=n_t$ as in \cref{rem:forms}.

\begin{theorem}[Insertions and deletions]
\label{thm:updates}
There is a deterministic online BST algorithm for searches, insertions and deletions that, starting from a tree that depends only on $S_1$, serves every sequence $X$ of $m$ operations at total cost
\begin{align*}
O\bigl(n_0+m+\UB(X)\bigr),
\end{align*}
and operation $t$ at cost $O(\log(n_t+2))$.
\end{theorem}

We do not know whether the term $n_0$ is needed, and we do not analyse the running time. Since the optimum costs at most as much as this algorithm, the theorem bounds the optimum.

\begin{corollary}[The optimum with insertions and deletions]
\label{cor:upd-opt}
For a sequence $X$ of $m$ searches, insertions and deletions, starting from a set of $n_0$ keys, let $\OPT(X)$ be the least total cost at which $X$ can be served in the model above, over all initial trees and over all algorithms, which may know $X$ in advance. Then $\OPT(X)=O(n_0+m+\UB(X))$.
\end{corollary}

\subsection{The idea}
\label{sec:upd-idea}

The algorithm is that of \cref{sec:algorithm}. Two things in it must change, the blocks and the time at which a block was last used, and the analysis needs one new argument. \Cref{app:updates} gives the algorithm and the complete proof.

\paragraph{Blocks that grow and shrink.} In \cref{sec:algorithm} a block is an interval of a grid that is fixed in advance. With insertions this is impossible. If a key smaller than all others is inserted, the rank of every key goes up by one, so every border of the grid would move by one key, and the tree would have to change everywhere. So the blocks are now intervals of the present keys, and an inserted key joins the block of the next larger key. A block of \emph{level} $k$, where $k$ is a power of two, has between $8^k$ and $4\cdot8^k$ keys, about $2^{3k}$, so it plays the part of a block of height $3k$: its keys lie at depth $O(k)$ when the block is kept. When it leaves this range, it is split into two blocks or merged with a neighbour, as the nodes of a B-tree are. For example, a block of level $1$ has $8$ to $32$ keys. Its $33$rd key splits it into blocks of $16$ and $17$ keys, and neither of them is split or merged again before it has gained at least $16$ keys or lost at least $9$. In general a new block of level $k$ must gain or lose a quarter of $8^k$ keys before it is repaired, so there are $O(m/8^k)$ repairs at level $k$ (\cref{lem:upd-blocks}). A repair is a split, a merge, or a merge followed by a split of the merged block. Each split or merge changes the regions at one place and retires or adds at most one block of level $k$, so a repair costs $O(k+1)$, and all repairs together cost $O(m)$.

\paragraph{The time of a block.} As before, the tree keeps at every level $k$ the $3\cdot8^k$ blocks that were used most recently. A search or an insertion uses its own key. A deletion should count as a use of the blocks of the deleted key, but the key is gone afterwards, and the time of the deletion would be gone with it: if the last operations in a block were deletions, the block would look as if it had not been used for a long time, and the tree would stop keeping it. So a deletion uses a neighbour of the deleted key in the same block of level $1$, its \emph{heir}. Then the time of a block, the last time that one of its keys was used, does not decrease while the block exists.

\paragraph{The analysis.} An operation costs as much as a search, plus $O(k+1)$ for every repair at a level $k$ (\cref{lem:upd-cost}). The comparison with the unified bound follows \cref{lem:keyloss} (\cref{lem:upd-keyloss}): an operation whose block of level $k$ was not used for a long time is charged to its chain of partners (\cref{lem:chain}), which lies near a border of this block. Two things are new. A block that a split or a merge has just created has no history, so an operation in it may have no chain to be charged to; but each block is in this situation for at most one operation, and these operations are paid for by the count of the repairs. And the borders of the blocks move as keys are inserted and deleted; but a present key never crosses the border of a block that survives, because an inserted key joins the block of the next larger key, and this is all that the charging needs.

\paragraph{Splay trees.} The insertions and deletions of splay trees \cite[Section~3]{st85} consist of steps of the model. A splay is a step whose set $U$ is the path that it follows. Adding a key, as a leaf or as the new root after a splay, and removing a node and joining its two subtrees are steps whose sets $U$ consist of nodes on the paths of the splays of the operation. So the cost in the model is at most a constant times the number of nodes that the splay tree visits. Call an algorithm \emph{dynamically optimal} for searches, insertions and deletions if it serves every sequence from every initial tree at cost $O(\OPT(X)+n_0)$. By \cref{cor:upd-opt}, such an algorithm costs $O(n_0+m+\UB(X))$. Bădoiu, Cole, Demaine and Iacono conjecture that splay trees satisfy their bound in the amortized sense \cite{bcdi07}. So if splay trees are dynamically optimal for all three operations, this conjecture holds for the total cost of every sequence that starts from the empty set, and for other sequences up to an additive $O(n_0)$ and the convention for keys that were never accessed.

\PaperPrintBibliography

\clearpage
\appendix
\section{Running time: the implementation and the proof}
\label{app:runtime}

This appendix proves \cref{thm:runtime}. \Cref{sec:runtime-lists} gives the records and the lists that find the victims and proves that they are correct (\cref{lem:lists}), and \cref{sec:runtime-trees} describes the rest of the implementation and adds up the time. The budget for search $t$, with $M_t$ the set of its missing scales, is
\begin{align}
\underbrace{O\bigl(\abs{M_t}+1\bigr)}_{\text{records and lists}}+\underbrace{\sum_{s\in M_t}O(s+1)}_{\text{retirements}}+\underbrace{O(\lambda(t))}_{\text{walk}}+\underbrace{O(\lambda(t))}_{\text{promotions}}=O(\lambda(t)),
\label{eq:runtime-budget}
\end{align}
where the first two terms are $O(\lambda(t))$ by \eqref{eq:heights}.

\subsection{Finding the victims}
\label{sec:runtime-lists}

The \emph{victim} of a scale is the kept block of its height that was searched least recently (\cref{sec:fingers}). This subsection describes the arrays, records and lists with which the implementation finds it in constant time.

\paragraph{Arrays and counters.} For every promoted scale $s$, an array has one entry for each block of height $s$, where the block $B_s(y)$ of a key $y$ has the number $\ceil{y/2^s}$; the entry points to the \emph{record} of the block if the block is kept. A counter holds the number of kept blocks of each promoted height. The arrays have $\sum_{s\in\mathcal S}\ceil{n/2^s}\le2n$ entries in all. They tell in constant time whether a block is kept and whether $2^s$ blocks of height $s$ are kept.

\paragraph{Children.} Let $\mathcal F$ be the set of the kept blocks at some moment. For $g\in\mathcal F$, let
\begin{align*}
{\downarrow}g\coloneqq\{c\in\mathcal F:c\subseteq g,\ h(c)\le h(g)\},\qquad\operatorname{kids}(g)\coloneqq\{c\in\mathcal F:\text{the parent of $c$ is }g\}
\end{align*}
be the kept blocks inside $g$ and the \emph{kept children} of $g$. Unlike the set ${\downarrow}_{\mathcal G}b$ in the proof of \cref{lem:kept}, the set ${\downarrow}g$ contains $g$ itself. The block $g$ is \emph{childless} if it contains no kept block of smaller height, and
\begin{align}
g\text{ is childless}\quad\Longleftrightarrow\quad{\downarrow}g=\{g\}\quad\Longleftrightarrow\quad\operatorname{kids}(g)=\emptyset.
\label{eq:childless}
\end{align}
The first equivalence holds because two blocks of the same height are disjoint. For the second, a kept child of $g$ is a kept block of smaller height inside $g$, and conversely, if $g$ contains kept blocks of smaller height, the highest of them has no kept block between itself and $g$, so its parent is $g$.

\paragraph{The records.} The record of a kept block $g$ holds the index of the block and two numbers: its \emph{date} $\operatorname{date}(g)$, which is the time of a search in the block, and the number $\abs{\operatorname{kids}(g)}$ of its kept children. For every promoted scale $s$, a doubly linked list, the \emph{list of the scale}, holds the records of the childless kept blocks of height $s$. Initially the only record is that of the top block, with the date $0$, and it is in its list. Three rules maintain the records and the lists, each in constant time.
\begin{itemize}
\item {\bf A block is added.} When search $t$ adds a block, its record gets the date $t$ and no kept children, and it is put at the front of the list of its scale. Its parent has one more kept child, and leaves its list if this is its first. So during the promotions of a search, an added block leaves its list again when the next lower block is added.
\item {\bf A search ends.} After its promotions, search $t$ gives the date $t$ to the lowest block $B_1(x_t)$ of its key and moves its record to the front of its list.
\item {\bf A block retires.} When a block $e$ retires, its record leaves its list and is deleted. Its parent $p$ has one kept child less and gets the date
\begin{align}
\operatorname{date}(p)\gets\max\bigl(\operatorname{date}(p),\operatorname{date}(e)\bigr).
\label{eq:inherit}
\end{align}
If $e$ was the last kept child of $p$, the record of $p$ is put at the front of the list of its scale.
\end{itemize}
The parent of a kept block of height $s$ is the kept block of height $2s$ around it, by \eqref{eq:parent}, so its record is found through the array of that scale. By \eqref{eq:childless}, the rules keep in the list of a scale $s$ exactly the childless kept blocks of height $s$.

For example, let $n=16$, so that the promoted scales are $1$, $2$ and $4$, and let the searches be $1,5,9,13$. Two blocks of height $1$ are kept at a time. Search $3$, for the key $9$, retires the block $\{1,2\}$, whose date is $1$. Its parent $\{1,\dots,4\}$, whose date is $1$ as well, enters the list of the scale $2$, which was empty. Search $4$, for the key $13$, retires $\{5,6\}$ in the same way, and its parent $\{5,\dots,8\}$ enters the same list at the front, with the date $2$. The list of the scale $2$ is now $\{5,\dots,8\},\{1,\dots,4\}$, which is the order of the last searches in these blocks. If a fifth search is for the key $2$, it retires $\{9,10\}$, whose parent enters the list at the front, and it adds $\{1,2\}$, so that $\{1,\dots,4\}$ has a kept child again and leaves the list.

The rules always put a record at the front. The next lemma shows that this is the right place: a parent that becomes childless was searched later than every block that is already in its list. As in \cref{sec:inv-fingers}, $\ell_t(g)$ is the last search in a block $g$ before $t$, and we let $\ell_t(g)\coloneqq0$ if $g$ is not searched before $t$. Recall that the victim of a scale $s$ in search $t$ is the block of height $s$ in $\mathcal F_t$ with the least $\ell_t$.

\begin{lemma}[The lists of the scales]
\label{lem:lists}
Let $t\in\{1,\dots,m+1\}$. When search $t$ begins and during its first phase, or after the last search if $t=m+1$,
\begin{itemize}
\item {\bf Part 1 (dates).} every kept block $g$ satisfies
\begin{align}
\max\bigl\{\operatorname{date}(c):c\in{\downarrow}g\bigr\}=\ell_t(g),
\label{eq:dates}
\end{align}
so a childless kept block $g$ has $\operatorname{date}(g)=\ell_t(g)$;
\item {\bf Part 2 (order).} in every list, the dates strictly decrease from the front to the end.
\end{itemize}
Hence, whenever the first phase of a search retires the victim of a scale, the victim is the last block of the list of this scale.
\end{lemma}

\begin{proof}
Part~1 is an induction over the changes of the kept blocks, in which each change leaves both sides of \eqref{eq:dates} as they are or sets both to $t$. Part~2 holds at all times; it reduces to one inequality between a block that enters a list and a block that is already there, and this inequality compares the last searches of two blocks through Part~1.

\emph{Part 1.} Write $D(g)$ for the left side of \eqref{eq:dates}. When the first search begins, the top block is the only kept block, and $D(g)=0=\ell_1(g)$ for it. There are two kinds of changes.

\emph{A retirement in the first phase of search $t$.} Let $e$ retire, with parent $p$, and let $g\ne e$ be a kept block. If $e\notin{\downarrow}g$, then $p\notin{\downarrow}g$ either, because $e\subseteq p$, so neither ${\downarrow}g$ nor a date in it changes. If $e\in{\downarrow}g$, then $g$ is a kept block that contains $e$ and is higher, and $p$ is the lowest such block, so $p\in{\downarrow}g$. By \eqref{eq:inherit}, the retirement replaces the two dates of $e$ and $p$ in ${\downarrow}g$ by their maximum. So in both cases $D(g)$ does not change, and $\ell_t(g)$ does not depend on the moment.

\emph{From the first phase of search $t$ to the beginning of search $t+1$.} In between, the kept blocks and the dates change only by the first two rules: blocks of $x_t$ are added, and blocks of $x_t$ get the date $t$. Let $g\in\mathcal F_{t+1}$. Then
\begin{align*}
x_t\notin g:&\quad D(g)\text{ does not change, and }\ell_{t+1}(g)=\ell_t(g);\\
x_t\in g:&\quad D(g)=t=\ell_{t+1}(g).
\end{align*}
In the first line, no block of $x_t$ lies in $g$, so ${\downarrow}g$ and its dates do not change, and $g$ was kept during the first phase because search $t$ adds only blocks of $x_t$. In the second line, $B_1(x_t)$ is a kept block inside $g$ with the date $t$ at the end of the search, and no date exceeds $t$.

For a childless block, ${\downarrow}g=\{g\}$ by \eqref{eq:childless}, so $D(g)=\operatorname{date}(g)$.

\emph{Part 2.} A record changes its date in two ways. It gets the date $t$ of the current search, and then it is put or moved to the front of its list. Or its block is a parent that takes the date of a retiring child by \eqref{eq:inherit}, and then it is not in a list, because it has a kept child. So it suffices to show that a record that is put at the front has a later date than every other record of its list. There are two cases.

\emph{The record gets the date $t$ in search $t$.} Its block contains $x_t$. Another block of the same height does not contain $x_t$, so it did not get the date $t$, and all other dates are earlier than $t$.

\emph{A parent becomes childless.} Let the first phase of search $t$ retire $e$, the last kept child of $p$. Let $s\coloneqq h(e)$, so that $h(p)=2s$, and let $q\ne p$ be a block in the list of the scale $2s$. The idea is that $q$ lost its last child at an earlier time, when this child was the least recently searched kept block of its height, while $e$ was searched more recently or was not kept yet. The block $q$ is not the top block, so it is searched before $t$. Let $c$ be the block of height $s$ inside $q$ that contains the key of the search $\ell_t(q)$, so that $\ell_t(c)=\ell_t(q)$. We show that
\begin{align}
\operatorname{date}(q)\overset{(a)}{=}\ell_t(q)=\ell_t(c)\overset{(b)}{<}\ell_t(e)\overset{(c)}{=}\operatorname{date}(e)\overset{(d)}{\le}\operatorname{date}(p),
\label{eq:list-order}
\end{align}
where $\operatorname{date}(p)$ is the date of $p$ after the retirement of $e$. Steps (a) and (c) are Part~1: the block $q$ is childless because it is in a list, and $e$ is childless when it retires (Part~5 of \cref{lem:kept}). Step (d) is \eqref{eq:inherit}.

For (b), let $u\coloneqq\ell_t(c)$. Search $u$ promoted $c$, so $c\in\mathcal F_{u+1}$. Now $c$ is not kept, because $q$ is childless, and a block leaves the kept blocks only as a victim. So $c$ was the victim of the scale $s$ in a search $t'$ with $u<t'\le t$. Here $t'\ne t$, because the victim of the scale $s$ in search $t$ is $e$, and $e\ne c$ since $e\subseteq p$ and $c\subseteq q$ are disjoint. No search between $u$ and $t$ lies in $c$, so $\ell_{t'}(c)=u$, and the victim $c$ has the least $\ell_{t'}$ among the blocks of height $s$ in $\mathcal F_{t'}$. The block $e$ is kept at time $t$. So
\begin{align*}
\ell_t(e)\ge\begin{cases}\ell_{t'}(e)>\ell_{t'}(c)=u,&\text{if }e\in\mathcal F_{t'},\\t'>u,&\text{if }e\notin\mathcal F_{t'},\end{cases}
\end{align*}
where in the second case $e$ was added by a search at the time $t'$ or later, and this search lies in $e$.

\emph{The victim.} Let the first phase of search $t$ retire the victim $e$ of a scale $s$. Then $e$ is childless (Part~5 of \cref{lem:kept}), so its record is in the list of the scale $s$. Every block $b$ of this list is a kept block of height $s$ with $\operatorname{date}(b)=\ell_t(b)$, by Part~1, and $e$ has the least $\ell_t$ among the kept blocks of height $s$. So $e$ has the earliest date of its list, and by Part~2 its record is the last one.
\end{proof}

\subsection{The rest of the implementation}
\label{sec:runtime-trees}

The proof of \cref{thm:runtime} goes through \cref{alg:search} line by line. For each line it says what the line needs besides the splits and joins of {\sc ChangeRegions}, where the implementation finds it, and how long this takes. \Cref{tab:runtime} collects the answers, and its last column adds up to the budget \eqref{eq:runtime-budget}.

\paragraph{What is stored.} The implementation stores the following.
\begin{itemize}
\item {\bf The trees.} The region tree and the inner trees are stored as globally biased trees with parent pointers and ranks. Every internal node stores its key, the key that it has in $T(\Rtree)$ by \cref{def:layout}. A leaf of the region tree stores the endpoints and the owner of its region and a pointer to its inner tree. The tree $T(\Rtree)$ is not stored separately: its nodes are the internal nodes of these trees and the last leaf, so a walk in it is a walk in them.
\item {\bf The kept blocks.} The arrays, the counters, the records and the lists of \cref{sec:runtime-lists}.
\item {\bf Three arrays on the keys.} For a key $y$, $\operatorname{first}[y]$ and $\operatorname{last}[y]$ are the leaves of the regions whose smallest key, or largest key, is $y$, if there are such regions, and $\operatorname{leaf}[y]$ is the leaf of $y$ in its inner tree.
\end{itemize}

\begin{table}[!ht]
\centering
\small
\begin{tabular}{@{}>{\raggedright\arraybackslash}p{0.2\linewidth}>{\raggedright\arraybackslash}p{0.22\linewidth}>{\raggedright\arraybackslash}p{0.31\linewidth}>{\raggedright\arraybackslash}p{0.18\linewidth}@{}}
\toprule
Step of {\sc Search}$(t)$&What it needs&Where it is found&Time\\
\midrule
\Lineref{line:missing}: the set $M$&whether $B_s(x_t)$ is kept&the array of the scale $s$, for $s=1,2,4,\dots$ up to the first kept block&$O(\abs{M_t}+1)$\\
\addlinespace
\Linerange{line:full}{line:victim}, for $s\in M_t$&whether $2^s$ blocks are kept; the victim&the counter of $s$; the last record of the list of $s$ (\cref{lem:lists})&$O(1)$\\
\addlinespace
{\sc Retire}$(e)$, for $s\in M_t$&the list $\mathcal Q$; the rule for a retiring block&$\operatorname{first}$, $\operatorname{last}$ and the test \eqref{eq:owner-test}&$O(1)$; its call of {\sc ChangeRegions} $O(s+1)$\\
\addlinespace
\Lineref{line:serve}: the walk&the keys of the nodes&stored at the nodes&$O(\lambda(t))$\\
\addlinespace
{\sc Add}$(g)$, for $s\in M_t$&the region $Q$, and $Q_L$, $Q_R$; the rule for an added block&the leaf of $x_t$, or the block added before; the endpoints of $Q$ and $g$&$O(1)$; all calls of {\sc ChangeRegions} $O(\lambda(t))$\\
\addlinespace
End of the search&the date of $B_1(x_t)$&the array of the scale $1$&$O(1)$\\
\addlinespace
A call of {\sc ChangeRegions}&ranks of the new leaves, keys of the new nodes, the three arrays&computed from the weights and the old keys, as explained below&$O(1)$ per split, join and new region\\
\bottomrule
\end{tabular}
\caption{The implementation of \cref{alg:search}: what every step needs besides its splits and joins, and the time to find it. The times of {\sc ChangeRegions} and of the walk are their costs in the model (\cref{lem:amortized}).}
\label{tab:runtime}
\end{table}

\begin{proof}[Proof of \cref{thm:runtime}]
We show that search $t$ takes the time \eqref{eq:runtime-budget}, by checking the rows of \cref{tab:runtime}. Their times add up to
\begin{align*}
O\bigl(\abs{M_t}+1\bigr)+\sum_{s\in M_t}O(s+1)+O(\lambda(t))+O(\lambda(t))=O(\lambda(t)),
\end{align*}
where the second term bounds the retirements, the third the walk and the fourth the promotions, as in the proof of \cref{thm:scales}, and the first covers everything that takes constant time for each missing scale. Splits and joins run in the time bounded in the proof of \cref{lem:opcost}, because the rank of a leaf is computed in constant time: the weights of the regions are powers of $4$ with exponents of size $O(K)$, and the keys have weight $1$. The rows are checked as follows.

\emph{The missing scales and the victims.} The missing scales are the promoted scales below some bound (Part~3 of \cref{lem:kept}). So the implementation looks up $B_s(x_t)$ for $s=1,2,4,\dots$ until it meets a kept block, and the scales that it has passed are $M_t$. At each of them whose counter is $2^s$, the victim is the last block of the list of the scale (\cref{lem:lists}).

\emph{{\sc Retire}$(e)$.} The keys of $e$ form one region, whose leaf is $\operatorname{first}[\min e]$, and its two neighbours are $\operatorname{last}[\min e-1]$ and $\operatorname{first}[\max e+1]$. The list $\mathcal Q$ of \lineref{line:list} consists of this region and of those neighbours that the parent $p$ of $e$ owns. Let $z\coloneqq\min e-1$, and let $\mathcal F$ be the set of the kept blocks at this moment. Then
\begin{align}
p\text{ owns the left neighbour of }e\quad\Longleftrightarrow\quad z\in p\ \text{ and }\ B_{h(e)}(z)\notin\mathcal F,
\label{eq:owner-test}
\end{align}
and the right neighbour is tested in the same way, with $\max e+1$. Indeed, if $z\in p$, the owner of $z$ is $p$ or a lower kept block that contains $z$. Such a block has a height of at most $h(e)$, since $h(p)=2h(e)$ is the next promoted height, and by the nesting of Part~5 of \cref{lem:kept} there is one exactly if $B_{h(e)}(z)$ is kept.

\emph{The walk.} The walk compares $x_t$ with the keys that the nodes store. It then goes on from the node of $x_t$ to the leaf of $x_t$ in the expanded tree, and so passes the leaf of the region of $x_t$. Both parts take $O(h(f_t)+1)=O(\lambda(t))$ time (\cref{lem:pathdepth,lem:serving}).

\emph{{\sc Add}$(g)$.} The blocks of $x_t$ at the missing scales are added from the highest to the lowest. The region $Q$ of \lineref{line:region} is the region of $x_t$, which the walk has passed, for the first of them, and the region of the block added just before for the others. The regions $Q_L$ and $Q_R$ of \lineref{line:replace} follow from the endpoints of $Q$ and of $g$.

\emph{{\sc ChangeRegions}.} Besides its splits and joins, a call must keep the stored keys and the three arrays right. The arrays $\operatorname{first}$ and $\operatorname{last}$ change at the endpoints of the at most three old and three new regions, and $\operatorname{leaf}$ never changes, because splits and joins neither create nor destroy leaves; a cut starts at $\operatorname{leaf}[y]$. For the keys, recall that an internal node has the key of the last leaf in its left subtree: it separates two consecutive leaves and has the key of the first. Three facts keep the keys right in constant time per node.
\begin{itemize}
\item {\bf Nodes that survive.} Splits and joins change them only by rotations, which keep the two leaves that a node separates. So a node that survives keeps its key.
\item {\bf A join.} It creates one node, which separates the last leaf of its first tree from the first leaf of its second, and this node gets the largest key of its first tree.
\item {\bf A split at the leaf of a key $y$.} It rebuilds its two trees by joins up its path. The new node at a path node separates the same two leaves, so it takes the key of the path node. Two path nodes are not needed for this, the lowest one at which the path turns right and the lowest one at which it turns left, with the key before $y$ and the key $y$: they separate the leaf of $y$ from its two neighbours. Their keys are the keys of the new nodes when a cut joins the leaf of $y$ back.
\end{itemize}

\emph{The total and the initialization.} Since $\lambda(t)\le K$, every search takes $O(\log n)$ time, and all searches together take $O(m+\UB(X))$ time by \cref{thm:lambda-ub}. The initial inner tree is built by joining the leaves in pairs, the resulting trees in pairs, and so on; for every $j\ge1$ there are at most $\ceil{n/2^j}$ joins of trees with at most $2^j$ keys, each in $O(j)$ time, so this takes
\begin{align*}
\sum_{j\ge1}\Bigl\lceil\frac n{2^j}\Bigr\rceil\cdot O(j)=O(n)
\end{align*}
time, where the sum ends at $j=K$. The arrays, the counters and the lists take $O(n)$ time and words.
\end{proof}

\section{Insertions and deletions: the algorithm and the proof}
\label{app:updates}

This appendix proves \cref{thm:updates}. The algorithm is that of \cref{sec:algorithm}, with the blocks and the time of a block changed as in \cref{sec:upd-idea}, and the proof has the same three parts as for searches. First, the repairs keep every block at the right size, and they are rare (\cref{sec:upd-blocks}). Second, the set $\mathcal F$ that the algorithm maintains is always the set of the kept blocks, both after an access and after a repair, so that the procedures of \cref{sec:algorithm} can be applied (\cref{sec:upd-time,sec:upd-algorithm}). Third, the cost is bounded, and compared with the unified bound (\cref{sec:upd-cost}). With $N_k$ the number of blocks of level $k$ that the repairs create, the total cost will be
\begin{align*}
\underbrace{\sum_{t=1}^mO\bigl(\lambda(t)\bigr)}_{\text{\cref{lem:upd-cost}}}+\underbrace{\sum_kO(k+1)N_k}_{\text{repairs, \cref{lem:upd-cost}}}\ =\ O\bigl(m+\UB(X)\bigr),
\end{align*}
where the first sum is $O(m+\UB(X))$ by \cref{lem:upd-keyloss} and the second is $O(m)$ by \cref{lem:upd-blocks}. Here $\lambda(t)$ is the locality scale of operation $t$, defined in \cref{def:upd-locality} as for searches.

Until the proof of \cref{thm:updates} we assume that $S_1$ is empty; the proof reduces the general case to this one.

\paragraph{What we use from the earlier sections.} The tree is built from the kept blocks as in \cref{sec:regions}: the \emph{owner} of a key is the lowest kept block that contains it, a \emph{region} is a maximal interval of keys with one owner, the regions are the leaves of the region tree, where a region of height $a$ has the weight $4^{-a}$, and the keys of each region form an inner tree. Four facts are used.
\begin{itemize}
\item {\bf Depth} (\cref{lem:pathdepth}). If the total weight is less than $2$ and a region of height $a$ has $O(2^a)$ keys, then its keys have depth $O(a+1)$.
\item {\bf Changing regions} (\cref{lem:opcost}). Under the same conditions, a call of {\sc ChangeRegions} that replaces at most three consecutive regions by at most three regions with the same union costs $O(a+1)$, where $a$ is the largest height of these regions.
\item {\bf Retiring and adding} (\cref{lem:regions}). {\sc Retire} removes a kept block $g$, and {\sc Add} adds a block $g$ to the kept blocks. Both require that $g$ contain no kept block of a lower height. Then each makes one call of {\sc ChangeRegions}, for regions owned by $g$ and by its \emph{parent} $p$, the lowest kept block above $g$, so each costs $O(h(p)+1)$.
\item {\bf Chains} (\cref{lem:chain}). Let $c\ge2$, let $t$ be a time with a partner and $\tau_t\le J-c$, and let $v<t$ be a time with $v=0$ or $\rho(v,t)\ge2^J-1$, such that every time $r$ between $v$ and $t$ with $\tau_r<J-c$ has a \emph{partner} $p(r)<r$ with $\rho(p(r),r)\le2^{\tau_r}-2$. The \emph{chain} of $t$ follows partners backwards, $s_0\coloneqq p(t)$ and $s_{i+1}\coloneqq p(s_i)$, until the numbers $2^{\tau_{s_i}}$ of its times sum to at least $2^{J-c}$. Then the chain is defined and lies after $v$, and
\begin{align*}
\rho(s,t)<P(s)<2^{J-c+1}\quad\text{for every time $s$ of the chain},\qquad\sum_i\min\bigl(1,2^{\tau_{s_i}+c-J}\bigr)\ge1,
\end{align*}
where $P(s)$ is the sum of $2^{\tau_t}$ and of the numbers $2^{\tau}$ of the times of the chain later than $s$.
\end{itemize}

\subsection{Blocks}
\label{sec:upd-blocks}

This subsection defines the blocks and the repairs that keep their sizes in range, and shows that the repairs are rare. A \emph{level} is a power of two, and $a_k\coloneqq8^k$. The levels take the place of the promoted scales of \cref{sec:algorithm}: a block of level $k$ will have about $8^k=2^{3k}$ keys, so it plays the part of a block of height $3k$. The base $8$ leaves room for a factor $4$ in the sizes of the blocks and a factor $3$ in the count of \cref{lem:upd-recency}. For every level $k$, the \emph{blocks of level} $k$ partition the current set $S$ into intervals of consecutive keys of $S$, and every block of level $2k$ is a union of blocks of level $k$, its \emph{children}; the children of a block of level $1$ are its keys. A level is \emph{trivial} if it has a single block, which then consists of all keys; like the root of a B-tree, this block may have fewer keys than the others. The levels above a trivial level are trivial, and the lowest level that is trivial is the \emph{top level} $\hat K$. Every block $g$ of a level $k$ satisfies
\begin{align}
a_k\le\abs g\le4a_k\quad\text{if level $k$ is not trivial},\qquad\abs g\le4a_k\quad\text{if it is}.
\label{eq:block-sizes}
\end{align}
Two blocks of different levels can consist of the same keys; we still regard them as different blocks. A block of level $k$ has the \emph{height} $3k$. If $\hat K\ge2$, the level $\hat K/2$ has two blocks of at least $a_{\hat K/2}$ keys, so $\hat K=O(\log(\abs S+2))$.

Initially every level has one empty block; only the levels up to the top level ever matter. An inserted key $x$ joins the block of level $1$ of its \emph{anchor}, which is its successor, or the largest key if $x$ has no successor, and $x$ has the blocks of its anchor at all levels. If there is no key, $x$ joins the empty blocks. A deleted key leaves its blocks. After an insertion or a deletion of $x$, the levels are \emph{repaired} in increasing order; at level $k$ only the block $g$ that contains or contained $x$ can violate \eqref{eq:block-sizes}.
\begin{itemize}
\item {\bf Split.} If $\abs g>4a_k$, then $g$ is split into two blocks, between two of its children and as evenly as possible, at the left one of two such places.
\item {\bf Merge.} If $\abs g<a_k$ and level $k$ is not trivial, then $g$ is merged with a neighbouring block of level $k$ in the same block of level $2k$, the left one if there are two, and the merged block is split if it has more than $3a_k$ keys.
\end{itemize}
A split or a merge destroys the blocks that it is applied to and creates new blocks; all other blocks remain the same blocks.

For example, a block of level $1$ has between $8$ and $32$ keys, unless it is the only one. Its $33$rd key splits it into blocks of $16$ and $17$ keys, and neither of them becomes too large or too small before it has gained $16$ keys or lost $9$. This is why splits and merges are rare.

The next lemma is what the analysis needs from the repairs: the sizes stay in range, and the number of blocks that the repairs create at level $k$ falls like $8^{-k}$, so that repairs costing $O(k+1)$ each cost $O(m)$ in all.

\begin{lemma}[Blocks]
\label{lem:upd-blocks}
After every operation, the blocks satisfy \eqref{eq:block-sizes}, and every block of level $2k$ is a union of blocks of level $k$. If $N_k$ is the number of blocks of level $k$ that the splits and merges create, then $\sum_k(k+1)N_k=O(m)$.
\end{lemma}

\begin{proof}
A block is repaired as soon as its size leaves the range \eqref{eq:block-sizes}, and the repair creates blocks well inside the range, which must gain or lose many keys before they are repaired again.

\emph{Sizes.} Consider a repair at level $k$, of a block $g$. Three facts determine the sizes of the new blocks.
\begin{itemize}
\item {\bf Children are small.} A child of a block of level $k$ has at most $a_k/2$ keys: it is a key, or a block of level $k/2$ with at most $4a_{k/2}\le a_k/2$ keys. So the two parts of a split, which cuts between two children as evenly as possible, differ by at most $a_k/2$ keys.
\item {\bf The size of $g$.} When level $k$ is repaired, $\abs g$ has changed by one. So $\abs g=4a_k+1$ in a split and $\abs g=a_k-1$ in a merge.
\item {\bf A merge has a neighbour.} Let $p$ be the block of level $2k$ around $g$. If level $2k$ is not trivial, then $\abs p\ge a_{2k}-1>\abs g$, because level $2k$ is repaired after level $k$, so that $p$ has lost at most the one key of the operation. If level $2k$ is trivial, $p$ consists of all keys, and level $k$ has a block other than $g$. In both cases $g$ has a neighbour $r$ of level $k$ inside $p$. The operation has not changed $r$, so $a_k\le\abs r\le4a_k$.
\end{itemize}
Hence the new blocks have the following numbers of keys.
\begin{align*}
\text{parts of a split of $g$:}&\quad\tfrac12(4a_k+1)\pm\tfrac14a_k,&&\tfrac74a_k\ \text{to}\ \tfrac94a_k+1,\\
\text{a merged block:}&\quad(a_k-1)+\abs r,&&2a_k-1\ \text{to}\ 5a_k-1,\\
\text{parts of a merged block that is split:}&\quad\text{half of it}\pm\tfrac14a_k,&&\tfrac54a_k\ \text{to}\ \tfrac{11}4a_k.
\end{align*}
The last column gives the least and the largest possible size. A merged block is split only if it has more than $3a_k$ keys, so one that is not split has at most $3a_k$ keys. So all new blocks satisfy \eqref{eq:block-sizes}. The nesting is kept, because a split cuts between two children and a merge stays inside one block of level $2k$.

\emph{Count.} By the three lines above, the size of a block of level $k$ satisfies
\begin{align*}
\text{when $b$ is created and not split at once:}&\quad\tfrac54a_k\le\abs b\le3a_k,\\
\text{when $b$ is the block $g$ of a repair:}&\quad\abs b\in\{a_k-1,\ 4a_k+1\}.
\end{align*}
So more than $a_k/4$ insertions and deletions change the size of $b$ in between. The same holds for the empty block of the level at the start, which is repaired only at $4a_k+1$ keys. An insertion or a deletion changes the size of one block of level $k$. Every repair has one block $g$, where a merge and the split of its merged block count as one repair, and a repair creates at most three blocks. Hence
\begin{align*}
\#\{\text{repairs at level }k\}<\frac m{a_k/4}=\frac{4m}{8^k},\qquad N_k\le\frac{12m}{8^k},\qquad\sum_k(k+1)N_k\le12m\sum_{k\ge1}\frac{k+1}{8^k}=O(m).
\end{align*}
\end{proof}

\subsection{The time of a block}
\label{sec:upd-time}

The tree keeps, at each level, the blocks whose keys were used most recently. This subsection defines when a key is used and the time of a block, its \emph{recency}, and proves the four facts about recencies on which the rest relies.

Throughout, a statement about the time $t$ refers to the moment before operation $t$. The \emph{blocks of} $x_t$ are the blocks that contain $x_t$, or for an insertion the blocks that $x_t$ joins. Operation $t$ \emph{accesses} one key, unless it deletes the only key. A search or an insertion accesses $x_t$. A deletion accesses the \emph{heir} of $x_t$: its predecessor if this key lies in the block of level $1$ of $x_t$, and its successor otherwise. One of the two lies there, because a block of level $1$ has at least $8$ keys or consists of all keys. So the accessed key lies in the blocks of $x_t$ and is present after the operation. The \emph{access time} $\mathrm{acc}(y)$ of a present key $y$ is the time of the last operation that accessed $y$ since $y$ was last inserted, and $0$ before the first such operation, and the \emph{recency} of a block $g$ is
\begin{align*}
\theta(g)\coloneqq\max\{\mathrm{acc}(y):y\in g\},
\end{align*}
or $0$ if $g$ is empty. While $S_1$ is empty, every present key was inserted, and its insertion accessed it, so every block that is not empty has a positive recency. The heir is what keeps a recency from falling. Suppose that a deletion accessed nothing, and that the last operations deleted keys of one block. Then the block would have the recency of some older key, as if these operations had never touched it, and the tree would have to stop keeping a block that is in use.

The next lemma collects these facts. Parts 1 and 2 say that the blocks of a level are ranked by their recency as in a cache with the least-recently-used rule, and that the ranking is compatible with the nesting of the levels. Part~3 bounds the number of blocks that compete with a recently used block, which keeps the block of $x_t$ in the tree at the levels from $\lambda(t)/3$ on. Part~4 says that the borders of a block move only with the block itself, which is what the comparison with the unified bound needs.

\begin{lemma}[Recency]
\label{lem:upd-recency}
\leavevmode
\begin{itemize}
\item {\bf Part 1 (order).} If a block $g$ lies in a block $g'$, then $\theta(g)\le\theta(g')$. Two blocks of one level with positive recency have different recencies.
\item {\bf Part 2 (one operation).} Unless operation $t$ deletes the only key, it gives the recency $t$ to the blocks of $x_t$ and changes the recency of no other block that exists before and after it. So a recency does not decrease while its block exists and the set is not empty.
\item {\bf Part 3 (few recent blocks).} Let $v<t$. Before operation $t$, at most $3\rho(v,t)$ blocks of a level have a recency larger than $v$.
\item {\bf Part 4 (keys stay on their side).} Let $g$ be a block that exists and is not empty from a time $s$ to a time $t$, and let $y$ be a key that is present from $s$ to $t$. Then $y$ is in $g$ at all of these times or at none. In the second case, $y$ is smaller than all keys of $g$ at all of these times, or larger than all keys of $g$ at all of these times.
\end{itemize}
\end{lemma}

\begin{proof}
\emph{Part 1.} A maximum over more keys is not smaller. An operation accesses one key, so two present keys with positive access times have different access times, and the blocks of a level are disjoint.

\emph{Part 2.} Let $y$ be the key that operation $t$ accesses. The operation changes the access times and the blocks in two ways only. It gives the access time $t$ to $y$, which lies in the blocks of $x_t$ and is present afterwards. And $x_t$ joins or leaves its blocks, if the operation is an insertion or a deletion.
So a block of $x_t$ has the recency $t$ afterwards, because it contains $y$ and no access time exceeds $t$. Another block that exists before and after the operation has the same keys with the same access times.

\emph{Part 3.} A block with a recency after $v$ holds a key that was accessed after $v$, and different blocks hold different keys. An accessed key is the key of an operation, or the heir of a deleted key, which is the nearest remaining key on one side of it. So every key of an operation accounts for at most three accessed keys: itself and one on each side. Precisely, let
\begin{align*}
D\coloneqq\{x_r:v<r<t\},\qquad\text{so that }\abs D=\rho(v,t),
\end{align*}
and for $z\in D$ let $L(z)$ and $R(z)$ be the next smaller and the next larger key than $z$ among the keys outside $D$ that are present before operation $v+1$. Let $b$ be a block of the level with $\theta(b)>v$, and let $y_b\in b$ be the key that was accessed at the time $u\coloneqq\theta(b)$, where $v<u<t$. We show that
\begin{align}
y_b\in D\cup\{L(z),R(z):z\in D\}.
\label{eq:upd-three}
\end{align}
The set on the right has at most $3\abs D$ keys, and different blocks $b$ have different keys $y_b$, which proves Part~3. For \eqref{eq:upd-three}, suppose that $y_b\notin D$. Then operation $u$ is a deletion and $y_b$ is the heir of $x_u\in D$, a neighbour of $x_u$ in $S_u$. No operation strictly between $v$ and $t$ inserts or deletes a key outside $D$. So
\begin{align*}
\{\text{keys outside $D$ that are present before operation }v+1\}=S_u\setminus D,
\end{align*}
and $y_b$ belongs to this set. A neighbour of $x_u$ in $S_u$ that lies in $S_u\setminus D$ is the nearest key of $S_u\setminus D$ on its side of $x_u$, so $y_b=L(x_u)$ or $y_b=R(x_u)$.

\emph{Part 4.} An inserted key joins the block of the next larger key, so it cannot push a present key across the border of a block. Precisely, a present key enters or leaves $g$ only when $g$ is created or destroyed, so $y$ is in $g$ at all of these times or at none. In the second case, let $z$ be a key that is inserted between $s$ and $t$, with anchor $q$.
\begin{itemize}
\item {\bf $y<\min g$ and $z<y$.} Then $q\le y$, so $q\notin g$, and $z$ does not join $g$.
\item {\bf $y>\max g$ and $z>y$.} Then $q>y$, or $q$ is the largest key, so again $q\notin g$, and $z$ does not join $g$.
\item {\bf Otherwise} $z$ lies on the same side of $y$ as $g$, so it does not change the side of $y$ even if it joins $g$.
\end{itemize}
A deletion does not change the side of $y$ either.
\end{proof}

\begin{definition}[Locality scale of an operation]
\label{def:upd-locality}
Let $v_k(t)$ be the recency of the block of $x_t$ of level $k$ before operation $t$. The \emph{locality scale} of operation $t$ is $\lambda(t)\coloneqq3k$ for the least level $k$ with
\begin{align*}
v_k(t)\ge1\qquad\text{and}\qquad\rho\bigl(v_k(t),t\bigr)+2\le8^k,
\end{align*}
and $\lambda(t)\coloneqq3$ if $S_t$ is empty.
\end{definition}

In words, $\lambda(t)/3$ is the least level at which the block of $x_t$ was accessed recently, where recently means that at most $8^k-2$ distinct keys had an operation since. This is \cref{def:locality} with the levels in place of the scales. If $S_t$ is not empty, every trivial level satisfies the condition, because its block contains the key that operation $t-1$ accessed. So $\lambda(t)$ is defined, and $\lambda(t)\le3\hat K_t$ for the top level $\hat K_t$ before operation $t$. By Part~1 of \cref{lem:upd-recency}, the condition then holds at every level from $\lambda(t)/3$ on.

\subsection{The algorithm}
\label{sec:upd-algorithm}

This subsection defines the kept blocks, states the algorithm, and proves that the set $\mathcal F$ that it maintains follows the kept blocks (\cref{lem:upd-kept}). An access changes the kept blocks as a search does in \cref{sec:algorithm}; what is new are the repairs.

For a level $k$ that is not trivial, the \emph{recency list} of level $k$ lists the blocks of level $k$ with a positive recency, by decreasing recency; by Part~1 of \cref{lem:upd-recency} their recencies are distinct. The \emph{kept blocks} of level $k$ are the first
\begin{align*}
C_k\coloneqq3\cdot8^k
\end{align*}
blocks of this list, or all of them if there are fewer. The block of a trivial level is kept at all times. A level is \emph{missing} for operation $t$ if the block of $x_t$ of this level is not kept before the operation. The access then moves this block to the front of the recency list, so it becomes kept, and if $C_k$ blocks were kept, the last of them is no longer kept; we call it the \emph{victim} of the level. Owners, regions, the region tree, the inner trees, the realization and the weights are those of \cref{sec:regions}, for the keys of $S$ and the heights above, with the largest key at the last leaf. The blocks of the levels above the top level own no keys.

\Cref{alg:update} processes an operation. Its first three phases are those of \cref{alg:search}, with the access in place of the search. \Lineref{line:upd-add} and \lineref{line:upd-remove} add or remove one leaf of an inner tree and change no recency. An insertion splits the inner tree at the leaf of the anchor $q$ and joins the two resulting trees, the leaf of $q$ and the new leaf, in key order, by three joins. A deletion splits the inner tree at the leaf of $x_t$ and joins the two resulting trees. As for the recencies, an inserted key has the access time $0$ until \lineref{line:upd-access}, and a deleted key leaves after its heir has received the time $t$. A region is a union of blocks of level $1$, so it does not become empty unless the set does, and the empty set has one region, whose inner tree has no leaf until a key is inserted. Only the levels up to the top level take part in an operation. The fourth phase repairs the blocks, and $\mathcal F$ must follow.

\paragraph{A split.} Let $g$ be split into $g_1$, which has the key that the operation accessed, and $g_2$. The obvious repair is to retire $g$ and to add its two parts. This does not work: {\sc Retire} needs a block that contains no kept block of a lower level, and for $k\ge2$ the block $g$ contains the lower blocks of the accessed key, which are kept. So $g$ is not retired. Its two parts take its place in $\mathcal F$ (\lineref{line:upd-split-F}). A key that $g$ owned is then owned by its part, at the same height, so the regions stay as they are, except that a region of $g$ that reaches across the cut is divided there (\lineref{line:upd-cut}). It remains to bring $\mathcal F$ back to the kept blocks. In the recency list of level $k$, the split has the effect
\begin{align*}
(g,b_2,b_3,\dots)\quad\longrightarrow\quad(g_1,b_2,\dots,b_j,\ g_2,\ b_{j+1},\dots):
\end{align*}
the part $g_1$ takes the place of $g$ at the front, and $g_2$ enters at the place of its own recency, or not at all if its recency is $0$. The kept blocks are the first $C_k$ blocks of the list. So at most one block of $\mathcal F$ is no longer kept: $g_2$, if it is not among the first $C_k$, or else the block that $g_2$ has pushed out of them. This block retires (\lineref{line:upd-split-retire}).

\paragraph{A merge.} Let $g$ be merged with $r$ into $\hat g$. This is the mirror image. If $r$ is not kept, it is added first (\lineref{line:upd-merge-add}), so that both blocks are in $\mathcal F$. Then $\hat g$ takes their place (\lineref{line:upd-merge-F}), the keys that they owned are owned by $\hat g$, and their two regions at the \emph{seam}, the border between $g$ and $r$, become one (\lineref{line:upd-seam}). In the recency list,
\begin{align*}
(g,b_2,\dots,b_j,\ r,\ b_{j+1},\dots)\quad\longrightarrow\quad(\hat g,b_2,\dots,b_j,b_{j+1},\dots):
\end{align*}
the block $\hat g$ takes the place of $g$, and $r$ is gone. So if $r$ was among the first $C_k$ blocks, the block after them moves up, and it is added (\lineref{line:upd-merge-enter}). If the merged block is then split, this is a split as above.

\begin{algorithm}[!t]
\caption{The online BST with insertions and deletions, procedure {\sc Operation}. It replaces the procedure {\sc Search} of \cref{alg:search}.}
\label{alg:update}
\begin{algorithmic}[1]
\DataStructure{OnlineBST} \Comment{continues \cref{alg:search}}
  \Procedure{Operation}{$t$} \Comment{operation $t$, with the key $x_t$}
    \State $M\gets$ the missing levels of operation $t$\label{line:upd-missing}
    \State \PhaseComment{Phase 1: make room at the missing levels}
    \For{each $k\in M$, by increasing $k$}
      \If{$\mathcal F$ has $C_k$ blocks of level $k$}
        \State \Call{Retire}{the victim of the level $k$} \Comment{cost $O(k+1)$}\label{line:upd-retire}
      \EndIf
    \EndFor
    \State \PhaseComment{Phase 2: go to the key}
    \State walk from the root to $x_t$, or along its search path in an insertion \Comment{cost $O(\lambda(t))$}\label{line:upd-walk}
    \If{operation $t$ is an insertion}
      \State $x_t$ joins the blocks of its anchor $q$
      \State add the leaf of $x_t$ to the inner tree of the region of $q$ \Comment{cost $O(\lambda(t))$}\label{line:upd-add}
    \EndIf
    \State \PhaseComment{Phase 3: access and promote}
    \State give the access time $t$ to $x_t$, or to its heir in a deletion\label{line:upd-access}
    \For{each $k\in M$, by decreasing $k$}
      \State \Call{Add}{the block of $x_t$ of level $k$} \Comment{cost $O(\lambda(t))$ for all of them}\label{line:upd-grow}
    \EndFor
    \If{operation $t$ is a deletion}
      \State $x_t$ leaves its blocks
      \State remove the leaf of $x_t$ from the inner tree of its region \Comment{cost $O(\lambda(t))$}\label{line:upd-remove}
    \EndIf
    \State \PhaseComment{Phase 4: repair the blocks; every split or merge at level $k$ costs $O(k+1)$}
    \For{each level $k$, by increasing $k$, and each split or merge at level $k$}
      \If{it splits $g$ into $g_1$ and $g_2$}
        \State $\mathcal F\gets(\mathcal F\setminus\{g\})\cup\{g_1,g_2\}$\label{line:upd-split-F}
        \State \Call{ChangeRegions}{the region that $g$ owned across the cut, its two parts} \Comment{if any}\label{line:upd-cut}
        \State \Call{Retire}{the block of level $k$ in $\mathcal F$ that is not kept} \Comment{if there is one}\label{line:upd-split-retire}
      \Else \Comment{it merges $g$ and $r$ into $\hat g$}
        \State \Call{Add}{$r$} \Comment{if $r\notin\mathcal F$}\label{line:upd-merge-add}
        \State $\mathcal F\gets(\mathcal F\setminus\{g,r\})\cup\{\hat g\}$\label{line:upd-merge-F}
        \State \Call{ChangeRegions}{the two regions at the seam, their union} \Comment{if $g$ and $r$ owned them}\label{line:upd-seam}
        \State \Call{Add}{the kept block of level $k$ that is not in $\mathcal F$} \Comment{if there is one}\label{line:upd-merge-enter}
      \EndIf
    \EndFor
  \EndProcedure
\EndDataStructure
\end{algorithmic}
\end{algorithm}

The next lemma is the counterpart of \cref{lem:kept}. Part~1 is the nesting of the kept blocks, Part~2 bounds the missing levels, and Part~3 says that $\mathcal F$ follows the kept blocks through accesses and repairs, so that {\sc Retire} and {\sc Add} are always applied as \cref{lem:regions} requires. Here $\mathcal F$ is the set that \cref{alg:update} maintains, and whether a block is kept is always judged by the blocks and the access times of the moment in question. Initially $\mathcal F$ consists of the kept blocks, which are the empty blocks of all levels; only the levels up to the top level ever take part in an operation.

\begin{lemma}[Kept blocks]
\label{lem:upd-kept}
\leavevmode
\begin{itemize}
\item {\bf Part 1 (nesting).} At every moment, the blocks of the higher levels around a kept block are kept.
\item {\bf Part 2 (missing levels).} The missing levels of operation $t$ are the levels below some bound, and they are less than $\lambda(t)/3$.
\item {\bf Part 3 (the steps).} The set $\mathcal F$ consists of the kept blocks when an operation begins, after its third phase, and after the lines of every split or merge. Every block that \cref{alg:update} retires or adds contains, at that moment, no block of $\mathcal F$ of a lower level. At every moment, $\mathcal F$ contains the blocks of the higher levels around each of its blocks and has at most $C_k+1$ blocks of each level $k$. In a split or a merge, the replacement of the blocks changes the regions only as \lineref{line:upd-cut} and \lineref{line:upd-seam} say.
\end{itemize}
\end{lemma}

\begin{proof}
As for searches, the kept blocks of each level form a cache with the least-recently-used rule, now of $C_k$ blocks. An access changes the recency lists exactly as a search does in \cref{lem:kept}, so the first three phases need no new argument. A split or a merge at level $k$ changes one recency list only locally, and as a result it changes the kept blocks of level $k$, apart from the blocks it destroys and creates, by at most one block. The lines of the repair replace the destroyed blocks and then retire or add this one block.

\emph{Part 1.} Let $g$ be a kept block of level $k$, and let $g'$ be the block of a higher level $k'$ around it. If level $k'$ is trivial, $g'$ is kept. Otherwise level $k$ is not trivial either, so $g$ is in its recency list, with $\theta(g)>0$. For a block $b$ with $\theta(b)>0$, let
\begin{align*}
\operatorname{Pre}(b)\coloneqq\{c:c\text{ is a block of the level of $b$ and }\theta(c)>\theta(b)\}
\end{align*}
be the set of the blocks that precede $b$ in its recency list, so that a block $b$ of a level $j$ that is not trivial is kept if and only if $\abs{\operatorname{Pre}(b)}<C_j$. We show that $g'$ has no more blocks before it than $g$ has, by mapping $\operatorname{Pre}(g')$ into $\operatorname{Pre}(g)$. For $c\in\operatorname{Pre}(g')$, let $y_c\in c$ be the key with the access time $\theta(c)$, and let $\iota(c)$ be the block of level $k$ that contains $y_c$. It lies in $c$ and contains $y_c$, so its recency is exactly $\theta(c)$, and by Part~1 of \cref{lem:upd-recency},
\begin{align*}
\theta\bigl(\iota(c)\bigr)=\theta(c)>\theta(g')\ge\theta(g),
\end{align*}
that is, $\iota(c)\in\operatorname{Pre}(g)$. The blocks of $\operatorname{Pre}(g')$ are disjoint, so $\iota$ is injective, and
\begin{align*}
\abs{\operatorname{Pre}(g')}\le\abs{\operatorname{Pre}(g)}<C_k\le C_{k'}.
\end{align*}
Since $\theta(g')\ge\theta(g)>0$, the block $g'$ is in its list, so it is kept.

\emph{Part 2.} The first claim follows from Part~1. Let $k\ge\lambda(t)/3$ be a level; we show that it is not missing. If it is trivial, the block of $x_t$ of level $k$ is kept. Otherwise $S_t$ is not empty and level $k$ satisfies the condition of \cref{def:upd-locality}. So the block of $x_t$ has the recency $v_k(t)\ge1$, and by Part~3 of \cref{lem:upd-recency} the number of blocks of level $k$ that precede it in the recency list is at most
\begin{align*}
3\rho\bigl(v_k(t),t\bigr)\le3\cdot(8^k-2)<C_k.
\end{align*}
So the block of $x_t$ is kept.

\emph{Part 3, the first three phases.} If the operation deletes the only key, all levels are trivial before and after it, and $\mathcal F$ does not change. Otherwise, by Part~2 of \cref{lem:upd-recency}, the access moves the blocks of $x_t$ to the front of their recency lists and changes nothing else; adding or removing the leaf of $x_t$ changes no recency. So at a missing level the block of $x_t$ becomes kept, and the victim ceases to be kept if $C_k$ blocks were kept. These are the changes of Part~4 of \cref{lem:kept}, with the levels in place of the promoted scales, and the first and third phases make them as in \cref{alg:search}. The proof of Part~5 of \cref{lem:kept} carries over word for word, with the following reading. The family $\mathcal F_t$ is the set of the kept blocks before the access and $\mathcal F_{t+1}$ the set after it; both are closed by Part~1. The set $V$ consists of the victims of the missing levels, $A$ of the blocks of $x_t$ at the missing levels, and $U$ of the other kept blocks. Facts (V) and (A) of that proof use only that $\mathcal F_t$ and $\mathcal F_{t+1}$ are closed, that $\mathcal F_{t+1}=U\cup A$, and that the blocks of $x_t$ at the levels that are not missing are kept and are not victims, all of which hold here. That $x_t$ joins or leaves its blocks during the operation changes neither which blocks exist nor how they are nested. So the proof gives the claims for these phases, with at most $C_k$ blocks of level $k$.

\emph{Part 3, a split or a merge at level $k$.} Let $\mathcal K^-$ and $\mathcal K^+$ be the sets of the kept blocks before and after the split or merge, which we call the \emph{event}. The proof has five steps.

\emph{(i) The kept blocks.} We show that
\begin{align}
\text{split:}\quad\mathcal K^+\cup E=\bigl(\mathcal K^-\setminus\{g\}\bigr)\cup\{g_1,g_2\},\qquad\text{merge:}\quad\mathcal K^+=\bigl(\mathcal K^-\setminus\{g,r\}\bigr)\cup\{\hat g\}\cup E,
\label{eq:upd-kept-change}
\end{align}
where $E$ is a set of at most one block of level $k$, which is not kept after a split and not kept before a merge. Only blocks of level $k$ are destroyed or created, so the recencies and the kept blocks of the other levels stay the same. At level $k$, the recencies change as follows.
\begin{itemize}
\item The block $g$ contains, or contained, $x_t$, so it contains the key that the operation accessed and is at the front of its list, with the recency $t$.
\item The new block that contains this key, $g_1$ or $\hat g$, has the recency $t$ and takes the front.
\item The part $g_2$ has the largest access time of its keys, and every other block of level $k$ keeps its recency.
\end{itemize}
This gives the two displays before the lemma. The kept blocks are the first $C_k$ blocks of the list, so \eqref{eq:upd-kept-change} holds with the following set $E$.
\begin{itemize}
\item {\bf Split.} If $g_2$ is not among the first $C_k$ blocks of the new list, then $E=\{g_2\}$. Otherwise $E$ consists of the block that $g_2$ has pushed out of them, if there is one.
\item {\bf Merge.} If $r$ was among the first $C_k$ blocks, then $E$ consists of the block that moves up into them, if there is one. Otherwise $E=\emptyset$.
\end{itemize} At a trivial level, where there is no recency list, \eqref{eq:upd-kept-change} holds with $E=\emptyset$: after a split of the only block, both parts are kept because $C_k\ge2$, and a merge of the last two blocks, which are both kept, gives the only block of its level.

\emph{(ii) The set $\mathcal F$.} By induction, $\mathcal F=\mathcal K^-$ before the event. The lines of the event then change $\mathcal F$ as follows, where $r$ is added at \lineref{line:upd-merge-add} only if $r\notin\mathcal K^-$, and the last step is void if $E=\emptyset$:
\begin{align}
\begin{split}
&\text{split:}\qquad\mathcal K^-\ \xrightarrow{\ \text{\lineref{line:upd-split-F}}\ }\ \mathcal K^+\cup E\ \xrightarrow{\ \text{\lineref{line:upd-split-retire}: retire }E\ }\ \mathcal K^+,\\
&\text{merge:}\qquad\mathcal K^-\ \xrightarrow{\ \text{\lineref{line:upd-merge-add}}\ }\ \mathcal K^-\cup\{r\}\ \xrightarrow{\ \text{\lineref{line:upd-merge-F}}\ }\ \mathcal K^+\setminus E\ \xrightarrow{\ \text{\lineref{line:upd-merge-enter}: add }E\ }\ \mathcal K^+.
\end{split}
\label{eq:upd-kept-path}
\end{align}
So $\mathcal F=\mathcal K^+$ after the event. Each family in \eqref{eq:upd-kept-path} has at most one block of level $k$ more than $\mathcal K^-$ or $\mathcal K^+$, so at most $C_k+1$.

\emph{(iii) The blocks that are retired or added contain no block of $\mathcal F$ of a lower level.} These blocks are the block of $E$ and $r$. Each of them is not kept at the moment in question: the block of $E$ after a split is not in $\mathcal K^+$, and $r$ and the block of $E$ in a merge are not in $\mathcal K^-$. The blocks of $\mathcal F$ of lower levels are kept, both before and after the event, because the lower levels do not change. A block that contains a kept block of a lower level is kept, by Part~1. So a block that is retired or added contains none.

\emph{(iv) Nesting.} We show that every family in \eqref{eq:upd-kept-path} contains, with each of its blocks, the blocks of the higher levels around it. For $\mathcal K^-$ and $\mathcal K^+$ this is Part~1. For the three others:
\begin{itemize}
\item $\mathcal K^+\setminus E$ arises from $\mathcal K^+$ by removing a block that contains no block of the family of a lower level, by (iii).
\item $\mathcal K^-\cup\{r\}$ and $\mathcal K^+\cup E$ with $E=\{g_2\}$ arise by adding $r$ or $g_2$. The block of level $2k$ around it is the block around $g$, which is kept because $g$ is (Part~1), and so are the higher blocks around it.
\item $\mathcal K^+\cup E$, where the block of $E$ was pushed out, arises by adding a block that was kept before the event, and so were the blocks around it.
\end{itemize}
In the last two cases the blocks around the added block are kept before the event, and the blocks of the levels above $k$ do not change.

\emph{(v) The regions.} The blocks $g_1$, $g_2$ and $\hat g$ have the height of $g$. So after a split, a key that $g$ owned is owned by the part that contains it. After a merge, a key that $g$ or $r$ owned is owned by $\hat g$; here we use that $r\in\mathcal F$ before \lineref{line:upd-merge-F}. Every other key keeps its owner. Hence, after a split, the only change of the regions is that a region of $g$ with keys in both parts is divided at the cut, and after a merge, that the two regions at the seam become one if $g$ and $r$ owned them.
\end{proof}

By Part~3, the parent of a block of level $k$ in $\mathcal F$ is always the block of level $2k$ around it, of height $6k$, as in \eqref{eq:parent}, and {\sc Retire} and {\sc Add} are applied only as \cref{lem:regions} requires.

\subsection{The cost and the unified bound with updates}
\label{sec:upd-cost}
\label{sec:upd-unified}

This subsection fills the two slots of the budget at the start of the appendix. \Cref{lem:upd-cost} bounds the cost of an operation by its locality scale and the levels of its repairs, and \cref{lem:upd-keyloss} bounds the sum of the locality scales by the unified bound.

\begin{lemma}[Cost of an operation]
\label{lem:upd-cost}
Operation $t$ costs $O(\lambda(t))$ for its first three phases and $O(k+1)$ for every split or merge at a level $k$, and it costs $O(\log(n_t+2))$ in all.
\end{lemma}

\begin{proof}
The budget is that of a search in \cref{thm:scales}, plus the leaf of $x_t$ and the repairs:
\begin{align*}
\underbrace{\sum_{k\text{ missing}}O(k+1)}_{\text{retirements and additions}}+\underbrace{O(\lambda(t))}_{\text{walk}}+\underbrace{O(\lambda(t))}_{\text{leaf of $x_t$}}+\underbrace{\sum_{\text{splits, merges}}O(k+1)}_{\text{repairs}}.
\end{align*}
The tools of \cref{sec:analysis} apply, with regions at most five times as large. They need two bounds, on the total weight $W$ of the regions and on the size of a region $Q$ of height $3k$:
\begin{align*}
W\overset{(a)}{\le}2\sum_k(C_k+1)\cdot4^{-3k}\le2\sum_k4\cdot8^k\cdot4^{-3k}=8\sum_{j\ge0}8^{-2^j}<2,\qquad\abs Q\overset{(b)}{<}5a_k=5\cdot2^{3k}.
\end{align*}
Here (a) is the proof of \cref{lem:weights}, with the at most $C_k+1$ blocks of height $3k$ in $\mathcal F$ (Part~3 of \cref{lem:upd-kept}), and (b) holds because $Q$ lies in a block of level $k$, which has fewer than $5a_k$ keys at every moment, also when it is a merged block before its split. So \cref{lem:pathdepth,lem:opcost} hold with $5\cdot2^a$ in place of $2^a$. Their proofs use only that the leaves of the region tree are the regions of one family of blocks before the call and of another after it, both of total weight less than $2$ and with regions of the stated size, where the heights of the old regions are read in the first family and those of the new regions in the second. This also holds for the calls of {\sc ChangeRegions} in \lineref{line:upd-cut} and \lineref{line:upd-seam}, with the family $\mathcal F$ before and after \lineref{line:upd-split-F} or \lineref{line:upd-merge-F}. Hence a call of {\sc ChangeRegions} for regions of height at most $a$ costs $O(a+1)$, and retiring or adding a block of level $k$ costs $O(k+1)$, because its parent has the height $6k$.
\emph{Retirements and additions.} The missing levels are powers of two below $\lambda(t)/3$ (Part~2 of \cref{lem:upd-kept}), so they sum to less than $\lambda(t)$, and as in the proof of \cref{lem:amortized} these steps cost $O(\lambda(t))$.

\emph{The walk.} The level $\lambda(t)/3$ is not missing, and the first phase retires no block of $x_t$. So during the walk the region of $x_t$, or of its anchor $q$ in an insertion, has a height of at most $\lambda(t)$, and its keys have depth $O(\lambda(t))$. In an insertion no key lies between $x_t$ and $q$, so a search for $x_t$ and a search for $q$ turn the same way at every node, and the search path of $x_t$ is the path to the leaf of $q$ in the expanded tree.

\emph{The leaf of $x_t$.} This is the one step that changes the set of keys. The model of \cref{sec:upd-model} asks for a top set $U$ that contains the search path of $x_t$ in an insertion, and the node of $x_t$ and that of its predecessor in a deletion. We take
\begin{align*}
U\coloneqq{}&\{\text{nodes that the split and the joins visit}\}\\
{}\cup{}&\{\text{nodes on the path from the root to the leaf of $x_t$, or of $q$ in an insertion}\},
\end{align*}
where the path is in the expanded tree and starts at the root of the region tree. If the largest key is deleted, $U$ also contains the path to the leaf of its predecessor, which becomes the last leaf. We check the size of $U$, that nothing outside $U$ changes, and that $U$ is as the model asks.

The set $U$ has $O(\lambda(t))$ nodes. Adding or removing the leaf is a split and at most three joins in an inner tree of fewer than $5\cdot2^{\lambda(t)}$ leaves, so it takes $O(\lambda(t))$ time and visits as many nodes. The paths have $O(\lambda(t))$ nodes by the bound on the walk; for the predecessor of the largest key this holds because it lies in the same region.

Nothing outside $U$ changes, for two reasons. The split and the joins change children only at nodes that they visit. And a node holds the key of the last leaf in its left subtree, so the only nodes whose key changes are those that hold the key $x_t$ or the key of its predecessor, and the node of a key lies above the leaf of this key, on the paths in $U$.

Finally, $U$ is a top set by \cref{lem:wc-visited}, because the visited nodes form a top set of the inner tree and the paths contain all nodes above it. In an insertion it contains the search path of $x_t$. In a deletion it contains the node of $x_t$ and the node of its predecessor, which separates the leaf of $x_t$ from the leaf before it and so lies above both.

\emph{Repairs.} A split or a merge at level $k$ makes one call of {\sc ChangeRegions}, for regions of height $3k$, and retires or adds at most one block of level $k$: in a merge, if $r$ is added at \lineref{line:upd-merge-add}, then $r$ was not among the first $C_k$ blocks, so $E=\emptyset$ and \lineref{line:upd-merge-enter} adds nothing. Each of these costs $O(k+1)$.

For the bound on one operation, a block of a level $k>\hat K_t$ has at most $n_t+1\le4a_k$ keys, so the repairs of operation $t$ are at most one merge and one split at each level up to $\hat K_t$. These levels sum to less than $2\hat K_t$, and $\lambda(t)\le3\hat K_t=O(\log(n_t+2))$.
\end{proof}

The next lemma is the counterpart of \cref{lem:keyloss}, and its proof has the same plan.

\begin{lemma}[Block boundaries]
\label{lem:upd-keyloss}
$\sum_{t=1}^m\lambda(t)=O\bigl(m+\UB(X)\bigr)$.
\end{lemma}

\begin{proof}
We follow the proof of \cref{lem:keyloss}: level by level, an operation with a large locality scale pays for itself or is charged to its chain of partners (\cref{lem:chain}), which stays near a border of its block, and every operation is on at most one chain per level. Two things are new. Operations in a block that a split or a merge has just created are paid for by \cref{lem:upd-blocks}. And the borders move, but a key never crosses the border of a block that survives (Part~4 of \cref{lem:upd-recency}).

\emph{Levels.} By \cref{def:upd-locality}, $\lambda(t)=3k'$ for a level $k'$, and a power of two is one more than the sum of the smaller powers of two, so
\begin{align}
\sum_{t=1}^m\lambda(t)=3m+3\sum_{k}k\cdot\#\{t:\lambda(t)>3k\},
\label{eq:upd-layers}
\end{align}
where $k$ ranges over the levels. Fix a level $k$, let $J\coloneqq3k$, and fix an operation $t$ with $\lambda(t)>J$. Then $S_t$ is not empty, and level $k$ is not trivial before operation $t$, because a trivial level satisfies the condition of \cref{def:upd-locality}. Let $g$ be the block of $x_t$ of level $k$ before operation $t$, and $v\coloneqq v_k(t)$ its recency. The block $g$ is not empty, so $v\ge1$ (\cref{sec:upd-time}). One of three cases holds.
\begin{itemize}
\item {\bf The operation pays for itself:} $\tau_t>J-5$. (The constant $5$ is the slack $c$ with which \cref{lem:chain} is applied below; it is chosen so that the chain stays within $3\cdot2^{J-4}$ keys of a border, less than half a block.) The levels $k$ with $3k<\tau_t+5$ sum to less than $2(\tau_t+5)/3$, so these cases contribute $O(m+\UB(X))$ to \eqref{eq:upd-layers}.
\item {\bf The block is new:} a split or a merge of an operation later than $v$ created $g$. Each block is in this case for at most one operation. Indeed, level $k$ is not trivial before operation $t$, and it becomes trivial only by a merge of its last two blocks, which destroys $g$. An empty set has only trivial levels, so the set is not empty while $g$ exists. So by Part~2 of \cref{lem:upd-recency} the recency of $g$ is at least $t$ after operation $t$ and for as long as $g$ exists. This is later than the creation of $g$, so $g$ is not new for any later operation. Hence, by \cref{lem:upd-blocks}, these cases contribute at most $3\sum_kkN_k=O(m)$.
\item {\bf A crossing:} $\tau_t\le J-5$, $v\ge1$, and $g$ exists from the end of operation $v$ to operation $t$.
\end{itemize}

It remains to charge the crossings. In a crossing, $x_t$ is close to the key of an operation that is not long ago, and yet its block $g$ has not been accessed for a long time. This is possible only if a border of $g$ lies between the two keys; \cref{fig:keyloss} shows the situation for searches, where the border is fixed; here it is the border of a block, and it moves with the keys. We charge the crossing to earlier operations near this border, and we show that no operation is charged twice at one level.

For every operation $r$ whose term is attained by a witness we fix one such witness $u$, its \emph{partner}; then $\rho(u,r)\le2^{\tau_r}-2$, because $\tau_r=\tau(u,r)$. Fix a crossing $t$ of level $k$, with $g$ and $v$ as above. Since $\lambda(t)>J$ and $v\ge1$, \cref{def:upd-locality} gives
\begin{align}
\rho(v,t)\ge2^J-1.
\label{eq:upd-cold}
\end{align}
Three more facts hold for the crossing.
\begin{itemize}
\item {\bf Large blocks.} From the end of operation $v$ to operation $t$, every block of level $k$ has at least $2^J$ keys. Indeed, level $k$ is not trivial at any of these times: otherwise $g$ would be its only block, and it would have to be split for the level to become non-trivial again. So \eqref{eq:block-sizes} gives at least $a_k=2^J$ keys.
\item {\bf Partners.} Every operation $r$ with $v<r\le t$ and $\tau_r\le J-5$ has a partner. Indeed, the set has at least $2^J$ keys at these times, so $\tau_r\le J-5<\log_2(n_r+2)$, and the term of $r$ is attained by a witness.
\item {\bf The chain.} By \cref{lem:chain} with $c=5$, whose hypotheses are \eqref{eq:upd-cold} and the fact about partners, the chain $s_0>\dots>s_{i^*}$ of $t$ is defined and later than $v$, and every operation $s$ of the chain satisfies
\begin{align}
\rho(s,t)<P(s)<2^{J-4}.
\label{eq:upd-chain}
\end{align}
Recall that $P(s)$ is the sum of the numbers $2^{\tau}$ of $t$ and of the operations of the chain that are later than $s$, and that the chain ends when the numbers of its operations reach $2^{J-5}$.
\end{itemize}
Three facts about the chain remain.

\emph{The partner of $t$ lies across a border of $g$.} Operation $u\coloneqq s_0$ accessed $x_u$, which is present until $t$, so
\begin{align*}
\mathrm{acc}(x_u)\ge u>v=\theta(g),\qquad\text{and hence}\qquad x_u\notin g.
\end{align*}
Since $d_t(x_u,x_t)<2^{J-5}$ and every block of level $k$ has at least $2^J$ keys, $x_u$ lies in a block next to $g$. The border of $g$ on this side is the \emph{cut} of the crossing. It has two properties.
\begin{itemize}
\item {\bf The cut exists as long as $g$ does.} At all these times $g$ has a block next to it on this side. Indeed, if $g$ were the last block of its level on this side at some time, it would remain the last one, because a key that is inserted beyond the last block of a level joins this block; but at time $t$ the key $x_u$ lies beyond $g$.
\item {\bf Keys stay on their side.} A key that is present from a time $s>v$ to $t$ is on the same side of the cut at both times (Part~4 of \cref{lem:upd-recency}). An inserted key counts on the side of the block that it joins.
\end{itemize}

\emph{The chain stays near the cut.} The keys of the chain form a path of short hops that starts next to the cut, so they cannot go far from it. Precisely, for every operation $s$ of the chain,
\begin{align}
\#\{\text{keys of $S_s$ between $x_s$ and the cut at time $s$}\}<3P(s)<3\cdot2^{J-4}.
\label{eq:upd-near}
\end{align}
To see this, let $H$ be the set of all keys, present or not, from the smallest to the largest of $x_t$ and the keys of the chain down to $x_s$, and let $w$ be the one of $x_t,x_u$ that lies on the other side of the cut than $x_s$, where the side of $x_s$ is taken at time $s$ and the sides of $x_t$ and $x_u$ at time $t$. A key of $S_s$ between $x_s$ and the cut is deleted between $s$ and $t$, or it stays on its side of the cut until $t$ and so lies between $x_s$ and $w$, in $H$. So the number of these keys is at most
\begin{align*}
\underbrace{\#\{\text{keys deleted between $s$ and $t$}\}}_{<P(s)\ \text{(a)}}+\underbrace{\#(S_s\cap H)}_{<2P(s)\ \text{(b)}}.
\end{align*}
\begin{itemize}
\item {\bf (a)} Every deleted key is the key of an operation between $s$ and $t$, and $\rho(s,t)<P(s)$ by \eqref{eq:upd-chain}.
\item {\bf (b)} The set $H$ is covered by the hops of the chain: the intervals between $x_r$ and $x_{r'}$, where $r$ is $t$ or an operation of the chain later than $s$, and $r'$ is its partner. A hop has at most $d_r(x_{r'},x_r)+1<2^{\tau_r}$ keys of $S_r$, and these bounds sum to $P(s)$. The keys of $S_s$ in a hop that are not in $S_r$ are deleted between $s$ and $t$, and there are fewer than $P(s)$ of these in all, by (a).
\end{itemize}

\emph{An operation is on one chain per level.} Suppose that an operation $s$ is on the chains of two crossings $t_1<t_2$ of level $k$, with blocks $g_1,g_2$ and partners $u_1,u_2$. We show that a key of $g_2$ is accessed at the time $s$ or later, and before $t_2$. By Part~2 of \cref{lem:upd-recency}, the recency of $g_2$ before operation $t_2$ is then at least $s$. This recency is $v_k(t_2)$, and the chain of $t_2$ lies after it, so
\begin{align*}
s\le\theta(g_2)=v_k(t_2)<s,
\end{align*}
a contradiction.

First, the two cuts are the same border. Both blocks exist at time $s$, since $v_k(t_1)<s$ and $v_k(t_2)<s$. By \eqref{eq:upd-near}, each cut has fewer than $3\cdot2^{J-4}$ keys of $S_s$ between itself and $x_s$, so
\begin{align*}
\#\{\text{keys of $S_s$ between the two cuts}\}<2\cdot3\cdot2^{J-4}+1<2^J.
\end{align*}
Two different borders have a whole block of level $k$ between them, which has at least $2^J$ keys. So the cuts are one border. Hence $g_1=g_2$, or $g_1$ and $g_2$ are the two blocks at this border. In the second case they remain the two blocks at this border while both exist, because a key that is inserted between them joins one of them.

Now a key of $g_2$ is accessed. If $g_1=g_2$, operation $t_1$ accesses a key of $g_2$, and $t_1>s$. Otherwise $x_{u_1}$ lies across the cut from $g_1$ at time $t_1$. The cut exists as long as $g_1$ does, and the block on its other side at time $t_1$ is $g_2$, which exists from before $s$ to $t_2$. So $x_{u_1}$ lies in $g_2$, and $\mathrm{acc}(x_{u_1})\ge u_1\ge s$, because the chain of $t_1$ starts at $u_1$ and runs backwards. Both times are before $t_2$.

\emph{Charging.} The operation $s_i$ of the chain of a crossing of level $k$ receives $\min(1,2^{\tau_{s_i}+5-3k})$. By \cref{lem:chain}, these amounts sum to at least $1$ for every crossing, and an operation $s$ receives such an amount at most once per level. In \eqref{eq:upd-layers} the amount is multiplied by $3k$. Let $k_0$ be the least level with $3k_0>\tau_s+5$, so that the levels from $k_0$ on are $2^jk_0$ for $j\ge0$. Then the operation $s$ receives, up to the factor $3$, at most
\begin{align*}
\sum_kk\cdot\min\bigl(1,2^{\tau_s+5-3k}\bigr)\le\underbrace{\sum_{k<k_0}k}_{<k_0}+\underbrace{\sum_{j\ge0}2^jk_0\cdot8^{-(2^j-1)k_0}}_{<2k_0}<3k_0\le2(\tau_s+5).
\end{align*}
The first sum is over powers of two below $k_0$. In the second, $2^{\tau_s+5-3\cdot2^jk_0}<2^{3k_0-3\cdot2^jk_0}$ by the choice of $k_0$, and each term is at most a quarter of the previous one. The last step holds because $k_0=1$, or $3k_0/2\le\tau_s+5$ by the choice of $k_0$. So an operation receives $O(\tau_s+1)$ in all, and \eqref{eq:upd-layers} gives
\begin{align*}
\sum_{t=1}^m\lambda(t)\le3m+\underbrace{\sum_tO(\tau_t+1)}_{\text{pays for itself}}+\underbrace{O(m)}_{\text{new blocks}}+\underbrace{\sum_sO(\tau_s+1)}_{\text{crossings}}=O\bigl(m+\UB(X)\bigr).
\end{align*}
\end{proof}

\begin{proof}[Proof of \cref{thm:updates}]
Let first $S_1$ be empty. The algorithm is \cref{alg:update}. Every split or merge at level $k$ creates a block of level $k$, so by \cref{lem:upd-cost,lem:upd-blocks,lem:upd-keyloss} the total cost is
\begin{align*}
O\Bigl(\sum_{t=1}^m\lambda(t)+\sum_{k}(k+1)N_k\Bigr)=O\bigl(m+\UB(X)\bigr),
\end{align*}
and operation $t$ costs $O(\log(n_t+2))$. For a set $S_1$ of $n_0$ keys, the initial tree is the tree of the algorithm after it has inserted the keys of $S_1$ in increasing order into the empty set. Let $X'$ consist of these $n_0$ insertions followed by $X$. Each of the insertions has the term $1$ in $\UB(X')$, with the insertion before it as a witness. An operation of $X$ has the same witnesses in $X'$ as in $X$, with the same values $\tau(u,t)$, and further ones. So $\UB(X')\le n_0+\UB(X)$, and the algorithm serves $X$ at a cost of $O(n_0+m+\UB(X))$.
\end{proof}

\section*{AI methodology statement}

We used Claude Opus 5.5, Claude Fable 5.1 and Claude Fable 5.2 (Anthropic) and GPT-6 Astra and GPT-5.6 Sol (OpenAI) to assist with literature review, the design and analysis of the data structure, independent checking of the proofs, and manuscript drafting. These tools materially affected all sections. The authors take full responsibility for the contents of this paper.

\PaperEndMatter

\end{document}